\documentclass[11pt]{article}
\pdfoutput=1
\usepackage{epsf,amsmath,amssymb,graphicx,scalefnt,rotating,enumitem,fancyhdr}
\usepackage{array}
\usepackage{filemod}
\usepackage{multirow}
\usepackage[affil-it]{authblk}
\usepackage[usenames,dvipsnames]{color}
\usepackage[pdftex, colorlinks=true, linkcolor=black, filecolor=black, citecolor=black, urlcolor=black, draft=false, bookmarks, bookmarksnumbered=true, plainpages=false, linktocpage={true}]{hyperref}
\usepackage[final]{showlabels}

\usepackage[normalem]{ulem}
\usepackage{cite}
\usepackage[tocflat]{tocstyle}
\usetocstyle{standard}
\newcommand{\pt}{p_T}
\newcommand{\vhnnlo}[1]{{\code vh@nnlo#1}}
\newcommand{\thdmc}{{\code 2HDMC}}
\newcommand{\fh}{{\code FeynHiggs}}
\newcommand{\dy}{\text{\abbrev{DY}}}
\newcommand{\ggzh}{gg\to Z\!H}
\newcommand{\ggzhsmall}{gg\to Z\!h}
\newcommand{\ggzphi}{gg\to Z \phi}
\newcommand{\bbzh}{b\bar{b}\to Z\!H}
\newcommand{\bbzhsmall}{b\bar{b}\to Z\!h}
\newcommand{\bbzphi}{b\bar{b}\to Z \phi}
\newcommand{\higgsstrahlung}{Higgs Strahlung}
\newcommand{\blockentry}[3]{{\tt #1(#2)#3}}
\newcommand{\cp}{\abbrev{CP}}
\newcommand{\dimsix}{dimension-6}
\newcommand{\dimfive}{dimension-5}
\newcommand{\vlq}{\abbrev{VLQ}}
\newcommand{\thdm}{\abbrev{2HDM}}
\newcommand{\citere}[1]{Ref.\cite{#1}}
\newcommand{\citeres}[1]{Refs.\cite{#1}}
\newcommand{\code}{\tt}
\newcommand{\abbrev}[1]{{\scalefont{.9}#1}}

\newcommand{\eqn}[1]{Eq.\,(\ref{#1})}

\newcommand{\fig}[1]{Fig.\,\ref{#1}}
\newcommand{\figs}[1]{Figs.\,\ref{#1}}
\newcommand{\tab}[1]{Table\,\ref{#1}}
\newcommand{\tabs}[1]{Tables\,\ref{#1}}
\newcommand{\sct}[1]{Sect.\,\ref{#1}}
\newcommand{\scts}[1]{Sects.\,\ref{#1}}

\newcommand{\order}[1]{{\cal O}(#1)}
\newcommand{\lhc}{\abbrev{LHC}}
\newcommand{\qcd}{\abbrev{QCD}}
\newcommand{\sm}{\abbrev{SM}}
\newcommand{\mssm}{\abbrev{MSSM}}
\newcommand{\susy}{\abbrev{SUSY}}
\newcommand{\bsm}{\abbrev{BSM}}
\newcommand{\pdf}{\abbrev{PDF}}
\newcommand{\lo}{\abbrev{LO}}
\newcommand{\nlo}{\abbrev{NLO}}
\newcommand{\nnlo}{\abbrev{NNLO}}

\newcommand{\slha}{\abbrev{SLHA}}
\newcommand{\msbar}{\ensuremath{\overline{\mbox{\abbrev{MS}}}}}

\newcommand{\muF}{\mu_\text{F}}
\newcommand{\muR}{\mu_\text{R}}
\newcommand{\wphi}{W\!\phi}
\newcommand{\zphi}{Z\!\phi}
\newcommand{\vphi}{V\!\phi}

\newcommand{\zh}{Z\!h}
\newcommand{\vH}{V\!H}
\newcommand{\wH}{W\!H}
\newcommand{\zH}{Z\!H}

\newcommand{\zA}{Z\!A}
\newcommand{\RHheaderline}{KA-TP-01-2018, TTK-17-47}
\fancypagestyle{firstpage}
{
  
  \fancyhead[R]{\RHheaderline}
}
\title{
\vhnnlo{-v2}: New physics in Higgs Strahlung
}
\author[1]{Robert V. Harlander}
\author[1]{Jonas Klappert}
\author[2]{Stefan Liebler}
\author[1]{Lukas Simon}
\affil[1]{Institute for Theoretical Particle Physics and Cosmology,\protect\\
RWTH Aachen University, D-52066 Aachen, Germany}
\affil[2]{Institute for Theoretical Physics, Karlsruhe Institute of Technology,\protect\\ D-76131 Karlsruhe, Germany}
\begin{document}
\date{}
\maketitle
\thispagestyle{firstpage}
\begin{abstract}
  Introducing version {\tt 2} of the code
  \vhnnlo{}~\cite{vhnnlowebpage}, we study the effects of a number of
  new-physics scenarios on the Higgs-Strahlung process. In particular,
  the cross section is evaluated within a general \thdm\ and the
  \mssm. While the Drell-Yan-like contributions are consistently taken
  into account by a simple rescaling of the \sm\ result, the
  gluon-initiated contribution is supplemented by squark-loop mediated
  amplitudes, and by the $s$-channel exchange of additional scalars
  which may lead to conspicuous interference effects. The latter holds
  as well for bottom-quark initiated \higgsstrahlung{}, which is also
  included in the new version of \vhnnlo.  Using an orthogonal rotation
  of the three Higgs \cp\ eigenstates in the \thdm\ and the \mssm,
  \vhnnlo\ incorporates a simple means of \cp\ mixing in these models.
  Moreover, the effect of vector-like quarks in the \sm{} on the
  gluon-initiated contribution can be studied.  Beyond concrete models,
  \vhnnlo{} allows to include the effect of higher-dimensional operators
  on the production of \cp-even Higgs bosons.  Transverse momentum
  distributions of the final state Higgs boson and invariant mass
  distributions of the $\vphi$ final state for the gluon- and
  bottom-quark initiated contributions can be studied. Distributions for
  the Drell-Yan-like component of \higgsstrahlung{} can be included
  through a link to \texttt{MCFM}.  \vhnnlo{} can also be linked to
  \fh{} and \thdmc{} for the calculation of Higgs masses and mixing
  angles. It can also read these parameters from an \abbrev{SLHA}-file
  as produced by standard spectrum generators.  Throughout the
  manuscript, we highlight new-physics effects in various numerical
  examples, both at the inclusive level and for distributions.
\end{abstract}
\newpage
\tableofcontents

%- }}}
%- {{{ body:

%- {{{ intro:

\section{Introduction}\label{sec:intro}

With the ongoing Run~II of the Large Hadron Collider (\lhc{}), more and
more properties of the Higgs boson $H$ with mass $125$\,GeV, discovered
in 2012~\cite{Aad:2012tfa,Chatrchyan:2012xdj}, are determined with
increasing precision~\cite{Aad:2015zhl,Khachatryan:2016vau}.  An
important process in this respect is \higgsstrahlung{}, i.e.\ the
production of the Higgs boson in association with a gauge boson~$V$,
which was used recently to access for the first time the decay rate
$H\to b\bar{b}$~\cite{Aaboud:2017xsd,Sirunyan:2017elk}, for example.

At tree-level, \higgsstrahlung{} is initiated through light quarks
(Drell-Yan (\dy{})-like contribution). At $\order{\alpha_s^2}$, a
loop-induced, gluon-initiated contribution enters $\zH$ production which
is particularly sensitive to new-physics effects such as modified Yukawa
couplings, new particles in the loop, or resonant additional Higgs
bosons.  In extended Higgs sectors, also the bottom-quark initiated
tree-level sub-process may become important for $\zH$ production.

In order to study such new-physics effects at the level of total as well
as differential cross sections, we extend the code
\vhnnlo{}~\cite{Brein:2012ne}, which in its previous public release only
included \higgsstrahlung{} within the Standard Model (\sm{}).  The main
purpose of \vhnnlo{} so far was the efficient calculation of the total
cross section for \higgsstrahlung{} including next-to-next-to-leading
order (\nnlo{}) quantum chromodynamics
(\qcd{})~\cite{Han:1991ia,Brein:2003wg} and next-to-leading order
(\nlo{}) electro-weak effects~\cite{Ciccolini:2003jy}. The \nnlo{}
\qcd{} corrections to the \dy{}-like terms were calculated by employing
\texttt{ZWPROD}'s implementation of the total \nnlo{} Drell-Yan cross
section~\cite{Hamberg:1990np}.  Soft-gluon resummation effects are small
compared to the \nnlo{} fixed-order result~\cite{Dawson:2012gs}, which
indicates a good convergence of the perturbative series.  The \sm{}
electro-weak effects to the \dy{}-like terms are implemented in terms of
numerical tables, obtained from \citere{Ciccolini:2003jy}.  Also
included are contributions to \dy-like production involving a closed
fermion loop, first calculated in \citere{Brein:2011vx} (see
\sct{sec:sm} for details).  The gluon-initiated contribution~$\ggzh{}$
was originally implemented with the help of {\tt
  FeynArts}~\cite{Hahn:2000kx} and {\tt FormCalc}~\cite{Hahn:1998yk} at
the level of squared amplitudes.  The current implementation, on the
other hand, is at the level of amplitudes: For the \sm, the
2-Higgs-Doublet Model (\thdm), and Minimal Supersymmetric Standard Model
(\mssm), we use the result of Kniehl and Palisoc\,\cite{Kniehl:2011aa};
other new-physics effects have been taken into account with the help of
\texttt{FormCalc}. For the \sm{}, the calculation of the gluon-initiated
contribution is supplemented by the \nlo{} \qcd{} $K$-factor as
described in \citere{Altenkamp:2012sx}.  \citere{Harlander:2014wda}
discussed the resummation of threshold effects for $\ggzh{}$ in the
\sm{}, which reduce the scale uncertainty substantially; they are not
included in the current version of \vhnnlo\ though.

As already reported in \citere{Harlander:2013mla}, \vhnnlo{} has been
extended to study \higgsstrahlung{} $pp\to \wphi$/$\zphi{}$ in a general
\thdm{}, one of the simplest extensions of the \sm{} Higgs sector, see
\citeres{Gunion:1989we,Akeroyd:1996he,Akeroyd:1998ui,Aoki:2009ha,Branco:2011iw,Craig:2012vn}.
Assuming the absence of tree-level flavor-changing neutral currents and
\cp\ violation, the two Higgs doublets form three physical, neutral
Higgs bosons~$\phi$: two \cp{}-even Higgs bosons~$h$ and $H^0$ with
$m_h<m_{H^0}$ and one \cp{}-odd Higgs boson~$A$.  The different ways to
couple the two Higgs doublets to fermions imply four types of \thdm{}s
(see \citere{Harlander:2013mla}, for example).  A brief report on
\higgsstrahlung{} within the \thdm{} is contained in
\citere{Djouadi:2005gj}; more detailed studies can be found in
\citeres{Harlander:2013mla,Hespel:2015zea}.  The \dy{}-like contribution
to the cross section in the \thdm\ can simply be obtained from the \sm{}
result by re-weighting with the corresponding $VV\phi$ couplings. The
gluon-initiated contribution, on the other hand, also depends on the
Yukawa couplings. Moreover, with an increased bottom-quark Yukawa
coupling, the bottom-quark initiated contribution, where the final state
Higgs~$\phi$ couples directly to the initial state bottom-quarks, is of
relevance. In addition, both latter processes involve contributions from
the $s$-channel exchange of a (resonant or virtual) scalar
$\phi'\neq\phi$, which may alter the cross section decisively.  The
program \vhnnlo{} takes them into account, including all interferences.
While these statements also hold for the \mssm, the latter involves
additional effects in the gluon-initiated component due to squark-loops
\cite{Kao:1991xg,Yin:2002sq,Kao:2003jw,Kao:2004vp,Yang:2003kr,
  Li:2005qna,Kniehl:2011aa}, albeit only for $\zA$ production. The
current release of \vhnnlo{} includes those as well.  Furthermore, both
in the \thdm\ and the \mssm, \vhnnlo\ allows for a mixing of the Higgs
\cp{} eigenstates to three neutral mass eigenstates.

The implementation of the $\ggzphi$ component at the amplitude level
clearly facilitates the inclusion of new-physics effects in
\vhnnlo{}. Therefore, aside from the \thdm{} and the \mssm{}, the new
version of \vhnnlo{} currently also includes vector-like quarks
(\vlq{}s) and effective Higgs couplings through \dimsix{} operators. The
latter were checked successfully against earlier work in the framework
of \texttt{MadGraph}\,\cite{Alwall:2014hca,Bylund:2016phk}.  For the
former, we follow the general parametrization of
\citere{Aguilar-Saavedra:2013qpa} of the seven relevant representations
of the \vlq{}s.

Whereas most of our discussion aims at the prediction of the inclusive
cross section for \higgsstrahlung{}, our implementation of the
$\ggzphi{}$ and the $\bbzphi{}$ components also allows for differential
quantities such as the Higgs boson's transverse momentum ($\pt$), and
the $\zH{}$ invariant mass distribution.  Within the \sm{}, fully
differential predictions for \higgsstrahlung{} at \nnlo{} \qcd{} were
provided in \citere{Ferrera:2011bk,Ferrera:2014lca,Campbell:2016jau}.
\nlo\ electro-weak effects were calculated in
\citere{Denner:2011id}. New-physics effects on the boosted regime
($p_T^H\gtrsim 150$\,GeV) for the gluon-initiated contribution were
first discussed in
\citere{Butterworth:2008iy,Englert:2013vua,Harlander:2013mla}.  In order
to allow for the prediction of kinematical distributions for
\textit{all} sub-contributions to \higgsstrahlung{}, the new release of
\vhnnlo{} provides a link to {\tt
  MCFM}~\cite{Campbell:2011bn,Campbell:1999ah,Campbell:2015qma,Boughezal:2016wmq}.

The paper is organized as follows: We start with an outline of the
Higgs-Strahlung process and its partonic sub-processes as they are
implemented in \vhnnlo\ in \sct{sec:setup}.
Section\,\ref{sec:genericoptions} describes the generic settings of
\vhnnlo{} which are relevant for every run of \vhnnlo{}.
Section\,\ref{sec:tmd} focuses on how to obtain kinematic distributions
in \vhnnlo. In \sct{sec:bsm}, we address the newly implemented
new-physics scenarios and consider their numerical effects through
various examples.  \sct{sec:conclusions} contains our
conclusions. Details about the installation and compilation of the code
can be found in Appendix~\ref{sec:installation}, and options for links
to external codes are described in Appendix~\ref{sec:externalcodes}.

%- }}}
%- {{{ section{General structure and SM mode}

\section{General structure and Standard Model mode}
\label{sec:setup}

%- {{{ subsection{Contributions to the cross section}

\subsection{Contributions to the cross section}

The implementation of the total inclusive cross section for
\higgsstrahlung{} in \vhnnlo{} has the form
\begin{align}
\label{eq:XSSM}
\sigma^{\vphi{}}=(1+\delta_\text{\abbrev{EW}}^{\vphi})
\sigma_{\dy{}}^{\vphi{}}+\sigma_{\rm
   I}^{\vphi{}} +\delta_{V\!Z}\left(\sigma_{\rm
   II}^{\zphi{}}+K^\infty_\text{\nlo}\sigma_{\ggzphi{}} +
 \sigma_{\bbzphi{}}\right)\,,
\end{align}
where $\sigma_{\dy{}}^{\vphi{}}$ denotes the $\dy{}$-like terms, whose
leading order diagram is shown in \fig{fig:lodiagrams}~(a).  By
definition, we only count diagrams as \dy-like if they can be obtained
from \fig{fig:lodiagrams}\,(a) by dressing it with real or virtual
gluons and quarks via strong interactions, and possibly crossing them
between the initial and final state.  Consequently, the structure of the
\qcd{} corrections to $\sigma_{\dy{}}^{\vphi{}}$ is the same for both
$\wphi{}$ and $\zphi{}$ production; it is determined by the \qcd{}
corrections to the \dy{} process $q'\overline{q}\to V^*$, provided
through \nnlo{} in \citere{Hamberg:1990np} (see \citere{Brein:2003wg}
for details). The other terms in \eqn{eq:XSSM} will be discussed in more
detail in subsequent parts of this paper.

%- }}}
%- {{{ subsection{Standard Model cross section}

\subsection{Standard Model cross section}\label{sec:sm}

In the \sm\ mode of \vhnnlo, all terms shown in \eqn{eq:XSSM} can be
included.  All but the first and the last one only contribute at \nnlo{}
\qcd{} and beyond, i.e.\ at $\mathcal{O}(\alpha_s^2)$, and involve loops
of top- or bottom-quarks. \eqn{eq:XSSM} assumes that electro-weak
effects
$\delta_\text{\abbrev{EW}}^{\vphi}$\,\cite{Ciccolini:2003jy,Denner:2011id}
to the \dy-like contribution fully factorize from the \qcd\ effects. The
term $\sigma_{\rm I}^{\vH{}}$ is similar for $\wH{}$ and $\zH{}$
production and involves diagrams where a top-quark loop is inserted into
a gluon line of the real and virtual \nlo{} \qcd{} diagrams to $pp\to
V^*$, from which the Higgs boson is radiated.  The effect of these
contributions on the inclusive cross section is at the 1--2\%
level\,\cite{Brein:2011vx}.  The contribution $\sigma_{\rm II}^{\zH{}}$,
which only exists for $\zH{}$ production, represents $q\bar q$-induced
diagrams where the $Z$ boson couples to a closed top- or bottom-quark
loop. Its contribution on the total rate is even smaller than
$\sigma_{\rm I}^{\vH{}}$\,\cite{Brein:2011vx}. We will refer to the
latter two contributions collectively as $\sigma_\text{I+II}^{\vH}$ in
what follows.

\begin{figure}[h]
  \begin{center}
    \begin{tabular}{ccc}
      \includegraphics[height=.14\textheight]{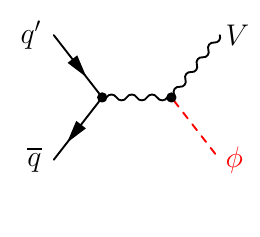} &
      \includegraphics[height=.14\textheight]{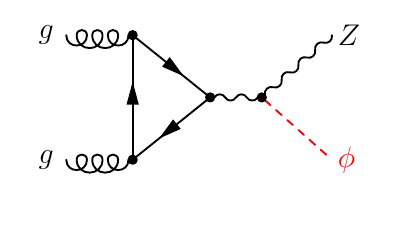} &
      \includegraphics[height=.14\textheight]{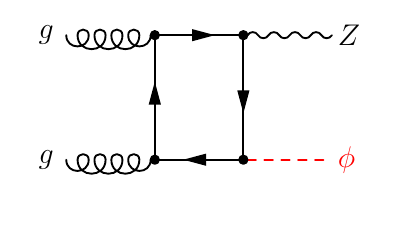}
      \\[-0.5cm]
      (a) & (b) & (c)
    \end{tabular}
    \parbox{\textwidth}{
      \caption[]{\label{fig:lodiagrams}\sloppy Leading-order Feynman
        diagrams to $\sigma^{\vphi{}}$, see \eqn{eq:XSSM}:
        (a)~\dy{}-like terms $\sigma_\text{\dy}^{\vphi}$,
        (b,c)~gluon-initiated contribution $\sigma_{\ggzphi}$.}}
  \end{center}
\end{figure}

Much more relevant for $\zH{}$ production is the gluon-initiated
contribution $\sigma_{\ggzh{}}$, whose generic \lo\ diagrams are
displayed in \figs{fig:lodiagrams}~(b) and (c).  Here, a closed fermion
loop is connected to two initial state gluons and the $Z$ boson is
attached to this loop, whereas the Higgs boson~$H$ is either radiated
off the $Z$ boson or off the closed fermion loop, leading to triangle
and box diagrams, respectively. It is well known that these two
contributions interfere destructively in the
\sm~\cite{Kniehl:1990iva,Dicus:1988yh}.  \citere{Altenkamp:2012sx}
provided the \nlo{} \qcd{} $K$-factor for the $\ggzh$ sub-process in the
heavy-top limit, denoted $K^\infty_\text{\nlo}$ in \eqn{eq:XSSM}, which
enhances this contribution by roughly a factor of two. Corrections
beyond the strict heavy-top limit were calculated in
\citere{Hasselhuhn:2016rqt}, and \citere{Harlander:2014wda} supplied the
soft-gluon resummation which leads to a decrease of the residual
renormalization scale dependence. As pointed out in \sct{sec:intro}, the
current implementation of $\sigma_{\ggzphi{}}$ in \vhnnlo{} uses the
results of \citere{Kniehl:2011aa} and the \nlo{} $K$-factor of
\citere{Altenkamp:2012sx}, as well as new-physics amplitudes obtained
with \texttt{FormCalc}\,\cite{Hahn:1998yk}. Soft-gluon
resummation\,\cite{Harlander:2014wda} is currently not included.

\begin{figure}[h]
  \begin{center}
    \begin{tabular}{ccc}
      \includegraphics[height=.14\textheight]{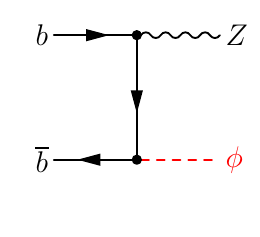} &
      \includegraphics[height=.14\textheight]{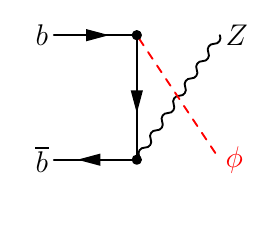} &
      \includegraphics[height=.14\textheight]{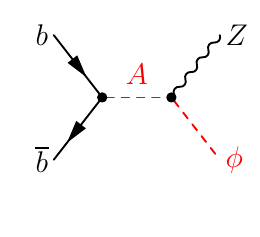}
      \\[-0.5cm]
      (a) & (b) & (c)
    \end{tabular}
    \parbox{\textwidth}{
      \caption[]{\label{fig:bottomdiagrams}\sloppy Feynman diagrams for
        bottom-quark initiated \higgsstrahlung{}, $\sigma_{\bbzphi}$.}}
  \end{center}
\end{figure}

\eqn{eq:XSSM} involves two bottom-quark initiated contributions. The
first one is contained in $\sigma_\text{\dy}^{\vphi}$ and depends only
on the gauge couplings of the bottom quarks. The second one,
$\sigma_{\bbzphi}$, is proportional to the bottom Yukawa coupling;
sample diagrams which contribute to the latter in the \sm\ are shown in
\fig{fig:bottomdiagrams}\,(a) and (b). Currently, $\sigma_{\bbzphi}$ is
implemented only at \lo\ \qcd.  Since we assume the initial-state
bottom-quarks to be massless (while keeping the Yukawa coupling
non-zero), there is no interference between $\sigma_{\bbzphi}$ and
$\sigma_\text{\dy}^{\vphi}$.  The numerical effect of $\sigma_{\bbzphi}$
becomes important in certain models beyond the \sm{} (\bsm{}), see
below.

In \vhnnlo{}, each of the terms in \eqn{eq:XSSM} is provided separately
for the total inclusive cross section, as will be explained in more
detail in \sct{sec:genericoptions}.

%- }}}

%- }}}
%- {{{ section{Generic options of \vhnnlo}

\section{Generic options of \vhnnlo{}}
\label{sec:genericoptions}

In this section we focus on the basic input for a
generic \vhnnlo{} run.  The installation of the code is described in
Appendix~\ref{sec:installation}.  The code itself can be downloaded
from \citere{vhnnlowebpage}.

Compared to version 1.0, \vhnnlo{} has significantly extended its
functionality and capability. This implies an extended set of input
parameters. The input of \vhnnlo{} is controlled through
\slha{}-inspired input blocks (see
\citeres{Skands:2003cj,Allanach:2008qq}).  We begin with the description
of the blocks {\tt VERSION}, {\tt VHATNNLO},\footnote{Read ``\abbrev{VH}
  at \nnlo''.} {\tt ORDER}, {\tt SCALES}, {\tt PDFSPEC}, and {\tt
  SMPARAMS}, which define the central parameters for each run of
\vhnnlo{} and need to be part of all input files.  The settings are
summarized in \tabs{tab:version}--\ref{tab:smparams}.  In these tables
and throughout the draft, the notation \blockentry{NAME}{$i$}{=$k$}
means that the entry $i$ of \texttt{Block NAME} is set to $k$.

For all input blocks which are required in the \sm\ mode
(i.e.\ \blockentry{VHATNNLO}{1}{=0}), \vhnnlo\ provides default entries
which are automatically set if the corresponding entry in the input file
is missing. Their values are specified in
Tables\,\ref{tab:version}--\ref{tab:smparams} and \ref{tab:distrib}.  In
contrast, blocks which specify new physics do not provide default
values, except for a few documented cases.  The units of the input
parameters are given in square brackets in the column ``meaning'' of
these tables. All dimensionful quantities are of data type ``real''.
For input values of data type ``integer'', the table lists the set of
allowed values in column ``range''. Since such restrictions are much
more difficult to define for most of the input parameters of type
``real'', we refrain from any explicit restrictions in this case and
appeal to the user's common sense when setting these parameters.

In order to check the consistency of the input file with the version of
\vhnnlo{}, the user needs to provide the version number in
\blockentry{VERSION}{1}{}.  Accordingly, input files of version 1.0 of
\vhnnlo{} will not run with the current version {\tt 2}.  The entries
$1$ and $2$ of {\tt Block VHATNNLO} define the model and the Higgs boson
type assumed in the calculation. In the \thdm{} or the \mssm{},
\blockentry{VHATNNLO}{2}{=11/12/21} selects $h/H^0/A$ as the final-state
Higgs boson, respectively. The only allowed setting for the \sm\ mode
(\blockentry{VHATNNLO}{1}{=0}) is \texttt{11}.  Entries $3$ and $4$ of
{\tt Block VHATNNLO} determine the type of collider ($pp$ or $p\bar p$)
and its center-of-mass energy.  \blockentry{VHATNNLO}{5}{} specifies the
vector boson in the final state ($Z\phi$, $W^+ \phi$, $W^-\phi$, or
$(W^++W^-)\phi$). Finally, \blockentry{VHATNNLO}{6}{} allows to
reduce/enhance the screen output of \vhnnlo{}.

\begin{table}[htb]
\begin{center}
\begin{tabular}{|>{\tt}c>{\tt}cl|}
  \hline
  \multicolumn{3}{|c|}{\tt Block VERSION}\\
  \hline
  \textrm{entry} & \textrm{default} & meaning\\
  \hline
  1 & 2.0 & version of \vhnnlo{}\\
  \hline
\end{tabular}
\end{center}
\vspace{-0.6cm}
\caption{Settings of {\tt Block VERSION}.}
\label{tab:version}
\end{table}

\begin{table}[htb]
\begin{center}
\begin{tabular}{|>{\tt}c>{\tt}c>{\tt}cl|}
  \hline
  \multicolumn{4}{|c|}{\tt Block VHATNNLO}\\
  \hline
  \textrm{entry} & \textrm{default} & \textrm{range} & meaning\\
  \hline
  1  & 0  & \{0,1,2\} & \{\sm, \mssm, \thdm\} --- model\\
  2 & 11 & \{11,12,21\} & \{\sm{}/$h$, $H$, $A$\} --- Higgs type\\  
  3 & 1 & \{1,2\} & \{$pp$, $p\bar p$\} --- collider type\\
  4 & 1.3d4 & & $\sqrt{s}$ [GeV] --- collider energy\\  
  5 & 1 & \{0,1,2,3\} & \{$W^++W^-$, $Z$, $W^-$, $W^+$\}$\phi$ --- final
  state\\
  6 & 1 & \{0,1\} & \{quiet, verbose\}\\
  \hline
\end{tabular}
\end{center}
\vspace{-0.6cm}
\caption{Basic run parameters of \vhnnlo.}
\label{tab:vhatnnlo}
\end{table}

{\tt Block ORDER} controls the inclusion of the various
sub-contributions to \higgsstrahlung{} and their corresponding order in
perturbation theory.  \blockentry{ORDER}{1}{=0/1/2} determines the order
of the \dy-like terms, \lo/\nlo/\nnlo; \blockentry{ORDER}{1}{=-1}
switches them off. The other sub-processes are unaffected by this
setting. Note that the order of the running of $\alpha_s$ for all
processes is always determined by the order of the \pdf{} set defined in
{\tt Block PDFSPEC}. The $\ggzh$ component is controlled by
\blockentry{ORDER}{2}{}: setting it to \texttt{0} or \texttt{1} includes
it at \lo\ or \nlo, i.e.\ $\order{\alpha_s^2}$ or $\order{\alpha_s^3}$,
respectively, while \blockentry{ORDER}{2}{=-1} switches it off. The
additional top-quark induced terms $\sigma_{\rm I}^{\vH{}}$ and
$\sigma_{\rm II}^{\zH{}}$ are disabled/enabled by setting
\blockentry{ORDER}{3}{=-1/0}, the bottom-quark induced contribution
$\bbzphi{}$ by \blockentry{ORDER}{4}{=-1/0}, and the electro-weak
correction factor~$\delta_{\text{EW}}^{\vH{}}$ by
\blockentry{ORDER}{5}{=-1/0}.  We note that the settings
\blockentry{ORDER}{2}{=1} and \blockentry{ORDER}{3,5}{=0} are only
allowed in the \sm{}.  The entry \blockentry{ORDER}{10}{=-1/0} toggles
$\Delta_b$ resummation in the bottom-Yukawa coupling for
$\tan\beta$-enhanced sbottom contributions, in case the link to \fh{} is
active.  We refer to \sct{sec:mssm} for details.

By default, the renormalization and factorization scale ($\muR$, $\muF$)
is set relative to the invariant mass of the $\vphi$ system through the
entries \blockentry{SCALES}{1,2}{}, respectively. This means that these
scales vary with the integration variable $\hat s$ when convolving the
partonic cross section with the parton densities.  With the exception of
the \dy-like terms, $\muR$ and $\muF$ can also be fixed in absolute
terms to the values given in \blockentry{SCALES}{3,4}{}, respectively,
by setting \blockentry{SCALES}{10}{=1}. The \dy-terms must be switched
off in this case, i.e.\ \blockentry{ORDER}{1}{=-1}.

\blockentry{PDFSPEC}{1}{} sets the name of the employed parton
distribution function (\pdf{}), see \tab{tab:pdfspec}.  Since \vhnnlo{}
needs to be linked to {\tt LHAPDF}~\cite{Buckley:2014ana}, all \pdf{}
sets of {\tt LHAPDF} can be loaded. For each run, only one \pdf{} set
can be specified, which will be used throughout the calculation of all
sub-processes.  \blockentry{PDFSPEC}{10}{} fixes the \pdf{} set
number. \vhnnlo\ will automatically use the value of $\alpha_s(M_Z)$
which corresponds to the specified \pdf{} set for the calculation of the
partonic cross section.

\begin{table}[htb]
\begin{center}
\begin{tabular}{|>{\tt}c>{\tt}c>{\tt}cl|}
  \hline
  \multicolumn{4}{|c|}{\tt Block ORDER}\\
  \hline
  \textrm{entry} & \textrm{default} & \textrm{range} & meaning\\
  \hline
  2 & 1 & \{-1,0,1,2\}    & $n$ ---
  include $\sigma_{\dy}^{\vphi}$ at order $\alpha_s^n$
  (\texttt{-1}=off)\\
  3 & 1 & \{-1,0,1\} & $n$ ---
  include $\sigma_{\ggzphi}$ at order $\alpha_s^{n+2}$
  (\texttt{-1}=off)\\
  4 & 1 & \{-1,0\} & include $\sigma_\text{I+II}^{\vH}$: \{no, yes\}\\
  5 & 1 & \{-1,0\} & include $\sigma_{\bbzphi}$:
  \{no, yes\}\\
  6 & 1 & \{-1,0\} & include $\delta_\text{\abbrev{EW}}^{\vphi}$: \{no, yes\}\\
  10 & 1 & \{-1,0\} & include $\Delta_b$ resummation: \{no, yes\}\\
  \hline
\end{tabular}
\end{center}
\vspace{-0.6cm}
\caption{Controlling the sub-contributions defined in \eqn{eq:XSSM}.}
\label{tab:order}
\end{table}

\begin{table}[htb]
\begin{center}
\begin{tabular}{|>{\tt}c>{\tt}c>{\tt}cl|}
  \hline
  \multicolumn{4}{|c|}{\tt Block SCALES}\\
  \hline
  \textrm{entry} & \textrm{default} & \textrm{range} & meaning\\
  \hline
  1 & 1. &  & renormalization scale $\muR/M_{\vphi}$\\
  2 & 1. & & factorization scale $\muF/M_{\vphi}$\\
  3 & 125. &  & fixed $\muR$ [GeV]\\
  4 & 125. &  & fixed $\muF$ [GeV]\\    
  10 & 0 & \{0,1\} & \parbox[t]{22em}{define $\muR$ and $\muF$ relative
    to $M_{\vphi}$:
    \{yes, no\};\\ if={\tt 0}, use \blockentry{SCALES}{1 \textrm{and} 2}{};\\
   if={\tt 1}, use \blockentry{SCALES}{3 \textrm{and} 4}{}}\\
  \hline
\end{tabular}
\end{center}
\vspace{-0.6cm}
\caption{Defining the unphysical scales. $M_{\vphi}$ is the invariant
  mass of the $\vphi$ system.}
\label{tab:scales}
\end{table}

\begin{table}[htb]
\begin{center}
\begin{tabular}{|>{\tt}c>{\tt}cl|}
  \hline
  \multicolumn{3}{|c|}{\tt Block PDFSPEC}\\
  \hline
  \textrm{entry} & \textrm{default} & meaning\\
  \hline
  1 & 'PDF4LHC15\_nnlo\_mc.LHgrid' & \pdf{} set name (from {\tt LHAPDF})\\
  10 & 0 & \pdf{} set number\\
  \hline
\end{tabular}
\end{center}
\vspace{-0.6cm}
\caption{Choosing a \pdf\ set.}
\label{tab:pdfspec}
\end{table}

\begin{table}[htb]
\begin{center}
\begin{tabular}{|>{\tt}c>{\tt}cl|}
  \hline
  \multicolumn{3}{|c|}{\tt Block MASS}\\
  \hline
  \textrm{entry} & \textrm{default} &  meaning\\
  \hline
  25 & 125. & Higgs mass (in the \sm{}) [GeV]\\ 
  \hline
\end{tabular}
\end{center}
\vspace{-0.6cm}
\caption{Setting the Higgs mass in the \sm. For the \thdm\ and the
  \mssm, see \tab{tab:block:2hdm}.}
\label{tab:mass}
\end{table}

\begin{table}[htb]
\begin{center}
\begin{tabular}{|>{\tt}c>{\tt}cl|}
  \hline
  \multicolumn{3}{|c|}{\tt Block SMPARAMS}\\
  \hline
  \textrm{entry} & \textrm{default} & meaning\\
  \hline
  2 & 1.16637d-5 &  Fermi's constant $G_F$ [GeV$^{-2}$]\\
  3 & 0.118 &  $\alpha_s(M_Z)$ [used by \fh/\thdmc]\\
  4 & 91.1876 &  $Z$-boson mass $M_Z$ [GeV]\\
  5 & 4.18 &  $\msbar$ $b$ mass $m_b(m_b)$ used by \fh/\thdmc,\\&& as well as
    $\bbzphi$ [GeV]\\
  6 & 172.5 &  on-shell $t$ mass [GeV]\\
  10 & 4.92 &  on-shell $b$ mass [GeV]\\
  11 & 80.385 &  $W$-boson mass $M_W$ [GeV]\\
  12 & 2.4952 &  $Z$-boson width $\Gamma_Z$ [GeV]\\
  13 & 2.085 &  $W$-boson width $\Gamma_W$ [GeV]\\
  14 & 0.05077 &  Cabbibo angle $\sin^2\theta_C$\\
  \hline
\end{tabular}
\end{center}
\vspace{-0.6cm}
\caption{Setting the \sm\ input parameters.}
\label{tab:smparams}
\end{table}

{\tt Block MASS} defines the masses of the Higgs bosons of the theory by
employing the corresponding \abbrev{PDG} codes; in the \sm\ mode, only
\blockentry{MASS}{25}{} is relevant.  If run with \fh\ or
\texttt{2HDMC}, this block is overwritten though, see \sct{sec:2hdm} and
\sct{sec:mssm}. In this case, the output file will also list the total
decay widths of the $h/H/A$ boson in \blockentry{MASS}{250/350/360}{} as
calculated by these external programs.

The block \texttt{SMPARAMS} defines numerical values for the central
\sm\ parameters.  \blockentry{SMPARAMS}{2}{} sets the numerical value
for Fermi's constant. \blockentry{SMPARAMS}{3}{} allows to provide a
value for $\alpha_s(M_Z)$ which is independent of the one associated
with the \pdf\ set specified in \texttt{Block PDFSPEC}, see below. It
will only be used by \fh{} or \thdmc{}, however; the partonic cross
section is always calculated using the value of $\alpha_s(M_Z)$ as taken
from the employed \pdf{} set.  In entries {\tt 6,10,5,4,11} of {\tt
  Block SMPARAMS}, the masses of the top-quark, the bottom-quark
(on-shell and $\msbar{}$ $m_b(m_b)$), the $Z$-boson and the $W$-boson
are defined, respectively. The on-shell bottom-quark mass $M_b$ is used
to define the \sm\ Yukawa coupling $Y_b^\text{OS}=M_b/v$ in the
$\ggzphi$ process, and also enters the associated loop integrals. The
$\msbar{}$ bottom-quark mass, on the other hand, enters as input for
\fh{}, \thdmc{}, and the sbottom-Higgs couplings, see \scts{sec:2hdm}
and \ref{sec:mssm}. For the Yukawa coupling $Y_b=m_b(\muR)/v$ which
occurs in $\bbzphi{}$, $m_b(\muR)$ is obtained from $m_b(m_b)$ through
four-loop renormalization group running.  \blockentry{SMPARAMS}{12,13}{}
contain the total widths of the gauge bosons.  \vhnnlo{} internally
calculates the weak mixing angle from $\sin^2\theta_W=1-M_W^2/M_Z^2$,
and the fine structure constant from
$\alpha=G_F\sqrt{2}M_W^2\sin^2\theta_W/\pi$, which are listed in entries
15 and 1 of {\tt Block SMPARAMS} in the output file, respectively.
Finally, the Cabbibo angle is specified in
\blockentry{SMPARAMS}{14}{}. It enters the calculation of the \dy{}-like
terms and top-quark induced contributions.

\begin{table}[htb]
\begin{center}
\begin{tabular}{|>{\tt}c>{\tt}cl|}
  \hline
  \multicolumn{3}{|c|}{\tt Block FACTORS}\\
  \hline
  \textrm{entry} & \textrm{default} & meaning\\
  \hline
       1 &    1.0   & factor for $ht\bar t$ coupling\\
       2 &    1.0   & factor for $hb\bar b$ coupling\\
       3 &    1.0   & factor for $hVV$ coupling\\
       111 &    entry 1   & factor for $ht\bar t$ coupling\\
       211 &    entry 2   & factor for $hb\bar b$ coupling\\
       311 &    entry 3   & factor for $hVV$ coupling\\
       121 &    entry 1   & factor for $At\bar t$ coupling\\
       221 &    entry 2   & factor for $Ab\bar b$ coupling\\
       321 &    entry 3   & factor for $AVV$ coupling\\
       112 &    entry 1   & factor for $Ht\bar t$ coupling\\
       212 &    entry 2   & factor for $Hb\bar b$ coupling\\
       312 &    entry 3  & factor for $HVV$ coupling\\
       \hline
\end{tabular}
\end{center}
\vspace{-0.6cm}
\caption{Modifying the Higgs couplings.}
\label{tab:factors}
\end{table}

Another potential input block available in all models is {\tt Block
  FACTORS} which allows to alter the couplings of all involved Higgs
bosons to quarks and gauge bosons.  The user can either specify common
factors identical for all Higgs bosons through entries {\tt 1}--{\tt 3}
for the couplings to the top quark, the bottom quark and the gauge
bosons, respectively.  Alternatively, the entries {\tt\{111, 211,
  311\}}, {\tt\{112, 212, 312\}} and {\tt\{121, 221, 321\}} separately
change the couplings of the light Higgs boson, the heavy Higgs boson and
the pseudoscalar, respectively (see \tab{tab:factors}). If not
specified, these entries are replaced by their default value
($=1.0$). The $AhZ$ and $AHZ$ coupling of the \thdm\ and the \mssm\ (see
\eqn{eq:gahz} below) are re-scaled in the same way as the $VVH$ and the
$VVh$ couplings, respectively.  Note that, since currently
\vhnnlo\ always assumes a vanishing $VVA$ coupling in accordance with a
\thdm\ with real parameters, \blockentry{FACTORS}{321} is
ineffective at the moment. The couplings of the \cp{}-even scalars to
squarks are currently fixed to their values determined by \susy\ and
cannot be changed.

Finally, {\tt Block VEGAS} allows to control the numerical integration
parameters, separately for the \dy{}-like, the gluon initiated, and
bottom-quark initiated contributions.  More details on this block can be
found in Appendix~\ref{sec:lhapdfcubalt}.

Before discussing the new-physics input, we shortly explain the output
file {\tt vh.out} of \vhnnlo{}.  It contains all input blocks, partially
extended by parameters calculated internally. In addition, {\tt Block
  MASS} and {\tt Block HIGGSCOUP} provide information about the Higgs
mass(es), their widths (if calculated by \thdmc{} or \fh) as
well as its/their coupling(s) to third generation quarks and gauge
bosons.  {\tt Block SIGMA} summarizes the calculated production cross
section, split into the individual sub-processes, including the
integration errors.  Its entries are presented in \tab{tab:sigma}.  The
total cross section presented in \blockentry{SIGMA}{1}{} is calculated
according to \eqn{eq:XSSM}.  {\tt Block SIGMA} also includes the value
of $\alpha_s(M_Z)$ associated with the employed \pdf{} set.

\begin{table}[htb]
\begin{center}
\begin{tabular}{|>{\tt}c>{\tt}cl|}
  \hline
  \multicolumn{3}{|c|}{\tt Block SIGMA}\\
  \hline
  \textrm{entry} & example & meaning\\
  \hline
         1  & \texttt{  8.68271087E-01  } & total cross section $\sigma^{\vphi}$ [pb]\\
       111  & \texttt{  2.31607286E-03  } & integration error on $\sigma^{\vphi}$ [pb]\\
        10  & \texttt{  1.18002301E-01  } & $\alpha_s(M_Z)$ (from \pdf{} set)\\
        11  & \texttt{  7.93648481E-01  } & $\sigma_{\dy}^{\vphi}$ [pb] (without $\delta_{\rm EW}^{\vH}$)\\
       110  & \texttt{  2.32596191E-03  } & integration error on $\sigma_{\dy}^{\vphi}$ [pb]\\
        12  & \texttt{  9.47250000E-01  } & $1+\delta_{\rm EW}^{\vH}$\\
        13  & \texttt{  5.14585867E-02  } & $\sigma_{\ggzphi}$ [pb] (without $K^\infty_\text{\nlo}$)\\
       130  & \texttt{  3.42836009E-04  } & integration error on $\sigma_{\ggzphi}$ [pb]\\
        14  & \texttt{  2.07452931E+00  } & $K$-factor for $\ggzh$, $K_\text{\nlo}^\infty$\\  
        15  & \texttt{  4.31358140E-04  } & $\sigma_{\bbzphi}$ [pb]\\
       150  & \texttt{  7.04767054E-07  } & integration error on $\sigma_{\bbzphi}$ [pb]\\
        16  & \texttt{  9.30486135E-03  } & $\sigma_\text{I+II}^{\vH}$ [pb]\\
       160  & \texttt{  6.30595571E-05  } & integration error on $\sigma_\text{I+II}^{\vH}$ [pb]\\ 
  \hline
\end{tabular}
\end{center}
\vspace{-0.6cm}
\caption{Results presented in {\tt Block SIGMA}.}
\label{tab:sigma}
\end{table}

%- }}}
%- {{{ section{Kinematic distributions}

\section{Kinematic distributions}
\label{sec:tmd}

Fully differential predictions for the \higgsstrahlung{} process within
the \sm{} through \nnlo{} \qcd{} have been available for some
time~\cite{Ferrera:2011bk,Ferrera:2014lca,Campbell:2016jau}. Since
recently, they are also publicly available in {\tt
  MCFM}~\cite{Campbell:2011bn,Campbell:1999ah,Campbell:2015qma,
  Boughezal:2016wmq}. Since the new-physics effects discussed here
affect the \dy{}-like component of the Higgs-Strahlung process only
through a global factor, we link {\tt MCFM} to \vhnnlo\ and rescale this
component accordingly.\footnote{{\tt MCFM} also includes the $\ggzh{}$
  component in the \sm{}, which we subtract within \vhnnlo{} before
  rescaling.  Currently, only version~8 of {\tt MCFM} can be linked to
  \vhnnlo, see subsequent paragraphs and Appendix~\ref{sec:mcfm} for its
  installation and necessary settings.}  The kinematic distributions of
the $\ggzphi$ and $\bbzphi{}$ components are calculated within \vhnnlo{}
at \lo{} \qcd{}, including all available new-physics effects. Here, one
can choose between $\pt$ or $M_{\vphi{}}$ distributions, where $\pt$
denotes the transverse momentum of the Higgs boson, and $M_{\vphi{}}$ is
the invariant mass of the $V\phi$ system.
\begin{table}[htb]
\begin{center}
\begin{tabular}{|>{\tt}c>{\tt}c>{\tt}cl|}
  \hline
  \multicolumn{4}{|c|}{\tt Block DISTRIB}\\
  \hline
  \textrm{entry} & \textrm{default} & \textrm{range} & meaning\\
  \hline
  1 & 1     & \{1,2\} & distribution type: \{$\pt$, $M_{\vphi{}}$\}\\
  2 & 0.    &   & minimal value of distribution parameter [GeV]\\
  3 & 1000. & &  maximal value of distribution parameter [GeV]\\
  4 & 10.   & & bin width [GeV]\\
  5 & 0    & \{0,1\} & run with {\tt MCFM}: \{no, yes\}\\
  6 & 0     & \{0,1\} & calculate $\ggzh{}$ distribution in the \sm:
  \{no, yes\}\\
  7 & 'undefined' &  & read histogram from {\tt MCFM} output file (character)\\
  8 & 0.       & & minimal invariant mass $M_{\vH{}}$ for {\tt MCFM} [GeV]\\
  9 & $\sqrt{s}$ & & maximal invariant mass $M_{\vH{}}$ for {\tt MCFM} [GeV]\\
  10 & 10      &   & iterations with {\tt VEGAS} for {\tt
    MCFM}\\
  11 & 10000   &  & number of calls per iteration for {\tt MCFM}\\  
  \hline
\end{tabular}
\end{center}
\vspace{-0.6cm}
\caption{Setting the parameters for kinematical distributions.}
\label{tab:distrib}
\end{table}

\begin{figure}[htb!]
  \begin{center}
    \begin{tabular}{cc}
   \includegraphics[width=0.49\textwidth]{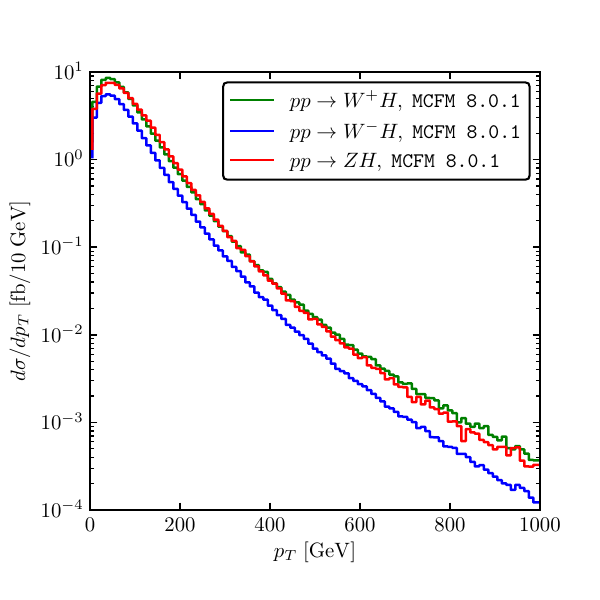} &
  \includegraphics[width=0.49\textwidth]{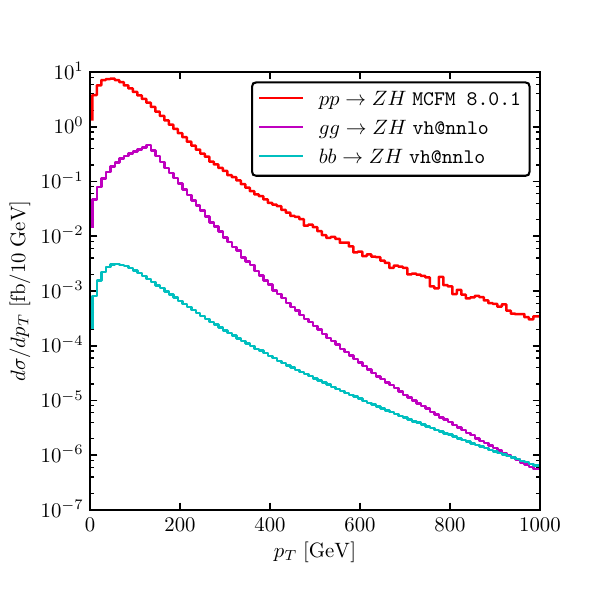}\\[-0.5cm]
   (a) & (b) \\[-0.1cm]
   \includegraphics[width=0.49\textwidth]{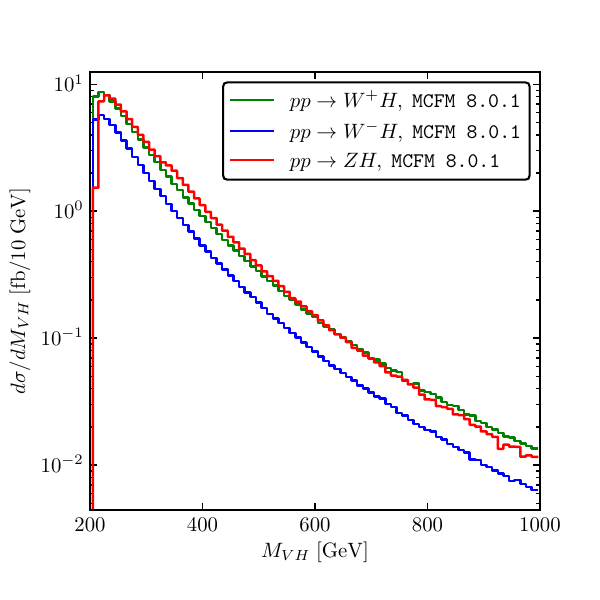} &
  \includegraphics[width=0.49\textwidth]{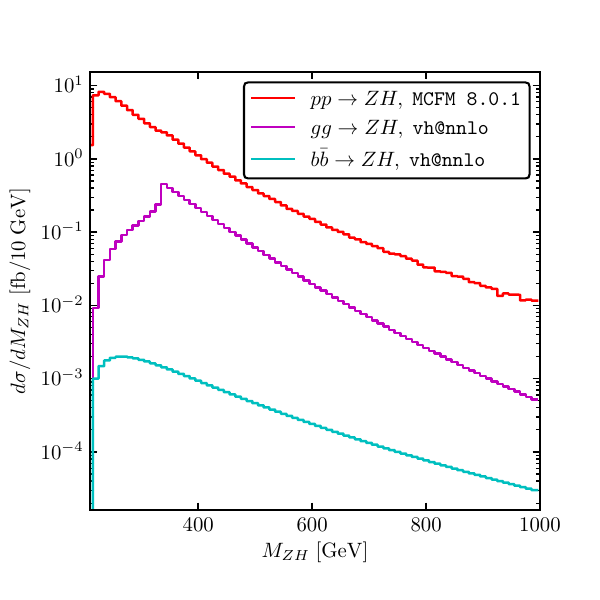} \\[-0.5cm]  
    (c) & (d) \\[-0.2cm]
    \end{tabular}
    \parbox{\textwidth}{
    \caption[]{\label{fig:mhv_pt_dist_example}\sloppy Example of
      transverse momentum distributions in (a) and (b) and invariant
      mass distributions in (c) and (d) for $\vH{}$ production in the
      \sm{} at the \lhc{} with $\sqrt{s} = 13\;\mathrm{TeV}$ and
      $M_\mathrm{H}=125\;\mathrm{GeV}$; (a,c) Comparison of $W^+H$
      (green), $W^-H$ (blue) and $\zH{}$ production (red) created with
         {\code MCFM 8.0.1}; (b,d) Comparison of full $\zH{}$ (red),
         gluon-induced (magenta) and bottom-induced $\zH{}$ (cyan)
         production. All results are obtained at $\order{\alpha_s^2}$,
         apart from $\bbzh{}$ which is calculated at \lo.}}
  \end{center}
\end{figure}

If the input file contains a {\tt Block DISTRIB} (see
\tab{tab:distrib}), \vhnnlo\ produces a kinematic distribution which is
written to the file \texttt{ptDist\_'outputfile'} or
\texttt{mhvDist\_'outputfile'}, where \texttt{outputfile} is the name of
the output file, see Appendix~\ref{sec:installation} for details.  The
prefix of the file name depends on the distribution type selected in
\blockentry{DISTRIB}{1}{}.  {\tt Block DISTRIB} also allows to set a
minimal and maximal value for the kinematic variable under consideration
in \blockentry{DISTRIB}{2/3}{} ({\tt histo start}/{\tt histo end}),
together with the bin width in \blockentry{DISTRIB}{4}{}. Unless
\vhnnlo{} is linked to {\tt MCFM}, the kinematical distribution only
includes the contributions from $\ggzphi$ and $\bbzphi$ (listed
separately). We emphasize that, rather than integrating the
distributions within each bin, \vhnnlo\ directly evaluates the
distributions $d\sigma/d\pt$ or $d\sigma/dM_{\vphi{}}$ for these
contributions at the center of the specified bins.  The \dy{}-like
contribution requires a link to \texttt{MCFM} and, if established, is
shown in a third column, potentially rescaled by the couplings of the
specified \bsm\ theory.  {\tt MCFM} needs a number of additional input
files to be put in the folder {\tt MCFM} of the \vhnnlo\ directory.  In
order to obtain results which are compatible with the settings of
\vhnnlo, the {\tt MCFM} flags {\tt removebr} and {\tt zerowidth} are set
to {\tt true} such that the Higgs and the $W/Z$ bosons do not
decay. \fig{fig:mhv_pt_dist_example} shows examples for $\pt$ and
$M_{\vH{}}$ distributions produced with {\tt MCFM} and \vhnnlo{} in the
\sm{}. Two well-known features are visible: The $\pt$ distributions of
quark- and gluon-induced components show different shapes and peak at
different positions in $\pt$; the $M_{\zH{}}$ distribution of the
$\ggzh$ contribution clearly shows the top-quark threshold at
$M_{\zH{}}=2m_t$.

%- }}}
%- {{{ section{Beyond the Standard Model}

\section{Beyond the Standard Model}\label{sec:bsm}

%- {{{ intro:

\begin{figure}[h]
  \begin{center}
    \begin{tabular}{cc}
      \includegraphics[height=.14\textheight]{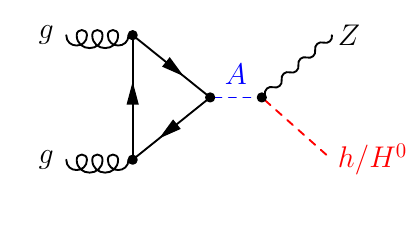} &
      \includegraphics[height=.14\textheight]{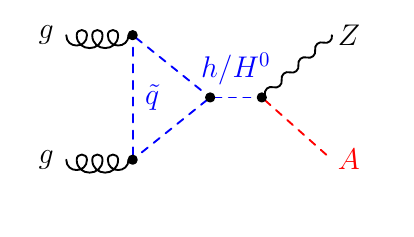}
      \\
      (a) & (b) \\
      \includegraphics[height=.14\textheight]{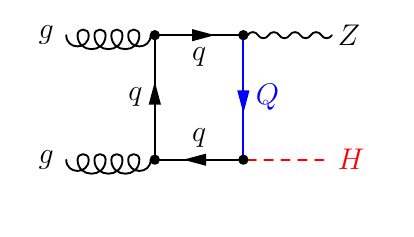} &
      \includegraphics[height=.14\textheight]{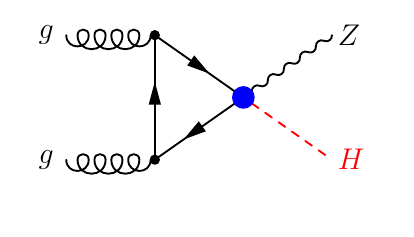} \\[-0.5cm]
       (c) & (d)
    \end{tabular}
    \parbox{\textwidth}{
      \caption[]{\label{fig:newphysicsdiagrams}\sloppy Selection of new contributions to the $\ggzphi$ amplitude in the \thdm{}/\mssm{} due to 
      (a) intermediate (pseudo)scalars and (b) squarks in addition; to the $\ggzh$ amplitude in the \sm{} due to
      (c) vector-like quarks and (d) \dimsix{} operators.}}
  \end{center}
\end{figure}

In the following sub-sections we list the different new-physics models
and contributions supported by \vhnnlo{}. A couple of sample diagrams
relevant for the $\ggzphi$ contribution are shown in
\fig{fig:newphysicsdiagrams}.  We start with the \thdm{} in
\sct{sec:2hdm} including a discussion of resonant (pseudo)scalars, and
continue with the \mssm{} in \sct{sec:mssm}. In both sub-sections we
explain the potential links to external codes for the calculation of a
consistent set of model parameters.  In \sct{sec:cpmix}, we address
\cp\ mixing among the three neutral \thdm{} scalars of these two models.
Section \ref{sec:vlq} describes the implementation of vector-like
quarks. Finally, in \sct{sec:dim6} we go beyond a concrete model
implementation and describe the incorporation of higher-dimensional
operators.  Since $\sigma_{\rm I+II}^{\vH{}}$ as well as the \nlo{}
$K$-factor to $\ggzh{}$ are only known for the \sm, they need to be
switched off whenever \abbrev{BSM} effects are requested.  Where
appropriate, we discuss numerical examples to highlight the new-physics
effects in \higgsstrahlung{}.

%- }}}
%- {{{ subsection{2HDM}

\subsection{2-Higgs-Doublet Model}
\label{sec:2hdm}

The program \vhnnlo{} allows to calculate the production cross section
for any of the three electrically neutral Higgs bosons in the
\thdm{}. The effects which distinguish this case from the \sm{}
calculation have been described in detail in
\citere{Harlander:2013mla}. Therefore, they shall be summarized only
briefly here. For once, the couplings of the \cp{}-even Higgs bosons to
the weak gauge bosons and the quarks are different from the \sm\ Higgs
couplings. While the former are the same for all \thdm{}s, the
Higgs-quark couplings depend on the specific realization of the
$\mathbb{Z}_2$ symmetry which is imposed in order to avoid tree-level
flavor-changing neutral currents. Specifically, for the gauge-Higgs
couplings we have
\begin{equation}
  \begin{split}
     g_{VV}^h = \sin(\beta-\alpha)\,,\qquad
     g_{VV}^{H}=\cos(\beta-\alpha)\,,\qquad
     g_{VV}^{A}=0\,,\qquad V\in\{W,Z\}\,,
     \label{eq:gvvh}
  \end{split}
\end{equation}
where $\alpha$ parametrizes the mixing of the \cp{}-even Higgs bosons
from the isospin to the mass eigenstates, and $\tan\beta$ is the ratio
of the vacuum-expectation values of the two Higgs doublets.  The
quark-Higgs couplings are summarized in \tab{tab:fcoup}. In types III
and IV, the quark-Higgs couplings equal the ones of types I and II,
respectively.  Only the couplings to leptons differ; this affects the
total widths of the Higgs bosons, and thus indirectly the
Higgs-Strahlung cross section through diagrams with Higgs bosons in the
$s$-channel.

%- {{{ tab:fcoup:

\begin{table}[htb]
\begin{center}
\begin{tabular}{|c|cccccc|}
\hline
& $g_u^h$ &  $g_u^H$ &  $g_u^A$ &  $g_d^h$ &  $g_d^H$ &  $g_d^A$ \\
\hline
Type I &
$\cos\alpha/\sin\beta$ &
$\sin\alpha/\sin\beta$ &
$\cot\beta$ &
$\cos\alpha/\sin\beta$ &
$\sin\alpha/\sin\beta$ &
$-\cot\beta$ \\
Type II &
$\cos\alpha/\sin\beta$ &
$\sin\alpha/\sin\beta$ &
$\cot\beta$ &
$\sin\alpha/\cos\beta$ &
$\cos\alpha/\cos\beta$ &
$\tan\beta$ \\
\hline
\end{tabular}
\end{center}
\vspace{-0.6cm}
\caption{\label{tab:fcoup} The Higgs-quark couplings $g_f^\phi$ for
  \thdm{} type I and II, see \fig{fig:frules}. For types III and IV we
  refer to the comment in the text.}
\label{tab:2hdmcouplings}
\end{table}

%- }}}  
%- {{{ fig:frules:

\begin{figure}[h]
  \begin{tabular}{rl}
\raisebox{-2.3em}{\includegraphics[viewport=280 545 370
    620,height=.1\textheight]{%
  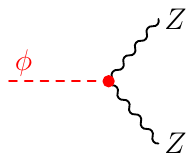}}
&$\displaystyle=i\frac{eM_Z}{c_Ws_W}g_{VV}^\phi g^{\mu\nu}$\qquad\qquad
\raisebox{-2.3em}{\includegraphics[viewport=280 545 370
    620,height=.1\textheight]{%
  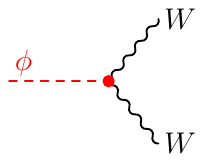}}
$\displaystyle=i\frac{eM_W}{s_W}g_{VV}^\phi g^{\mu\nu}$\\
    \raisebox{-2.3em}{\includegraphics[viewport=280 545 370
        620,height=.1\textheight]{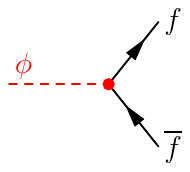}}
    &$\displaystyle=-\frac{m_f}{v}\{ig_f^h,ig_f^{H},g_f^A\gamma_5\}$
\text{for}\quad $\phi\in\{ h,H,A\}$\\
\raisebox{-2.3em}{\includegraphics[viewport=280 545 370
    620,height=.1\textheight]{%
  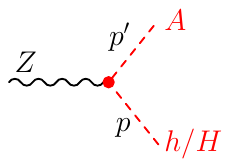}}
&$\displaystyle=\frac{e}{2c_Ws_W}g_Z^{A\phi}(p-p')^\mu$\\
  \end{tabular}
\caption{\label{fig:frules}Feynman rules for the Higgs couplings in the
  \thdm. All momenta are ingoing.}
\end{figure}

%- }}}

Figure\,\ref{fig:frules} and \tab{tab:fcoup} also show the couplings of
the \cp-odd Higgs boson $A$ to quarks. Its coupling to the weak gauge
bosons vanishes, but there are tri-linear couplings to $\zh{}$ and
$\zH{}$, also included in \fig{fig:frules}, where
\begin{equation}
  \begin{split}
     g_{Z}^{Ah} = \cos(\beta-\alpha)\,,\qquad g_{Z}^{AH} =
     -\sin(\beta-\alpha)\,.
     \label{eq:gahz}
  \end{split}
\end{equation}
These latter couplings imply new contributions to the Higgs-Strahlung
process, with virtual Higgs bosons in the $s$-channel.  When Higgs-quark
couplings are neglected for the first two quark generations, such
contributions can appear only in the partonic processes $\bbzphi{}$ and
$\ggzphi{}$. Their numerical effects have been studied in detail in
\citere{Harlander:2013mla}. They can be huge, for example for $gg\to
A\to \zh{}$ in the wrong-sign limit ($g_d^h=-1$) of the
\thdm{}~\cite{Ferreira:2017bnx}.  We note that using {\tt Block
  FACTORS}, see \sct{sec:genericoptions}, implies that the relative
factors of $g_Z^{Ah}$ and $g_Z^{AH}$ are set to the ones of $g_{VV}^H$
and $g_{VV}^h$, respectively.

If the mass $M_{\phi'}$ of the $s$-channel Higgs boson $\phi'$ is larger
than the threshold for the $Z\phi$ production process under
consideration, the integration over the parton density functions would
encounter a pole at $\hat s=M^2_{\phi'}$ which, however, is regulated by
the width $\Gamma_{\phi'}$ of $\phi'$ through the replacement
\begin{align}
 \frac{1}{\hat{s}-m_{\phi'}^2}\rightarrow
 \frac{1}{\hat{s}-m_{\phi'}^2+im_{\phi'}\Gamma_{\phi'}}\quad.
\end{align}
The numerical value for $\Gamma_{\phi'}$ is required by \vhnnlo{} as
input. If \fh{} or \thdmc{} are linked to
\vhnnlo{}, $\Gamma_{\phi'}$ will be taken from their output though.  The
same effect occurs in the bottom-quark initiated contribution
$\sigma_{\bbzphi}$ due to \fig{fig:bottomdiagrams}\,(c).  We emphasize
that our implementation allows to take into account the interferences of
the $s$-channel diagrams $gg/b\bar {b}\to A\to \zh{}$ or $gg/b\bar
{b}\to \{h,H\}\to \zA{}$ with all other Feynman diagrams. Until now,
searches for heavy pseudoscalars decaying to $\zh{}$ have applied the
narrow-width approximation, which means that such interference effects
have been neglected.  This allowed to include higher-order corrections
to the production mechanisms $gg/b\bar b\to A$, as implemented in {\tt
  SusHi}~\cite{Harlander:2012pb,Harlander:2016hcx}, which can also be
linked to {\tt 2HDMC} to obtain the branching ratio $A\to \zh{}$.
However, it is well-known from decays of heavy scalars into a pair of
gauge bosons~$V\!V$~\cite{Jung:2015sna,Greiner:2015ixr} or a pair of
Higgs bosons~$h\!h$~\cite{Hespel:2014sla} that interference effects in a
\thdm{} can be large.  Thanks to the option to obtain the invariant mass
distribution~$M_{\zphi{}}$, \vhnnlo{} allows to study the shape of the
interferences around the mass of the intermediate resonance.  We will
show an example at the end of this section.

One option to evaluate the cross section in the \thdm\ with \vhnnlo\ is
to define the three neutral Higgs boson masses and their total decay
widths in {\tt Block MASS}, see \tab{tab:block:2hdm}.  The other
relevant \thdm{} parameters need to be provided in {\tt Block 2HDM},
also described in \tab{tab:block:2hdm}. Recall that various
contributions to the cross section can be switched on and off in {\tt
  Block ORDER}, see \tab{tab:order}. \tab{tab:2hdmcouplings} may serve
as definition of $\tan\beta$ and the \cp{}-even Higgs mixing angle
$\alpha$ (up to shifts of $\alpha$ by $\pm\pi$, which leave the cross
section invariant).  As explained before, \thdm{} types I and III as
well as II and IV have identical quark-Higgs couplings.  Thus, using
{\tt Block 2HDM} will reveal identical results for
\blockentry{2HDM}{2}{=1} and {\tt 3} as well as {\tt =2} and {\tt 4},
provided the total Higgs widths are left unchanged.

%- {{{ Block MASS and 2HDM:

\begin{table}[h]
  \begin{center}
\begin{tabular}{|>{\tt}c>{\tt}cl|}
  \hline
  \multicolumn{3}{|c|}{\tt Block MASS}\\
  \hline
  \textrm{entry} & \textrm{default} &  meaning\\
  \hline
  25 & 125. & $m_h$ [GeV]\\ 
  35 &  & $m_H$ [GeV]\\ 
  36 &  & $m_A$ [GeV]\\ 
  250 &  & $\Gamma_h$ [GeV]\\ 
  350 &  & $\Gamma_H$ [GeV]\\ 
  360 &  & $\Gamma_A$ [GeV]\\ 
  \hline
  \hline
\multicolumn{3}{|c|}{\tt Block 2HDM}\\
\hline
\textrm{entry} & \textrm{range} & meaning\\
\hline
2 &   \{1,2,3,4\} & \thdm{} type I\ldots IV\\
3 &    & $\tan\beta$\\
4 &   \textrm{see text}  & \cp\ even Higgs mixing angle $\alpha$ \\
\hline
\end{tabular}
\caption{\label{tab:block:2hdm} Defining the masses and widths of the
  light and heavy \cp\ even Higgs $h$, $H$, and of the \cp-odd Higgs $A$.}
  \end{center}
\end{table}

%- }}}

Another option to define the \thdm{} input parameters is through the
link of \vhnnlo{} to \thdmc{}~\cite{Eriksson:2009ws,Eriksson:2010zzb}
(see Appendix \ref{sec:thdmcfhfs} how to
establish the link). If \vhnnlo{} finds the {\tt Block 2HDMC} in its
input file, it will read \thdmc{} input from this block, and ignore the
block {\tt 2HDM}. Linking to \thdmc{} gives you access to the features
of this program within \vhnnlo{}, meaning that you can use different
parameterizations of the \thdm{}, check for theoretical consistency of
the \thdm{}, and get consistent numerical values for the decay widths of
the Higgs particles.

The program \thdmc{} produces its own additional output file, named {\tt
  2HDMC.out}, which contains all relevant information about the Higgs
spectrum and decay widths.  The definition of the entries in {\tt Block
  2HDMC} is described in \tab{tab:block:2hdmc}. The specific \thdm{}
parameterization is set by \blockentry{2HDMC}{1}{}, which currently can
assume the values 1, 2, and 3, corresponding to the ``$\lambda$ basis'',
the ``physical basis'', and the ``$H_2$ basis'', respectively (see
\citeres{Eriksson:2009ws,Haber:2015pua} for the definition of these
bases). The relevant parameters for these bases are given in
\blockentry{2HDMC}{11..17}{}, \blockentry{2HDMC}{21..27}{}, and
\blockentry{2HDMC}{31..36}{}, respectively. For example, if
\blockentry{2HDMC}{1}{=1}, only \blockentry{2HDMC}{11..17}{} is used for
defining the \thdm{}, while \blockentry{2HDMC}{21..27}{} and
\blockentry{2HDMC}{31..36}{} are ignored.  Concerning the Higgs mixing
angle $\alpha$, the convention of \thdmc{} in the physical and the $H_2$
basis assumes $-\tfrac{\pi}{2}\leq \beta-\alpha\leq \tfrac{\pi}{2}$.  If
the user works with the convention $0\leq \beta-\alpha\leq \pi$, such
that $\cos(\beta-\alpha)$ covers the whole range $[-1,1]$ and
$\sin(\beta-\alpha)\in [0,1]$, then for values between
$\tfrac{\pi}{2}\leq \beta-\alpha\leq \pi$, the \thdmc{} convention is
obtained by shifting $\beta-\alpha$ by $-\pi$.  The other parameters are
in principle not constrained, but \thdmc{} can check whether a parameter
point obeys stability of the vacuum, unitarity and perturbativity
requirements.  If \blockentry{2HDMC}{10}{=1}, \vhnnlo{} will terminate
if a \thdm{} parameter point does not obey all of these criteria. This
is quite useful for large parameter scans.  In contrast to the run with
{\tt Block 2HDM}, the link to \thdmc{} will reveal differences among the
\thdm{} types I and III as well as II and IV. This is due to the fact
that the decay widths obtained from \thdmc{} depend on all couplings,
including the Higgs-lepton couplings.

%- {{{ tab:block:2hdmc:

\begin{table}
  \begin{center}
\begin{tabular}{|r>{\tt}c>{\tt}cl|}
  \hline
\multicolumn{4}{|c|}{\tt Block 2HDMC}\\
\hline
& \textrm{entry} &  \textrm{range} & meaning\\
\hline
&1 &  \{1,2,3\} & \thdmc{} input key (basis choice)\\
&2 &  \{1,2,3,4\} & \thdm{} type \{I, II, III, IV\}\\
&3 &   & $\tan\beta$\\
&4 &   & $m_{12}$ [GeV] \\
&10 & \{0,1\} & ignore theory inconsistencies: \{yes, no\}\\
\hline
\thdmc{} input key & \textrm{entry} & & meaning\\
\hline
\blockentry{2HDMC}{1}{=1} & 11 & & $\lambda_1$ \\
$\lambda$ basis & 12 & & $\lambda_2$ \\
& 13 & & $\lambda_3$\\
& 14 & & $\lambda_4$\\
& 15 & & $\lambda_5$\\
& 16 & & $\lambda_6$\\
& 17 & & $\lambda_7$\\
\hline
\blockentry{2HDMC}{1}{=2} &21 & & $M_h$ [GeV]\\
physical basis &22 & & $M_H$ [GeV]\\
&23 &  & $M_A$ [GeV]\\
&24 &   & $M^\pm$ [GeV]\\
&25 &  & $\sin(\beta-\alpha)$\\
&26 &  & $\lambda_6$\\
&27 &  & $\lambda_7$\\
\hline
\blockentry{2HDMC}{1}{=3} &31 & & $M_h$ [GeV]\\
$H_2$ basis &32 & & $M_H$ [GeV]\\
&33 &   & $\sin(\beta-\alpha)$\\
&34 & & $Z_4$\\
&35 & & $Z_5$\\
&36 & & $Z_7$\\
\hline
\end{tabular}
\caption{\label{tab:block:2hdmc} Input parameters of {\tt Block 2HDMC}.
  Either entries 11-17, 21-27 or 31-36 have to be specified, in
  accordance to \blockentry{2HDMC}{1}.  For details on the different
  bases, we refer to \citeres{Eriksson:2009ws,Haber:2015pua}.}
\end{center}
\end{table}

\begin{figure}[htb!]
  \begin{center}
   \includegraphics[width=0.49\textwidth]{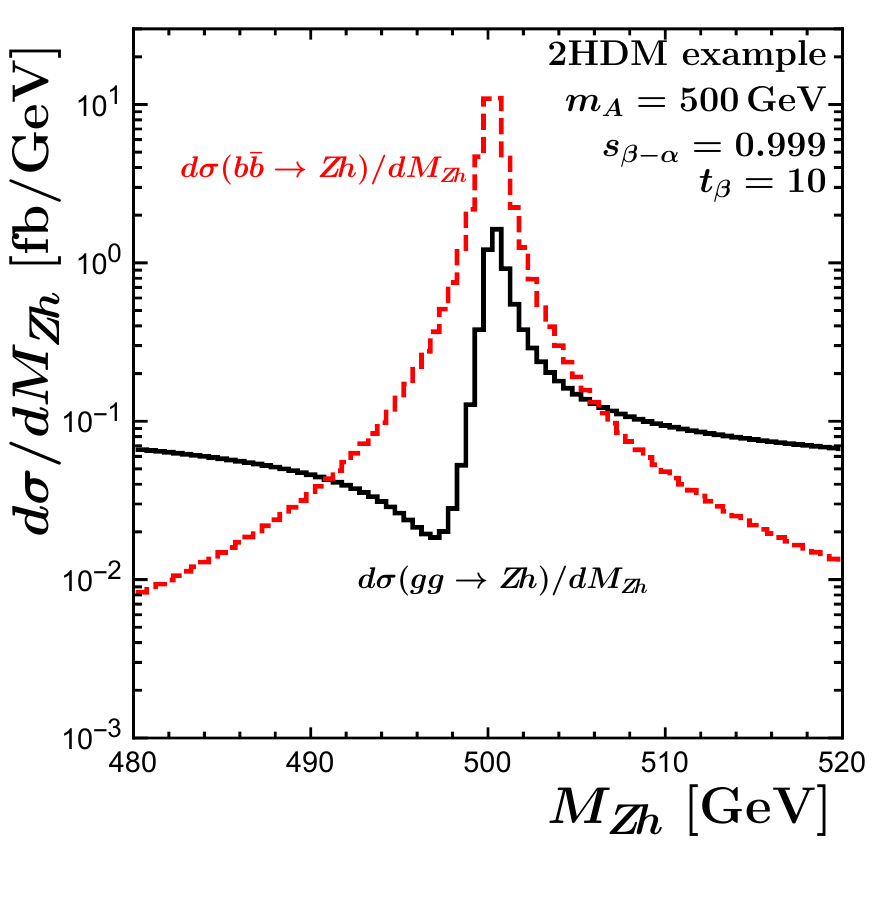}
    \parbox{\textwidth}{
    \caption[]{\label{fig:mhv_2hdm_int}\sloppy 
    The cross sections $gg\to \zh{}$ (black) and $b\bar b\to \zh{}$ (red)
    as a function of the invariant mass of the $\zh{}$ system for the \lhc{} at $13$\,TeV.
    The shapes are determined by the interference of the resonant
    $gg/b\bar b\to A\to \zh{}$ contributions with all other diagrams.}}
  \end{center}
\end{figure}

A thorough discussion of \higgsstrahlung\ in the \thdm\ can be found in
\citere{Harlander:2013mla}. We therefore focus only on a single
numerical example which shows the interference effects of intermediate
(pseudo)scalars with non-resonant diagrams as discussed above.  In
\fig{fig:mhv_2hdm_int}, we exemplify the two cross sections $gg\to
\zh{}$ and $b\bar b\to \zh{}$ as a function of the invariant mass of the
$\zh{}$ system around the pseudoscalar mass $m_A=500$\,GeV.  The light
Higgs mass is chosen to be $m_h=125$\,GeV, and the other relevant
parameters are fixed to $\tan\beta=10$ and $\sin(\beta-\alpha)=0.999$.
The total pseudoscalar width obtained by \thdmc{} is
$\Gamma_A=1.13$\,GeV.  In this region of the parameter space, the
``signal'' process $gg\to A\to \zh{}$ yields a small cross section and
induces large interference effects with all other diagrams as shown in
\fig{fig:mhv_2hdm_int}.  For $b\bar b\to A\to \zh{}$, the effect is not
as pronounced, since also the contribution of the non-resonant diagrams
is small.  We leave a thorough analysis of such interference effects to
future work. A generic study using \vhnnlo{} was already carried out in
\citere{Brooijmans:2018xbu}. An effective coupling of the pseudoscalar to two
gluons through the operator ${\cal{L}}_6$, see \sct{sec:dim6}, was
employed therein.

%- }}}

%- }}}
%- {{{ subsection{MSSM}

\subsection{Minimal Supersymmetric Standard Model}
\label{sec:mssm}

Concerning the quark-loop contributions to the $\ggzphi{}$ process, the
\mssm{} amplitudes are identical to those of a type-II \thdm{}. As
explained in \citere{Kniehl:2011aa}, the squark-loop contributions to
\cp{}-even Higgs production vanish, and of the squark-loop contributions
to \cp{}-odd Higgs production, only triangle-type diagrams are
non-vanishing. We compared the results for the quark- and the
squark-loop diagrams for both \cp{}-even and -odd Higgs production from
\citere{Kniehl:2011aa} to our own calculation of the $\ggzphi{}$ cross
section based on \texttt{FormCalc}\,\cite{Hahn:1998yk} and found
agreement after fixing some typos in \citere{Kniehl:2011aa}.

Similar to the \thdm, the parameters for the \mssm\ mode
\blockentry{VHATNNLO}{1}{=1} can be defined in several ways. The first
one is through the blocks {\tt 2HDM}, {\tt MASS}, and \texttt{SQUARK}.
The block {\tt SQUARK} is only required for \cp-odd Higgs production,
i.e.\ for \blockentry{VHATNNLO}{2}{=21}.  As described in detail in
\tab{tab:block:squark}, its entries define the masses and mixing of
stops and sbottoms.  For the squark mixing angle~$\theta_q$
($q\in\{t,b\}$) both conventions $\cos\theta_q\in [-1,1]$ and
$\sin\theta_q\in [0,1]$ or vice versa are possible.  If $\sin\theta_q$
is not specified, the convention with $\cos\theta_q\in [-1,1]$ is
assumed, and $\sin\theta_q\geq 0$ is obtained from
$\sqrt{1-\cos^2\theta_q}$.  Since we work at leading order for what
concerns squark effects, the renormalization scheme of these parameters
is arbitrary. Furthermore, \vhnnlo\ does not restrict the range of
values for $A_t$, $A_b$ and $\mu$.  However, we urge the user to choose
these parameters in accordance to the squark masses and mixings, which
limits their ranges.  The Feynman rules for the squark couplings can be
found in \citere{Harlander:2012pb}. They involve also the quark masses
$m_b$ and $m_t$, for which we insert the \msbar\ value $m_b(m_b)$
provided in \blockentry{SMPARAMS}{5}{}, and the on-shell value $m_t$ of
\blockentry{SMPARAMS}{6}{}, respectively.

Another way to provide the \mssm{} parameters is through the link of
\vhnnlo{} to the code~\fh{}~\cite{Heinemeyer:1998yj,Heinemeyer:1998np,Degrassi:2002fi,Frank:2006yh,Hahn:2013ria,Bahl:2016brp,Bahl:2017aev}
(see Appendix \ref{sec:thdmcfhfs} on how to establish
this link). If the input file contains the {\tt Block FEYNHIGGS},
\vhnnlo{} will ignore the blocks {\tt 2HDM}, {\tt MASS}, and {\tt
  SQUARK}, but rather call the program \fh{} to determine the spectrum
and the couplings of the specific \mssm{} parameter point. A number of
flags for \fh{} can be accessed through the block {\tt
  FEYNHIGGSFLAGS}. A detailed description of the two blocks is given in
\tab{tab:block:feynhiggs} and
\ref{tab:block:feynhiggsflags}.\footnote{We note that the notation
  differs from the one of {\tt SusHi}, which also uses blocks {\tt
    MINPAR} and {\tt EXTPAR}.}  The blocks associated with the link to
\fh{} generally understand all input parameters as
on-shell. Accordingly, stop and sbottom masses returned from \fh{} are
renormalized on-shell.
In case the link to \fh{} is active, \vhnnlo{} also allows to
incorporate the resummation of $\tan\beta$-enhanced sbottom
contributions in the bottom-quark Yukawa coupling, known as $\Delta_b$
resummation~\cite{Banks:1987iu,Hall:1993gn,Hempfling:1993kv,
Carena:1994bv,Carena:1999py,Carena:2000uj}.
Setting \blockentry{ORDER}{10}{=1} will
activate the so-called ``full resummation'' as presented in
Eqs.\,(16c)--(16e) of \citere{Harlander:2012pb}.  Note that one may also
define an independent value of $\Delta_b$ by adjusting the bottom-quark
Yukawa coupling through \texttt{Block FACTORS}, see \tab{tab:factors}.

A third option to provide the relevant \mssm{} input is through an
additional {\tt SLHA}-style input file, referred to as ``spectrum file''
in what follows, which contains the \mssm{} particle spectrum. The
relevant content and a list of potential spectrum generators providing a
spectrum file are given in Appendix~\ref{sec:thdmcfhfs}. This mode of
operating \vhnnlo\ is activated by including a {\tt Block SPECTRUMFILE}
in the actual input file of \vhnnlo, and specifying the path to the
spectrum file in \blockentry{SPECTRUMFILE}{1}{}. Apart from that, the
\vhnnlo{} input file only requires the generic blocks described in
\sct{sec:genericoptions}, as well as the widths of the Higgs bosons in
entries {\tt 250,350} and {\tt 360} in {\tt Block MASS}.

\begin{table}
  \begin{center}
\begin{tabular}{|>{\tt}cl|}
  \hline
\multicolumn{2}{|c|}{\tt Block SQUARK}\\
\hline
\textrm{entry} & meaning\\
\hline
1 & mass of $\tilde t_1$, $\tilde m_{t1}$ [GeV]\\
2 & mass of $\tilde t_2$, $\tilde m_{t2}$ [GeV]\\
3 & stop mixing angle $\cos\theta_t$\\
4 & stop mixing angle $\sin\theta_t$\\
5 & trilinear coupling $A_t$ [GeV]\\
11 & mass of $\tilde b_1$, $\tilde m_{b1}$ [GeV]\\
12 & mass of $\tilde b_2$, $\tilde m_{b2}$ [GeV]\\
13 & sbottom mixing angle $\cos\theta_b$\\
14 & sbottom mixing angle $\sin\theta_b$\\
15 & trilinear coupling $A_b$ [GeV]\\
20 & gluino mass $|M_3|$ [GeV]\\
21 & $\mu_{\rm \susy{}}$ [GeV]\\
\hline
\end{tabular}
\caption{\label{tab:block:squark} Input parameters of {\tt Block SQUARK}. For details, see main
  text.}
  \end{center}
\end{table}

\begin{table}
  \begin{center}
\begin{tabular}{|>{\tt}ccl|}
  \hline
\multicolumn{3}{|c|}{\tt Block FEYNHIGGS}\\
\hline
\textrm{entry} &  \textrm{default} &  meaning\\
\hline
0 & & $\tan\beta$\\
1 & & soft-breaking mass $M_1$ [GeV]\\
2 & & soft-breaking mass $M_2$ [GeV]\\
3 & & soft-breaking mass $M_3$ [GeV]\\
\hline
10 &       & generic trilinear coupling $A_x$ [GeV]\\
11-19 & $A_x$ & trilinear couplings $A_t,A_b,A_{\tau},A_c,A_s,A_\mu,A_u,A_d,A_e$ [GeV]\\
\hline
 23 & & $\mu_{\rm \susy{}}$ [GeV]\\
 26 & & pseudoscalar Higgs mass $m_A$ [GeV]\\
\hline
30 &                & generic soft-breaking mass $M_{\rm \susy{}}$ [GeV]\\
31-36 & $M_{\rm \susy{}}$ & soft-breaking masses
\texttt{M1SL,M2SL,M3SL,M1SE,M2SE,M3SE} [GeV]\\
\hline
41-49 & $M_{\rm \susy{}}$ & soft-breaking masses
\texttt{M1SQ,M2SQ,M3SQ,M1SU,M2SU,M3SU,}\\
& & \texttt{M1SD,M2SD,M3SD} [GeV]\\
\hline
\end{tabular}
\caption{\label{tab:block:feynhiggs} Input parameters of {\tt Block FEYNHIGGS}. The entries
\blockentry{FEYNHIGGS}{11-19}{} and \blockentry{FEYNHIGGS}{31-49}{}
are automatically set to the generic values \blockentry{FEYNHIGGS}{10}{}
and \blockentry{FEYNHIGGS}{30}{}, if they are not specified. The notation
of the soft-breaking masses follows the notation of \fh{}.}
\end{center}
\end{table}

\begin{table}
  \begin{center}
\begin{tabular}{|>{\tt}c>{\tt}c>{\tt}l|}
  \hline
\multicolumn{3}{|c|}{\tt Block FEYNHIGGSFLAGS}\\
\hline
\textrm{entry} & \textrm{default} & \textrm{meaning}\\
\hline
1  &   4   & mssmpart\\
2  &   0   & fieldren \\
3  &   0   & tanbren \\
4  &   2   & higgsmix \\
5  &   4   & p2approx \\
6  &   2   & looplevel \\
7  &   1   & runningMT \\
8  &   1   & botResum \\
9  &   0   & tlCplxApprox\\
10  &   3   & loglevel\\
\hline
\end{tabular}
\caption{\label{tab:block:feynhiggsflags} Input parameters of {\tt Block FEYNHIGGSFLAGS}
as they are compatible with {\tt FeynHiggs 2.13.0} and {\tt 2.14.0}. For the meaning and ranges (dependent on the \fh{} version)
of the settings we refer to the \fh{} webpage~\cite{fhwebpage}. If they are not specified they are automatically
set to the default values of \fh{} {\tt 2.13.0}. The settings {\tt fieldren} and {\tt tanbren} were removed in \fh{} {\tt 2.14.0}.}
\end{center}
\end{table}

\begin{figure}[htb!]
  \begin{center}
    \begin{tabular}{cc}
      \includegraphics[width=.45\textwidth]{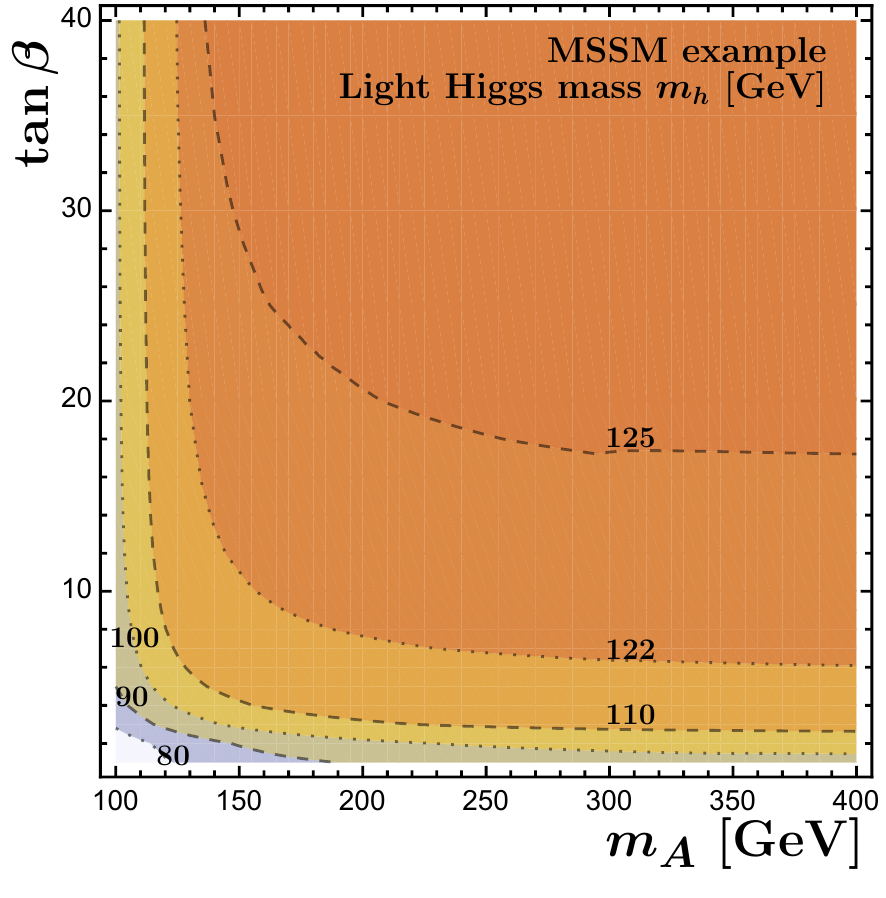} &
      \includegraphics[width=.45\textwidth]{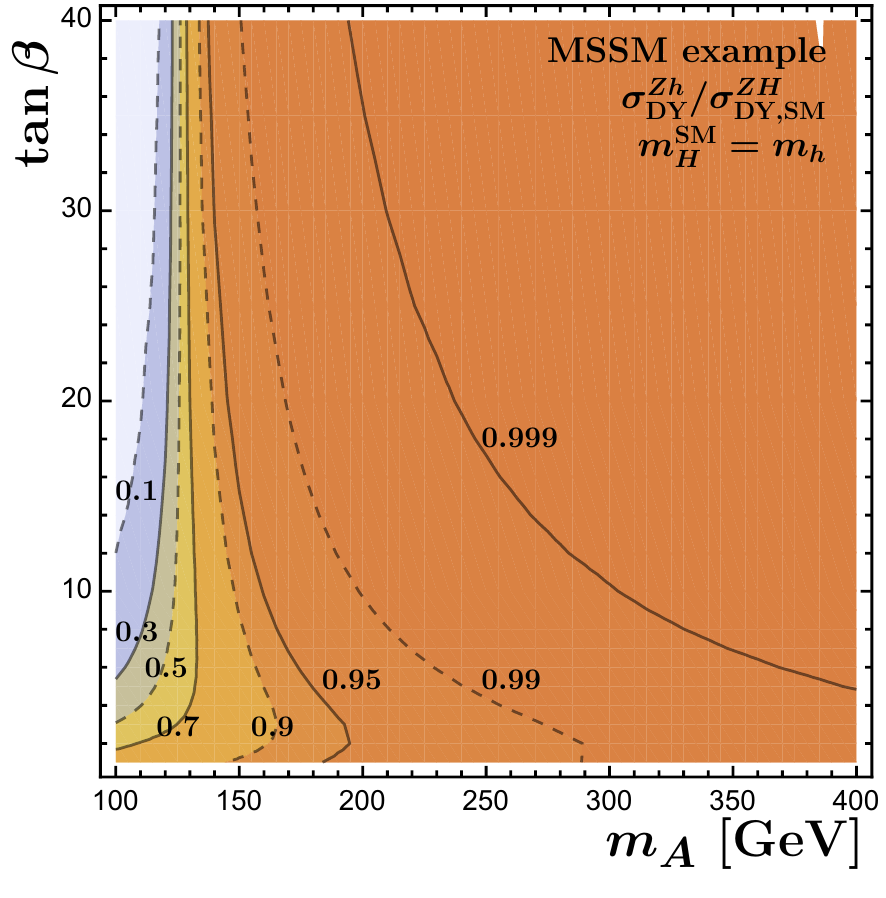} \\[-0.5cm]
      (a) & (b)\\
      \includegraphics[width=.45\textwidth]{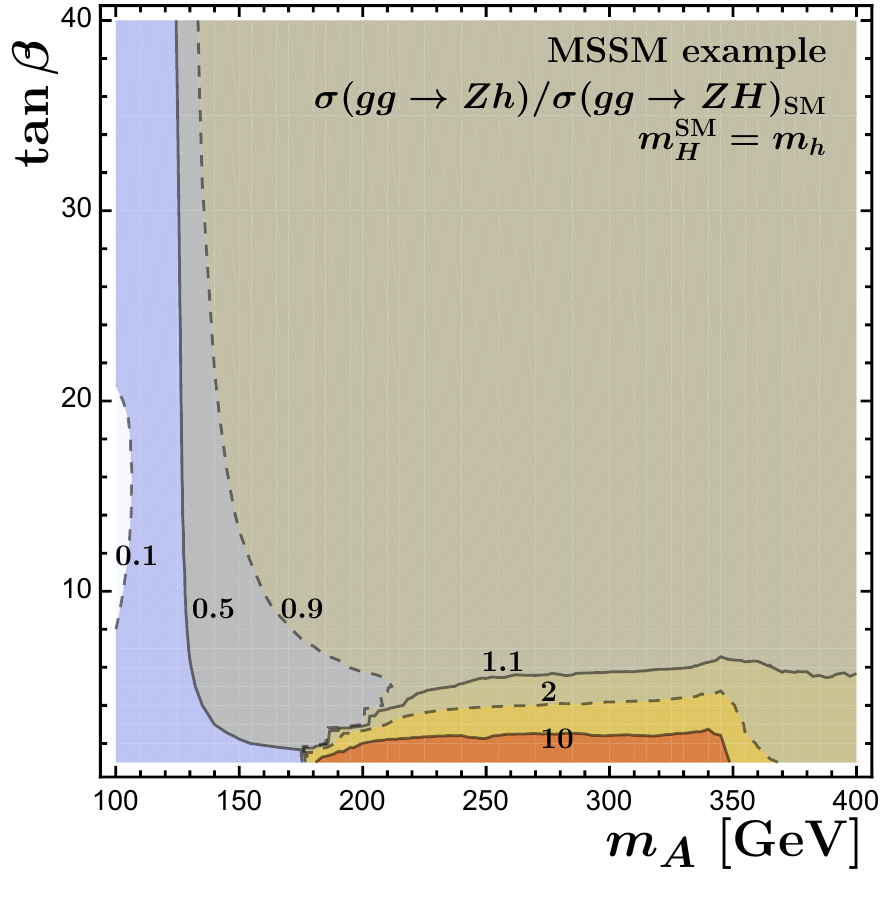} &
      \includegraphics[width=.45\textwidth]{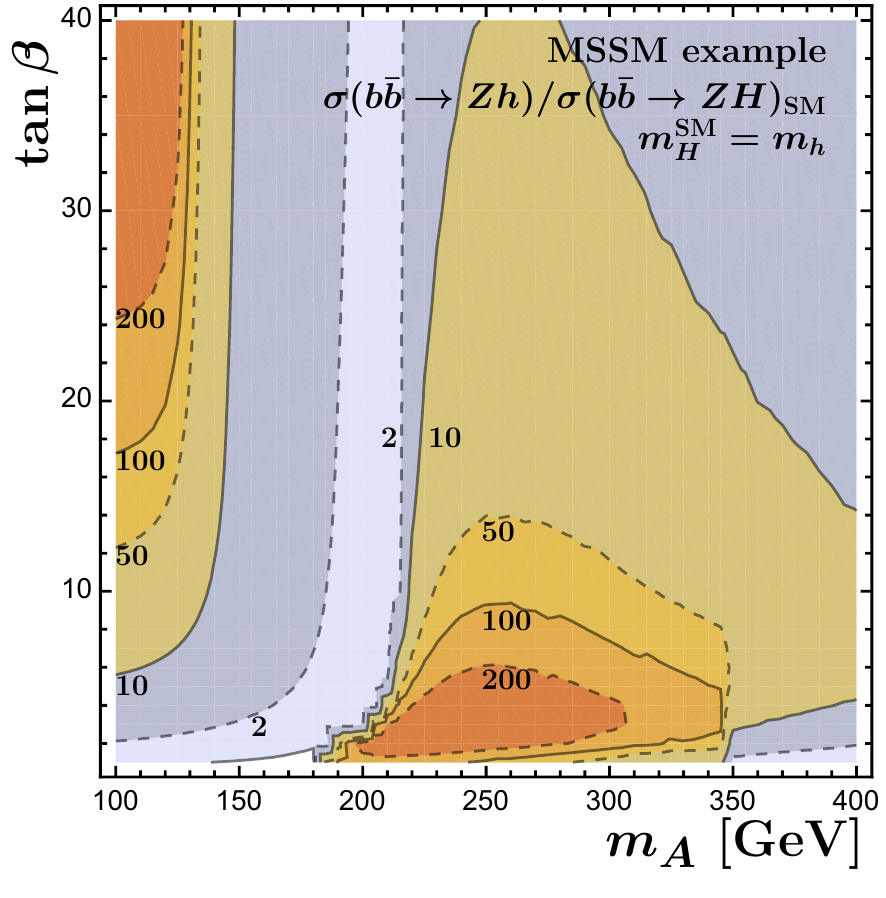}\\[-0.5cm]
      (c) & (d)
    \end{tabular}
    \parbox{\textwidth}{
      \caption[]{\label{fig:mssmlighthiggs}\sloppy Normalized production
        of a light Higgs boson in the \mssm{} as a function of $m_A$ in
        GeV and $\tan\beta$ showing (a) the Higgs mass in GeV; (b) the
        \dy{}-like cross section; (c) the gluon-induced cross section
        and (d) the bottom-quark initiated cross section.  All cross
        sections are normalized to the \sm{} Higgs cross section
        obtained for the same Higgs mass.}}
  \end{center}
\end{figure}

We present a numerical example for light Higgs production in association
with a $Z$ boson in the \mssm{} in \fig{fig:mssmlighthiggs} for the
$13$\,TeV \lhc{}. The Higgs mass and mixing is obtained from \fh{} {\tt
  2.13.0} for an \mssm{} scenario with relatively heavy \susy{} masses,
$M_{\text{SUSY}}=2$\,TeV, $m_{Q3}=m_{U3}=m_{D3}=1.5$\,TeV, $M_1=M_2=1$\,TeV, $M_3=2.5$\,TeV,
$\mu=1$\,TeV and $A_t=A_b=A_\tau=(2.7+1.5/\tan\beta)$\,TeV. Since there are no
squark contributions to $\zh{}$ production at order $\mathcal{O}(\alpha_s^2)$, the
features correspond to the well-known decoupling behavior for the light
Higgs boson. Since all \susy{} states are above $1$\,TeV, they do not
enter the pseudoscalar total decay width, which is relevant for the
sub-processes $\ggzhsmall{}$ and $\bbzhsmall{}$.

\fig{fig:mssmlighthiggs}~(a) shows the light Higgs mass in the
$m_A$-$\tan\beta$-plane as obtained by \fh.  Only for large enough
values of $m_A$ and $\tan\beta$, a light Higgs mass $m_h$ in the range
of $(125\pm 3)$\,GeV is reached through higher-order corrections to the
light Higgs boson mass, which at tree-level is bound to be below the
$Z$-boson mass $m_Z$.  This is the decoupling region, i.e.\ all light
Higgs boson couplings to \sm{} particles are close to their \sm{} value.
This explains the behavior of the \dy{}-like component to
\higgsstrahlung{}, shown in \fig{fig:mssmlighthiggs}~(b), since it
directly follows the squared coupling $(g_{VV}^h)^2$ of the light Higgs
boson~$h$ to gauge bosons~$V\in\{W,Z\}$.  All components of the cross
sections in \fig{fig:mssmlighthiggs}~(b)--(d) are normalized to their
corresponding \sm{} results, assuming $m_H^\text{\sm}=m_h$ for each
parameter point.

The gluon-fusion induced component $\ggzhsmall{}$, depicted in
\fig{fig:mssmlighthiggs}~(c), reveals a richer structure. For low values
of $m_A\lesssim 150$\,GeV, it is $g_{VV}^h\ll 1$, so that the
contribution from triangle diagrams is small, whereas the box diagrams
proportional to the Higgs Yukawa couplings depend on
$\tan\beta$. In fact, the minimal cross section is obtained for
intermediate values of $\tan\beta$, where the coupling $g_t^h$ to top
quarks is reduced w.r.t.\ the \sm, and the coupling $g_b^h$ to bottom
quarks is not significantly enhanced.  Towards larger values of $m_A$,
decoupling is reached rather quickly. However, for $m_A\gtrsim m_h+m_Z$,
the contribution $gg\to A\to \zh{}$ can be kinematically resonant. It
plays an important role for low values of $\tan\beta$, where it enhances
the cross section relative to the \sm{} Higgs cross section by more than
a factor of $10$.  The fact that this effect extends to values of $m_A$
below $200$\,GeV is due to the reduced light Higgs boson mass
$m_h<100$\,GeV at low values of $\tan\beta$.  For larger values of
$\tan\beta$, the coupling $g_{VV}^h$ quickly approaches $1$, while the
coupling $g_Z^{Ah}$ almost vanishes. Thus the contribution $gg\to A\to
\zh{}$ is only relevant for very low values of $\tan\beta$, where at the
same time the Yukawa coupling of the pseudoscalar $g_t^A$ is large.
Another kinematic threshold is observed at $m_A\approx 2m_t$, where the
decay $A\to t\bar t$ opens up. The increased pseudoscalar width
$\Gamma_A$ reduces the contribution due to $gg\to A\to \zh{}$.

Finally, we present the component $\bbzhsmall{}$ in
\fig{fig:mssmlighthiggs}~(d). The \mssm\ cross section exceeds the
\sm\ cross section (which is tiny) by one to two orders of magnitude
almost throughout the entire $m_A$-$\tan\beta$-plane.  For $m_A\lesssim
150$\,GeV, i.e.\ outside the decoupling region, the light-Higgs-boson
bottom-quark Yukawa coupling $g_b^h$ is strongly dependent on
$\tan\beta$, which explains the steep increase of the cross section with
$\tan\beta$.  Relevant Feynman diagrams are the $t$- and $u$-channel
contributions shown in \fig{fig:bottomdiagrams}~(a) and (b).  For
$m_A\gtrsim 200$\,GeV, those Feynman diagrams are less relevant, since
the light-Higgs-boson bottom-quark Yukawa coupling $g_b^h$ approaches
its \sm{} value---though slightly slower than the top-quark Yukawa
coupling $g_t^h$.  On the other hand, the pseudoscalar contribution
$b\bar b\to A\to \zh{}$ is manifest (see \fig{fig:bottomdiagrams}~(c)),
since the pseudoscalar can be resonant.  Here the strong enhancement of
the pseudoscalar bottom-quark Yukawa coupling $g_b^A$ with increasing
$\tan\beta$ leads to observable differences also at high values of
$\tan\beta$ despite the fact that $g_Z^{Ah}$ almost vanishes,
i.e.\ decoupling is reached.  We note that for this example we did not
include $\Delta_b$ resummation.

\begin{table}
\begin{center}
\begin{tabular}{|c|cccc|}
\hline
\susy{} scenario &  $m_{Q3}=m_{U3}=m_{D3}$ & $X_t$ [GeV] & $(m_{\tilde t_1},m_{\tilde t_2})$ & $(m_{\tilde b_1},m_{\tilde b_2})$\\
 & [GeV] & $A_b=A_t$ & [GeV,GeV] & [GeV,GeV]\\\hline
 1 & $3000$ & $3500$ & $(2902,3103)$ & $(2999,3003)$\\
 2 & $3000$ & $0$    & $(3004,3004)$ & $(2998,3004)$\\
 3 & $1200$ & $2300$ & $(1035,1365)$ & $(1196,1210)$\\
 4 & $1200$ & $0$    & $(1211,1211)$ & $(1194,1212)$\\
 5 & $600$  & $1000$ & $(464,748)$ & $(589,622)$\\ 
 6 & $600$  & $0$    & $(622,623)$ & $(586,624)$\\\hline
\end{tabular}
\caption{\label{tab:squarkmasses} Choice of squark masses for the $\pt{}$ and $M_{\zphi{}}$ distributions shown in \fig{fig:mssmpt}.}
\end{center}
\end{table}

\begin{figure}[htb!]
  \begin{center}
    \begin{tabular}{cc}
      \includegraphics[width=.45\textwidth]{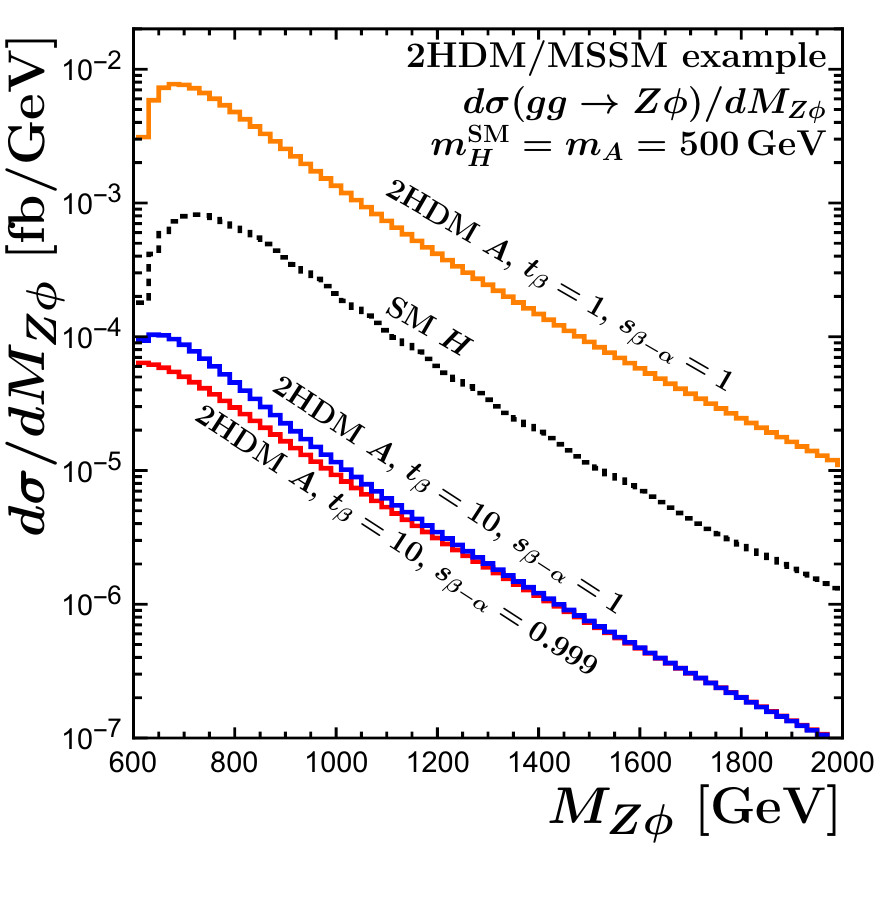} &
      \includegraphics[width=.45\textwidth]{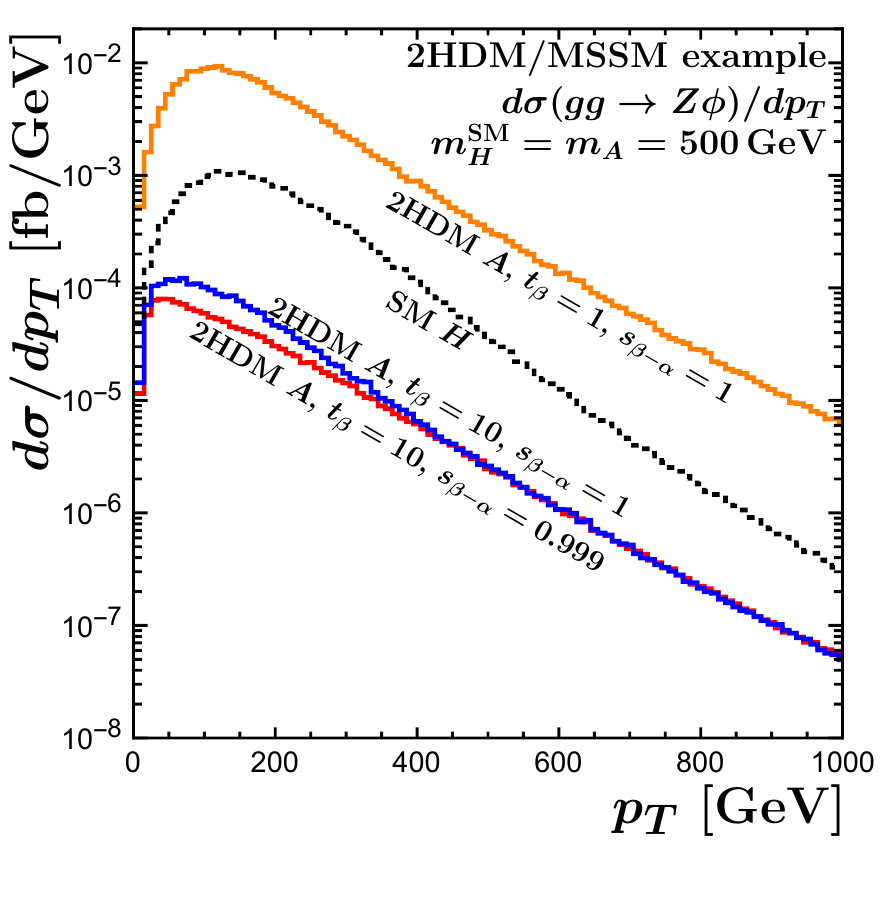} \\[-0.5cm]
      (a) & (b)\\
      \includegraphics[width=.45\textwidth]{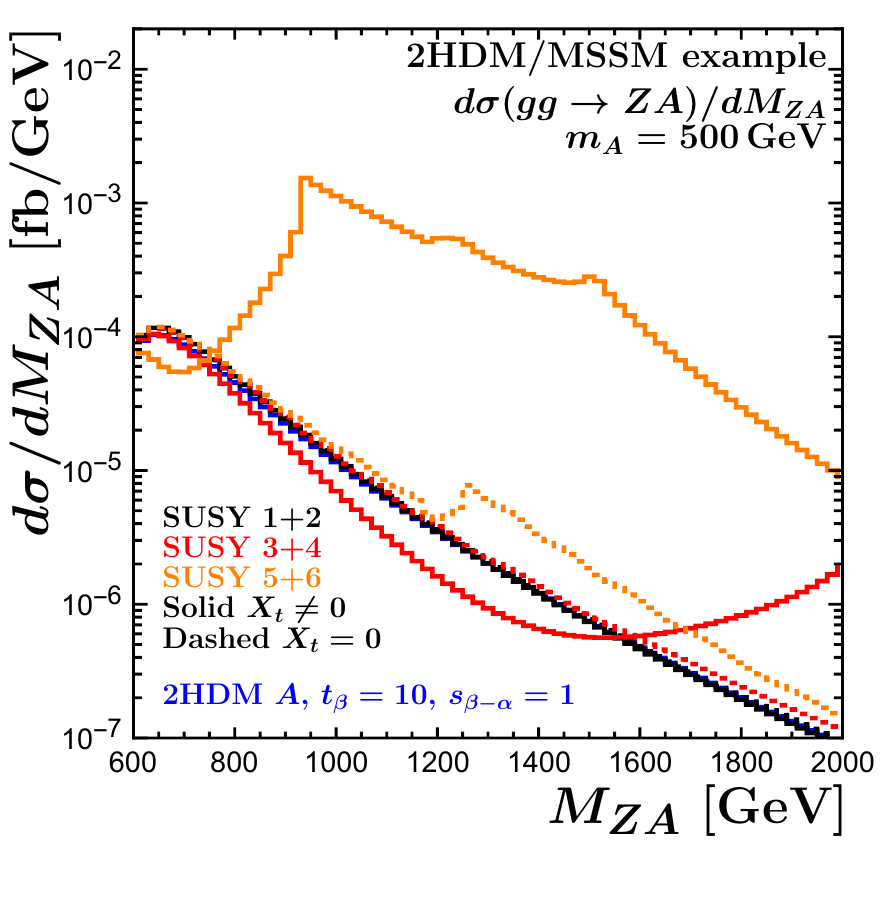} &
      \includegraphics[width=.45\textwidth]{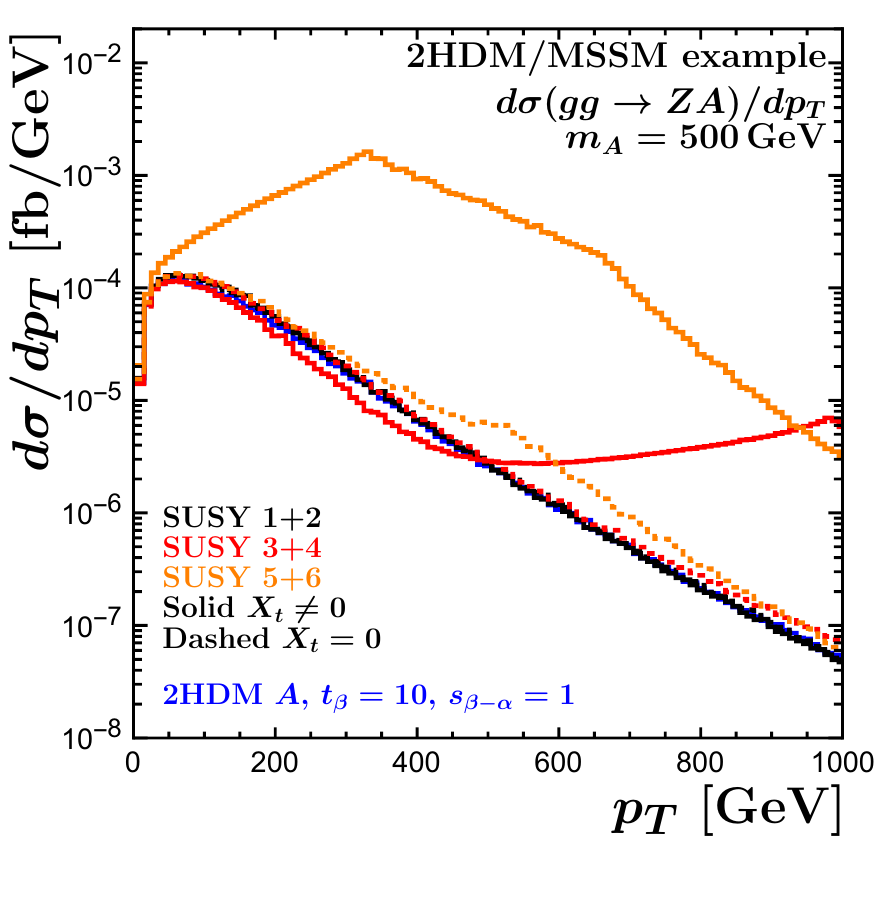}\\[-0.5cm]
      (c) & (d)
    \end{tabular}
    \parbox{\textwidth}{
      \caption[]{\label{fig:mssmpt}\sloppy Invariant mass $M_{\zphi{}}$
        and $\pt{}$ distributions for a pseudoscalar $A$ with mass
        $m_A=500$\,GeV produced in $gg\to \zA{}$ at the $13$\,TeV
        \lhc{}: (a) and (b) compare the distributions of a \sm{} Higgs
        (with $m_H=500$\,GeV) and different \thdm{} scenarios; (c) and
        (d) compare the pure \thdm{} case with $\tan\beta=10$ and
        decoupling $\sin(\beta-\alpha)=1$ with additional squark
        contributions. The \susy{} scenarios are defined in
        \tab{tab:squarkmasses}.}}
  \end{center}
\end{figure}

Before we close our discussion on the \mssm{}, we show $\pt{}$
distributions and $M_{\zphi{}}$ distributions for the gluon-induced
production of a pseudoscalar with a mass of $500$\,GeV to demonstrate
the influence of squark contributions.  The cross sections are produced
for the $13$\,TeV \lhc{}.  We choose $\tan\beta=10$ and $\mu=1$\,TeV and
pick six choices of squark masses shown in \tab{tab:squarkmasses}. Again
we link to \fh\ version {\tt 2.13.0} to obtain the light Higgs mass as
well as the squark masses.  Only scenarios $1$ and $3$ yield a light
Higgs mass in accordance with experimental data, i.e.\ within $125\pm
3$\,GeV. Scenario $6$ yields the lowest mass with $110$\,GeV.  Such
scenarios can be considered pedagogical examples to demonstrate squark
mass thresholds and the influence of $X_t$.  The process $gg\to \zA{}$
also has a light and heavy Higgs contribution $gg\to \{h,H\}\to \zA{}$,
where the former vanishes for exact decoupling $g_{VV}^h=1$,
i.e.\ $g_Z^{Ah}=0$.  The squark contributions in exact decoupling are
thus relevant only for $gg\to H\to \zA{}$.  In \fig{fig:mssmpt}~(a) and
(b) we first show the influence of the choice of $\tan\beta$ in a
\thdm{} of type II and compare it with the distribution of a ``\sm{}
Higgs'' with the same mass of $500$\,GeV.  Also the influence of $gg\to
\{h,H\}\to \zA{}$ can be inferred from the difference between
$\sin(\beta-\alpha)=1$ and $\sin(\beta-\alpha)=0.999$.  The choice of
$\tan\beta=10$ minimizes the cross section, since again $g_t^A$ is
reduced, while $g_b^A$ is only moderate.  Therefore one expects large
relative squark effects in this region of the parameter space, which we
show in \fig{fig:mssmpt}~(c) and (d).  Indeed, in the invariant mass
distribution, the squark mass thresholds are clearly visible in
scenarios 5+6.  Given that the partonic center-of-mass energy
$\sqrt{\hat s}$ has to fulfill $\hat{s}\geq
m_\phi^2+m_Z^2+2p_T^2+2\sqrt{(m_Z^2+p_T^2)(m_\phi^2+p_T^2)}$, also the
$\pt{}$ distribution in scenario 5 probes the squark mass thresholds
$2m_{\tilde t_1}$ and $2m_{\tilde t_2}$ at around
$\pt{}\approx 320-330$\,GeV and $\pt{}\approx 660-670$\,GeV, respectively.  Also the inclusive
cross section in scenario 5 is enhanced by a factor of more than $20$,
due to resonant squark contributions.  It is remarkable that even larger
squark masses still lead to a significant deformation at low $\pt{}$ and
$M_{\zphi{}}$ of the distributions, in particular in case of large squark mixing $X_t\neq 0$.
However, the cross sections are very small and hard to access
experimentally.

%- }}}
%- {{{ subsection{\cp\ mixing among Higgs bosons}

\subsection{\cp\ mixing among Higgs bosons}
\label{sec:cpmix}

By default, \vhnnlo\ assumes the Higgs \cp\ eigenstates $(h,H,A)$ of the
\thdm\ and the \mssm\ to be identical to the mass eigenstates
$(h_1,h_2,h_3)$.  However, it does provide the option for a naive mixing
according to
\begin{equation}
  \begin{split}
    \left( \begin{array}{c} h_1 \\ h_2 \\ h_3 \end{array} \right) = R
    \left( \begin{array}{c} h \\ H \\ A \end{array} \right)\,,
  \end{split}
\end{equation}
where the orthogonal rotation matrix~$R$ is parametrized in the form
\begin{equation}
  \begin{split}
\qquad R
    = \begin{pmatrix} c_{12}c_{13} & s_{12}c_{13} & s_{13}
      \\ -s_{12}c_{23}-c_{12}s_{23}s_{13} &
      c_{12}c_{23}-s_{12}s_{23}s_{13} & s_{23}c_{13}
      \\ s_{12}s_{23}-c_{12}c_{23}s_{13} &
      -c_{12}s_{23}-s_{12}c_{23}s_{13} &
      c_{23}c_{13} \end{pmatrix}
    \label{eq:1}
  \end{split}
\end{equation}
with $c_{ij}=\cos\theta_{ij}$ and $s_{ij} =\sin\theta_{ij}$. The three
independent angles $\theta_{12}$, $\theta_{13}$, and $\theta_{23}$ can
be specified in the input file. For all but the
$\ggzphi$ and the $\bbzphi$ contributions, this simply results in a
modification of the $VV\phi$ couplings according to
\begin{equation}
  \begin{split}
    g^{i}_{VV} = \sum_{\phi\in\{h,H,A\}}R_{i\phi}\,g^{\phi}_{VV}\,,
    \label{eq:2}
  \end{split}
\end{equation}
where $R_{ih}, R_{iH}, R_{iA}$ refer to the first, second, and third
column of the matrix $R$, respectively, the $g^{\phi}_{VV}$
($V\in\{W,Z\}$) are the couplings of the specified model
\textit{without} \cp\ mixing (see \eqn{eq:gvvh} and \fig{fig:frules}),
and $g^{i}_{VV}$ is the coupling constant for the $h_iVV$ vertex.  Note
that, up to the sign, $\theta_{12}$ has the same meaning as $\alpha$,
defined in \eqn{eq:gvvh} for the \thdm{}, since both angles determine
the rotation among the two \cp-even Higgs eigenstates. We keep this
redundancy in \vhnnlo{} for the sake of a transparent implementation of
the mixing.

Finally, we note that the rotation defined through \texttt{Block HCPMIX}
is applied \textit{after} a possible rescaling of the couplings by the
entries in \texttt{Block FACTORS} (see \tab{tab:factors} in
\sct{sec:genericoptions}).

%- {{{ tab:hcpmix:

\begin{table}
\begin{center}
\begin{tabular}{|ccl|}
\hline
\multicolumn{3}{|c|}{\tt Block HCPMIX}\\
\hline
\textrm{entry} & \textrm{default} & meaning\\
\hline
 1 & $0.$ & $\theta_{12}$\\
 2 & $0.$ & $\theta_{13}$\\
 3 & $0.$ & $\theta_{23}$\\
\hline
\end{tabular}
\caption[]{\label{tab:hcpmix} The mixing angles for \cp\ mixing in radian.}
\end{center}
\end{table}

%- }}}

The $\ggzphi$ component of the partonic cross section for a specific
$\phi$ involves also the other two Higgs bosons through $s$-channel
exchange. \vhnnlo\ combines the corresponding helicity amplitudes by
assuming a \cp-even and a \cp-odd coupling of each Higgs mass eigenstate
to the top and the bottom quarks, according to
\begin{equation}
  \begin{split}
    {\cal L}_\text{Y} = - \sum_{j=1}^3\sum_{q\in\{b,t\}}\frac{m_q}{v}\bar
    q\left[\sum_{\phi\in\{h,H\}}R_{j\phi}g_q^\phi
      - i R_{jA}g_q^A\gamma_5\right]qh_j
    \label{eq:3}
  \end{split}
\end{equation}
and using the appropriate $h_ih_jZ$ couplings
\begin{equation}
  \begin{split}
    g^{ij}_Z = \sum_{\phi\in\{h,H\}}
    \left(R_{iA}R_{j\phi}-R_{i\phi}R_{jA}\right)g^{A\phi}_Z\,,
    \label{eq:4}
  \end{split}
\end{equation}
where $g^\phi_q$ and $g^{A\phi}_Z$ are again the $q\bar q\phi$ and
$ZA\phi$ coupling constants of the specified model \textit{without}
\cp\ mixing (see \tab{tab:2hdmcouplings}, \eqn{eq:gahz}, and
\fig{fig:frules}).  Currently, the $\bbzphi$ component cannot be
evaluated with \cp\ mixing within \vhnnlo, so \blockentry{ORDER}{4}{}
needs to be set to \texttt{-1} when \cp\ mixing is switched on.  Mixing
is activated by including the \texttt{Block HCPMIX} in the input file,
where the mixing angles are specified, see \tab{tab:hcpmix}. In order to
obtain the cross section for $h_1$, $h_2$, or $h_3$ production, one
needs to set \blockentry{VHATNNLO}{2}{} to \texttt{11}, \texttt{12}, or
\texttt{21}, respectively. Similarly, the masses $m_{h_1}$, $m_{h_2}$,
and $m_{h_3}$ are defined through \blockentry{MASS}{25,35,36}{},
respectively.

Instead of a consistent implementation of \cp\ violation in the
\mssm\ through complex parameters, \vhnnlo\ simply assumes that the
quark and vector-boson couplings of the Higgs are modified as described
above, and in addition, the Higgs-squark couplings are affected by an
analogous rotation:
\begin{equation}
  \begin{split}
    g^{q,i}_{kl} = \sum_{\phi\in
      \{h,H,A\}}R_{i\phi}g^{q,\phi}_{kl}\,,\qquad k,l\in\{1,2\}\,,
    \label{eq:5}
  \end{split}
\end{equation}
where $g^{q,\phi}_{kl}$ is the coupling constant of $\phi$ to the
squarks $\tilde q_k$ and $\tilde q_l$ in the \cp-conserving \mssm, see
\citere{Harlander:2012pb}.  This approach misses contributions due to
\cp\ violation in the squark sector, which induces new couplings of the
three neutral eigenstates $\{h,H,A\}$ to squarks.

%- {{{ fig:hcpmix

%
\begin{figure}[h!]
  \begin{center}
    \begin{tabular}{cc}
      \includegraphics[height=.4\textheight]{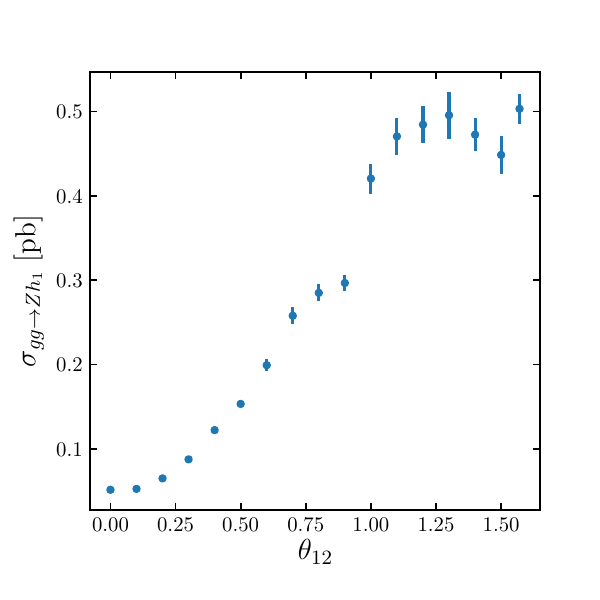}
      &
      \includegraphics[height=.4\textheight]{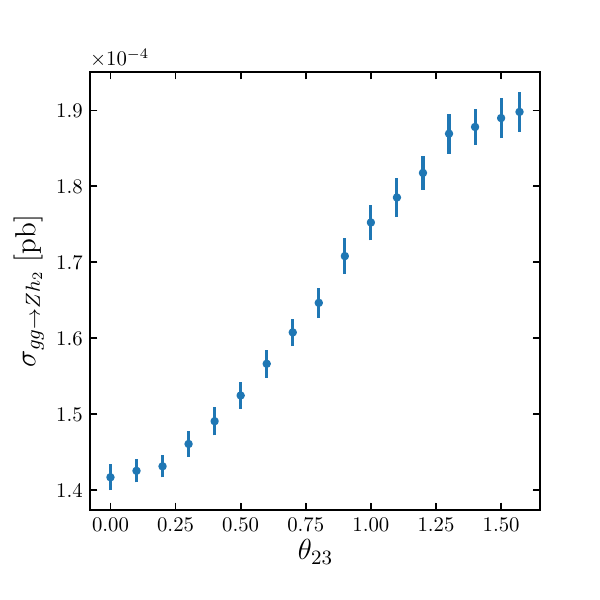}
      \ \\[-0.5cm]      (a) & (b)\\[-0.1cm]
      \multicolumn{2}{c}{\includegraphics[height=.4\textheight]{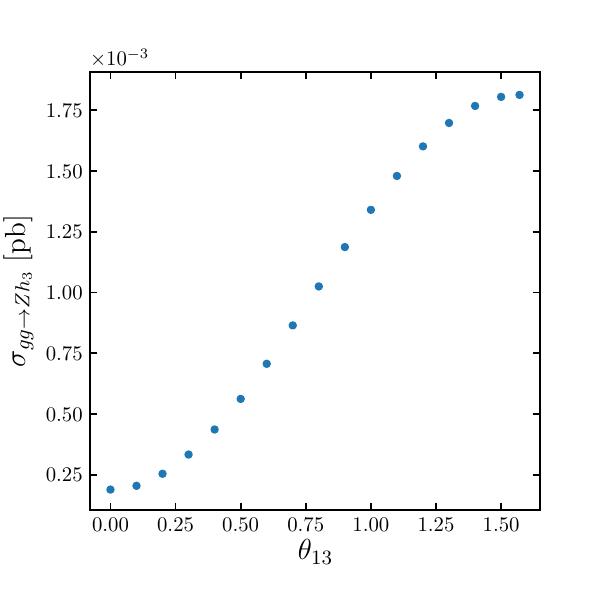}}
       \ \\[-0.5cm]   \multicolumn{2}{c}{(c)} 
    \end{tabular}
    \parbox{\textwidth}{
      \caption[]{\label{fig:hcpmix}\sloppy The $\ggzphi$ component of $\zh$
        production including \cp\ mixing for the production of $h_1$,
        $h_2$, and $h_3$. The left end of each plot corresponds, from
        $(a)$--$(c)$, to the production of $h$, $H$, and $A$ within a
        type-II \thdm\ with $\tan\beta=20$ and $\sin(\beta-\alpha)=1$. The Higgs
        masses are set to $m_{h_1}=125$\,GeV and 
        $m_{h_2}=m_{h_3}=300$\,GeV.}}
  \end{center}
\end{figure}
%

%- }}}

As an example, \fig{fig:hcpmix} shows the dependence of the $\ggzphi$
contribution to the cross section on one of the relevant mixing
parameters successively for $\phi=\phi_1$, $\phi_2$, and $\phi_3$. No
mixing corresponds to a type-II \thdm\ with $\tan\beta=20$ and $\sin(\beta-\alpha)=1$
in this figure. The Higgs boson masses are $m_{h_1}=125$\,GeV and $m_{h_2}=m_{h_3}=300$\,GeV.
As the mixing angle $\theta_{ij}$ increases, the
produced Higgs boson $h_i$ gradually assumes the \cp\ properties and the
couplings of what is $h_j$ in the unmixed case, while the mass remains
unchanged.

%- }}}  
%- {{{ subsection{Vector-like quarks}

\subsection{Vector-like quarks}
\label{sec:vlq}

Vector-like quarks (\vlq{}s) impact the gluon-initiated contribution
to \higgsstrahlung{}. We adopt the general parametrization of
\citere{Aguilar-Saavedra:2013qpa} for adding a single multiplet of
vector-like quarks (\vlq{}s) to the \sm{} Lagrangian. There are seven
possible representations which can mix with the \sm{} quarks through
Yukawa interactions\,\cite{delAguila:2000aa,delAguila:2000rc}. They all
involve vector-like quarks $T$ and/or $B$ with electric charges $+2/3$
and $-1/3$, respectively, and possibly one of $X$ and $Y$, with electric
charges $+5/3$ and $-4/3$. In this way, one arrives at two
singlets, three doublets, and two triplets:
  \begin{align}
    \text{singlets}:&& 1&: (T)\,,
    &2&: (B)\,,&&\nonumber\\
    \text{doublets}:&&
    3&: (XT)\,,
    &4&: (TB)\,,
    &5&: (BY)\,,
    \label{eq:vlqreps}
    \\
    \text{triplets}:&&
    6&: (XTB)\,,
    &7&: (TBY)\,.&&\nonumber
  \end{align}
Sizable mixing is only assumed between $T$ and $t$, and $B$ and $b$,
where $t$ and $b$ are the \sm{} top and bottom quark. In each case, this
mixing is parametrized by an angle $\theta^q$ and a phase $\phi^q$:
\begin{equation}
  \begin{split}
    \left(\begin{matrix}q\\Q\end{matrix}\right)
    =
    \left(
    \begin{matrix}
      \cos\theta^q & -e^{i\phi^q}\sin\theta^q\\
      e^{-i\phi^q}\sin\theta^q & \cos\theta^q
    \end{matrix}
    \right)
    \left(\begin{matrix}q^0\\ Q^0\end{matrix}
    \right)\,,
  \end{split}
\end{equation}
where the superscript ``0'' denotes the weak eigenstates in order to
distinguish them from the mass eigenstates. This parametrization of the
mixing applies to $(q,Q)=(b,B)$ and $(q,Q)=(t,T)$, and separately to
their left- and right-handed components. The left- and right-handed
mixing angles are related by
\begin{align}
    &\tan\theta^q_R = \frac{m_q}{m_Q}\tan\theta_L^q &&\qquad \text{for
    singlets and triplets, and}\\ &\tan\theta^q_L =
  \frac{m_q}{m_Q}\tan\theta_R^q &&\qquad \text{for doublets.}
\label{eq:vlq_right_handed_mixing_angle}
\end{align}
The phases $\phi^q$ are the same for the left- and right-handed
components. For the triplet representations, one finds an additional
constraint of the form
\begin{equation}
  \begin{split}
    \sin 2\theta_L^b = a\frac{m_T^2-m_t^2}{m_B^2-m_b^2}\sin 2\theta_L^t\,,
  \end{split}
   \label{eq:vlq_triplet_mixing}
\end{equation}
where $a=\sqrt{2}$ for the triplet $(X,T,B)$, and $a=1/\sqrt{2}$ for the
triplet $(T,B,Y)$. Since the $T$ and $B$ quark share the same mass term
in the Lagrangian, there is another constraint on the mass splitting of
the form
\begin{subequations}
  \begin{align}
  m_T^2 \cos^2\theta_R^t + m_t^2\sin^2\theta_R^t &= m_B^2
  \cos^2\theta_R^b
  + m_b^2\sin^2\theta_R^b&&\text{for $(TB)$},
  \label{eq:vlq_mass_splitting_tb}\\
  m_T^2 \cos^2\theta_L^t + m_t^2\sin^2\theta_L^t &= m_B^2
  \cos^2\theta_L^b
  + m_b^2\sin^2\theta_L^b&&\text{for triplets}.
  \label{eq:vlq_mass_splitting_trip}
  \end{align}
\end{subequations}%
Without mixing, the \vlq{}s as introduced here would not couple to the
Higgs boson, and their couplings to the $Z$ boson would be purely vector
like. This means that they would not contribute to the $\ggzh$ process
at \lo{} due to charge conjugation invariance\footnote{The box
  contributions would vanish also due to the absence of \vlq{}-Higgs
  couplings.} (Furry's theorem). In fact, the \lo{} amplitude for
$\ggzh$ does not involve $X$ and $Y$ for this reason.

%- {{{ fig:vlq
%
\begin{figure}
  \begin{center}
    \begin{tabular}{ccc}
      \includegraphics[height=.14\textheight]{feynmandiagrams/ggZH_VLQ1.pdf} &
      \includegraphics[height=.14\textheight]{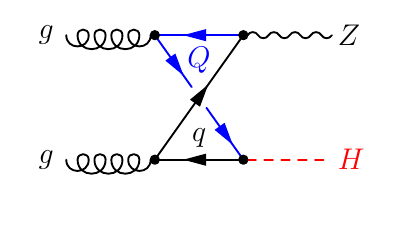} &
      \includegraphics[height=.14\textheight]{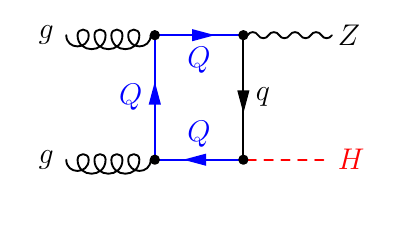} \\[-0.5cm]
      (a) & (b) & (c)\\
      \includegraphics[height=.14\textheight]{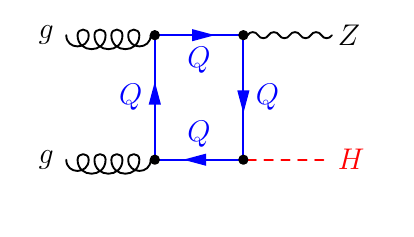} &
      \includegraphics[height=.14\textheight]{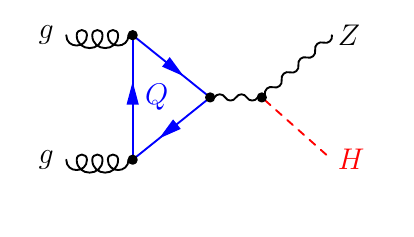} \\[-0.5cm]
      (d) & (e)
    \end{tabular}
    \parbox{\textwidth}{
      \caption[]{\label{fig:vlq}\sloppy New contributions to the $\ggzh$
        amplitude due to \vlq{}s. Here, $Q\in \{B,T\}$, and $q$ is the
        \sm{} quark which mixes with $Q$.  }}
  \end{center}
\end{figure}
%
%- }}}

%- {{{ fig:vlq_vertices:

\begin{figure}[h!]
  \begin{tabular}{rl}
    \raisebox{-2.3em}{\includegraphics[height = 0.115\textheight]{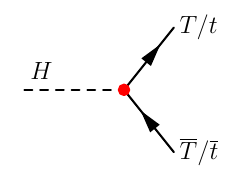}}
    &$\displaystyle=-i\frac{e m_{T/t} Y_{TT/tt}}{2 M_W s_W}$\\
    \raisebox{-2.3em}{\includegraphics[height=.115\textheight]{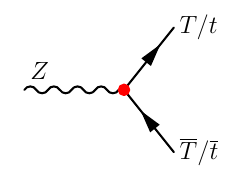}}
    &$\displaystyle=i\frac{e}{2c_Ws_W}\gamma^\mu\left(2Q_{t}s_W^2-\frac{X^L_{TT/tt}
      + X^R_{TT/tt}}{2} +
    \frac{X^L_{TT/tt}-X^R_{TT/tt}}{2}\gamma_5\right)$\\
    \raisebox{-2.3em}{\includegraphics[height=.115\textheight]{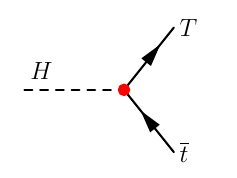}}
    &$\displaystyle=-i\frac{em_T}{2M_W}\left(\frac{Y^L_{tT}+Y^R_{tT}}{2}+\frac{Y^R_{tT} - Y^L_{tT}}{2}\gamma_5\right)$\\
    \raisebox{-2.3em}{\includegraphics[height=.115\textheight]{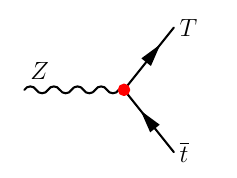}}
&$\displaystyle=-i\frac{e}{2c_Ws_W}\gamma^\mu\left(\frac{X^L_{tT}+X^R_{tT}}{2} + \frac{X^R_{tT}-X^L_{tT}}{2}\gamma_5\right)$\\
  \end{tabular}
  \caption{\label{fig:vlq_vertices}Feynman rules for the Higgs and $Z$ couplings due to a vector-like $T$ quark calculated with {\code FeynRules}~\cite{Christensen:2008py,Alloul:2013bka}.}
\end{figure}

%- }}}

However, aside from modified $t\bar{t}Z$, $b\bar{b}Z$, $t\bar{t}H$, and
$b\bar{b}H$ coupling constants with respect to their \sm{} values, the
mixing in the top and bottom sector leads to $t\bar{T}Z$, $b\bar{B}Z$,
$t\bar{T}H$, and $b\bar{B}H$ vertices, an axial-vector component of the
$T\bar{T}Z$ and $B\bar{B}Z$ coupling, and non-vanishing $B\bar{B}H$ and
$T\bar{T}H$ couplings. The \lo{} effects of the \vlq{}s on the $\ggzh$
process are therefore as follows: (i)~a modification of the \sm{} box
contributions; (ii)~additional box contributions involving one to four
$T$ or $B$ propagators, see \fig{fig:vlq}\,(a)--(d); (iii)~additional
triangle contributions involving $T$ or $B$, see \fig{fig:vlq}\,(e). The
Feynman rules corresponding to the vertices of two $T$ quarks with a
Higgs or $Z$, and a single $T$ quark with a top quark, a Higgs or a $Z$
boson are listed in \fig{fig:vlq_vertices} with generic couplings. Here,
$Q_{t}$ is the electric charge of the $t$ (and the $T$) quark,
$X^{L/R}_{TT}$ and $X^{L/R}_{tt}$ are the left- and right-handed
couplings of the $T$ and $t$ quark to the $Z$ boson, respectively;
$Y_{TT}$ and $Y_{tt}$ are the Yukawa couplings of the $T$ and $t$ quark;
and $X^{L/R}_{tT}$ and $Y^{L/R}_{tT}$ are the left- and right-handed
couplings of a $T$ and $t$ quark to a $Z$ boson and a Higgs.  Note that
vertices for a vector-like $B$ quark can be obtained by introducing an
additional minus sign to the couplings and changing $T\to B$, $t\to
b$. The generic couplings assume values which are specific to the
various representations. A full list can be found in
\citere{Aguilar-Saavedra:2013qpa}.  The couplings of \vlq{}s to gluons
are the same as for \sm{} quarks.  The amplitudes including \vlq{}s have
been evaluated with the help of {\tt FormCalc}.

The input block for \vlq{}s in \vhnnlo{} is described in
\tab{tab:vlq}. \blockentry{VLQ}{1}{} fixes the representation for the
\vlq{}s according to \eqn{eq:vlqreps}; \vhnnlo\ sets the generic
couplings discussed above according to this representation;
\blockentry{VLQ}{2}{} and \blockentry{VLQ}{3}{} set the mass of the $T$
and $B$ quark, respectively; the left- and right-handed mixing angles
for the bottom sector are set in \blockentry{VLQ}{4}{} and
\blockentry{VLQ}{5}{}, respectively, and in \blockentry{VLQ}{7}{} and
\blockentry{VLQ}{8}{} for the top sector; finally, the phases for the
bottom and top mixing can be set in \blockentry{VLQ}{6}{} and
\blockentry{VLQ}{9}{}. Note that there are conditions like
\eqn{eq:vlq_triplet_mixing} which constrain these parameters; the input
will be checked for consistency by \vhnnlo{}. In fact, most conveniently
the user may provide only an independent subset of the parameters, and
let \vhnnlo{} determine the remaining parameters from the theoretical
constraints described above. Specifically, for the $(TB)$ doublet, it is
sufficient to provide one of the following sets:
\begin{enumerate}
\item $m_B$, $m_T$, and one mixing angle;
\item $m_B$, ($\theta_L^b$ or $\theta_R^b$) and $\theta_R^t$;
\item $m_T$, ($\theta_L^t$ or $\theta_R^t$) and $\theta_R^b$.
\end{enumerate}
For the triplets $(TBY)$ and $(XTB)$, possible input sets are:
\begin{enumerate}
\item $m_B$ and ($\theta_L^b$ or $\theta_R^b$);
\item $m_T$ and ($\theta_L^t$ or $\theta_R^t$).
\end{enumerate}
Note, however, that the left- or right-handed mixing angle will always
be calculated for any representation, if only one of these mixings and
the corresponding mass of the vector-like quark is set as input. The
complete set of parameters, including the coupling factors appearing in
\fig{fig:vlq_vertices}, are listed in \texttt{Block VLQ} in the output file.

\begin{table}
  \begin{center}
\begin{tabular}{|>{\tt}c>{\tt}cl|}
  \hline
  \multicolumn{3}{|c|}{Block VLQ}\\
  \hline
  \textrm{entry} & \textrm{range} & meaning\\
  \hline
  1&  \{1,$\ldots$,7\} & \vlq{} representation\\
  2&    & $m_B$ [GeV]\\
  3&    & $m_T$ [GeV]\\
  4&    & $\theta_L^b$\\
  5&    & $\theta_R^b$\\
  6&    & $\phi^b$\\
  7&    & $\theta_L^t$\\
  8&    & $\theta_R^t$\\
  9&    & $\phi^t$\\
  \hline
\end{tabular}
  \end{center}
  \caption{\label{tab:vlq}Setting the parameters for vector-like
    quarks. The number of the representation refers to \eqn{eq:vlqreps}.}
\end{table}

To study the impact of \vlq{}s on the total cross section, we define the
ratios $R_{\theta}$ and $R_{\phi}$,
\begin{equation}
R_{\theta} = \frac{\sigma_{\ggzh{},\text{\sm{}} + \text{\vlq{}s}}}{\sigma_{gg\to \zH,\text{\sm{}}}}\Bigg|_{\phi^{t/b} = 0},\quad
R_{\phi} = \frac{\sigma_{\ggzh{},\text{\sm{}} + \text{\vlq{}s}}}{\sigma_{gg\to \zH,\text{\sm{}}+\text{\vlq{}s}}|_{\phi^{t/b} = 0}}\Bigg|_{\theta_{L/R}^{t/b} \neq 0}.
\end{equation}
In $R_\theta$ we normalize the total cross section of $\ggzh{}$ with
\vlq{}s at a given value of $\theta_{L/R}^{t/b}$ to the total cross
section of the gluon induced contribution of the \sm{}. Moreover, we
restrict ourselves to cases with $\phi^{t/b}=0$. For $R_\phi$ we choose
$\theta_{L/R}^{t/b} \neq 0$ since otherwise the \vlq{} will have no
effect on the cross section. Thus, instead of normalizing this result to
the \sm{} alone, we add the contributions of the \vlq{}s for $\phi^{t/b}
= 0$ in the denominator. Additionally, for $R_\phi$ we set
$\phi^t=\phi^b$. We choose the masses and the dominant mixing angle of
the vector-like $T$ partner as our input scheme for singlets and
doublets, and $m_T$ with $\theta_L^t$ as the input for the triplets.

\begin{figure}[h]
  \begin{center}
    \begin{tabular}{cc}
  \includegraphics[width=0.49\textwidth]{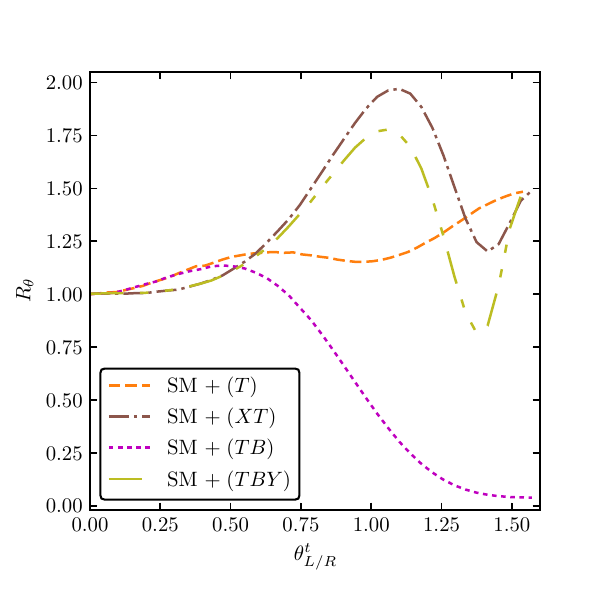} &
  \includegraphics[width=0.49\textwidth]{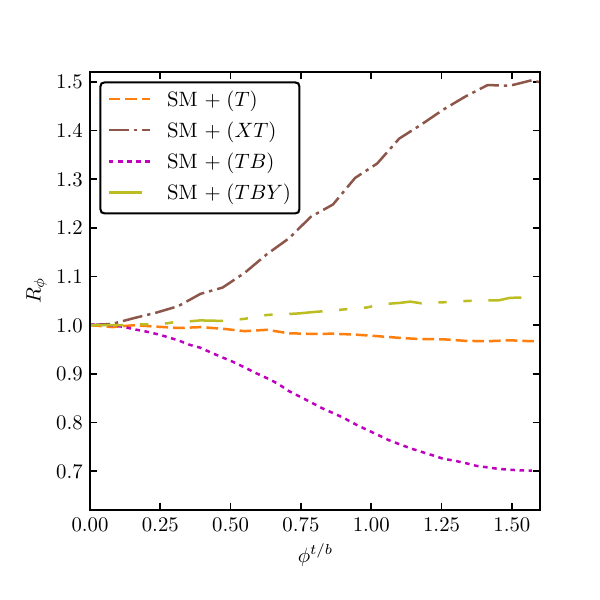}\\[-0.5cm]
   (a) & (b) 
    \end{tabular}
    \parbox{\textwidth}{
    \caption[]{\label{fig:vlq_total_theta_phi}\sloppy Total gluon induced cross section for $\zH{}$ production with additional \vlq{} multiplets
  normalized to the gluon induced production in the \sm{}
  for different mixing angles $\theta_{L/R}^{t}$ (a) and different phases $\phi^t=\phi^b$ (b). The vector-like quark masses are set to $m_T=m_B=600$\,GeV.
  In the case of the triplet $(TBY)$, only $m_T=600$\,GeV is assumed and the mixing angle $\theta_L^{b}$ together with the
  mass $m_B$ are calculated accordingly to \eqn{eq:vlq_triplet_mixing} and \eqn{eq:vlq_mass_splitting_trip}.}}
  \end{center}
\end{figure}

\fig{fig:vlq_total_theta_phi} shows these ratios for four different
representations of \vlq{}s by varying $\theta_{L}^t$ (for singlets and triplets)
or $\theta_R^t$ for doublets in (a) and
$\phi^{t/b}$ in (b). The other settings are described in the caption
of the figure. We first discuss \fig{fig:vlq_total_theta_phi}~(a)
in detail: All curves start at the \sm{} cross section
for zero mixing angle, where the \vlq{}s have no couplings to the \sm{}
particles. For the vector-like top singlet (T), the total cross section
for $\ggzh{}$ steadily increases with the mixing angle, until it reaches
a maximum of about $1.5$ times the \sm{} cross section at maximal mixing,
i.e.\ $\theta_L^{t} = \tfrac{\pi}{2}$, where the couplings of the $t$
and the $T$ are exactly interchanged w.r.t.\ the \sm{}. Therefore, the
cross section at $\theta_L^{t}=\tfrac{\pi}{2}$ is the \sm\ value for a
hypothetical top quark mass of $600$\,GeV.

\begin{figure}[h]
  \begin{center}
    \begin{tabular}{cc}
  \includegraphics[width=0.49\textwidth]{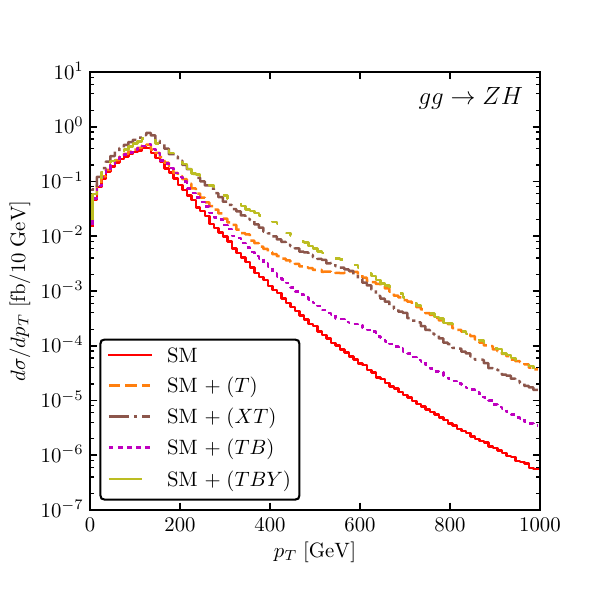}&
  \includegraphics[width=0.49\textwidth]{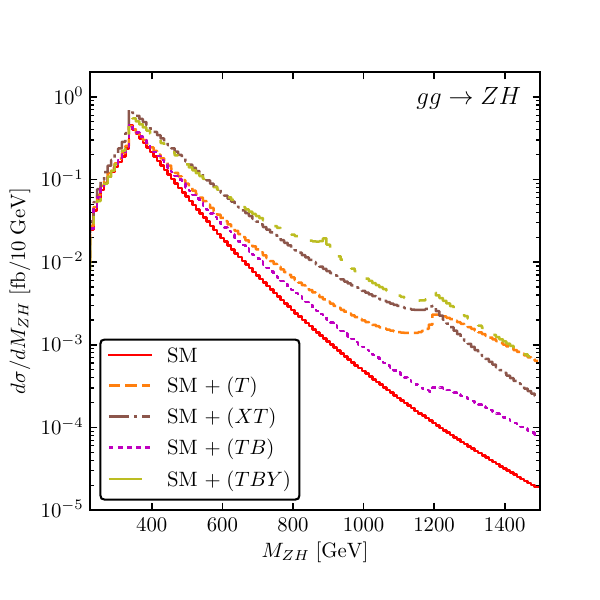}\\[-0.5cm]
   (a) & (b) 
    \end{tabular}
    \parbox{\textwidth}{
    \caption[]{\label{fig:vlq_pt_mhz}\sloppy Transverse momentum (a) and
      invariant mass (b) distributions in the \sm{} for the gluon
      induced process compared to the \sm{} with an additional \vlq{}
      multiplet The input parameters are $m_T=m_B=600$\,GeV for all
      representations other than $(TBY)$, where $m_T=600$\,GeV is chosen
      and $m_B = 447.08$\,GeV. The mixing angles are $\theta_L^t=0.5$
      for $(T)$, $\theta_R^t=1.02$ for $(XT)$, $\theta_R^t = 0.471$ for
      $(TB)$, and $\theta_L^t= 1.06$ for $(TBY)$.}}
  \end{center}
\end{figure}

This is also true for the $(XT)$ doublet, where the contributing
diagrams are the same as for the $(T)$ singlet. Since the couplings are
different, however, and the dominant mixing is the right-handed one, the
intermediate evolution of the cross section differs from the $(T)$
singlet. It reaches almost $\sim 200\%$ of the \sm\ cross section at
about $\theta_R^t=1.1$, which is mainly due to the enhancement of the
coupling to $t$ in this region (as can be seen
in \fig{fig:vlq_pt_mhz}).

For the $(TB)$ doublet, the behavior is strikingly different.  While it
resembles the behavior of a $(T)$ singlet at low mixing angles, which is
due to the dominance of the top quark in this regime, it quickly drops
down to very small values as the mixing increases towards
$\tfrac{\pi}{4}$, before it nearly vanishes at maximal mixing, where $T$
and $B$ assume the couplings of their \sm{} partners, while those of $t$
and $b$ vanish. Although both \vlq{}s carry the same mass, at maximal
mixing their couplings to the \sm{} bosons slightly differ due to
\eqn{eq:vlq_right_handed_mixing_angle}. If they coupled equally, their
amplitudes would be of equal magnitude but of opposite sign for maximal
mixing, such that they would completely cancel in the cross section. The
same behavior would be observed in the \sm{}, if the $b$ and the $t$
quark had identical masses.

Finally, the contributing diagrams for the $(TBY)$ triplet are the same
as for the $(TB)$ doublet, but with different couplings. We therefore do
not see the same cancellation as for the ($TB$) doublet, which is even
more suppressed since only at minimal and maximal mixing the $T$ and $B$
masses are degenerate. Instead, one observes similar features as in
$(XT)$, with the maximum shifted towards $\theta_R^t= 1.1$. At maximal
mixing, the cross section again assumes the value found for $(T)$ and
$(XT)$. In this case, however, all \sm{} and $T$ quark contributions
vanish, and the $B$ quark is the only contribution left. Since for
maximal mixing one finds $m_B=m_T=600$\,GeV and the couplings are the
same as for the $(T)$ singlet, their results coincide.

In \fig{fig:vlq_total_theta_phi} (b) we can observe the impact of
$\phi^{t/b}$ for the four different representations. The mixing angle of
the top and bottom sector is set to $\theta_L^{t/b} = \tfrac{\pi}{16}$,
where in the case of the triplet $\theta_L^b$ and $m_B$ are calculated
according to \eqn{eq:vlq_triplet_mixing} and
\eqn{eq:vlq_mass_splitting_trip}. Since the phase $\phi^{t/b}$ only
appears in mixed couplings of quarks, vector-like-quarks and \sm{}
bosons, one can probe the magnitude of the contributions in which these
mixed vertices appear. For the case of $\ggzh{}$ this only affects box
diagrams. Here, we set $\phi^t=\phi^b$ for doublets and triplets which
contain both $T$ and $B$ quarks. The shape is determined by the
dependence on $\exp(i\phi^{t/b})$ in the couplings, and varying the
phase can lead to an enhancement of up to $50\%$ for the $(XT)$ doublet
and a reduction of $-30\%$ for the $(TB)$ doublet. For the studied $(T)$
singlet and the $(TBY)$ triplet the effect of the phases is rather mild.

Kinematic distributions are typically much more sensitive to effects of
new physics than the total cross section. Thus we study the $p_T$ and
$M_{\zH{}}$ distributions with an additional multiplet. Again we
emphasize that $\pt$ refers to the transverse momentum of the Higgs
boson.  For the $(T)$, $(XT)$ and $(TB)$ multiplets, we fix
$m_T=m_B=600$\,GeV, set the mixing angle such that we are in a maximum
of \fig{fig:vlq_total_theta_phi}, and calculate the remaining input
parameters with \vhnnlo{}. For the $(TBY)$ triplet, we only fix $m_T =
600$\,GeV and $\theta_L^t = 1.06$. Moreover, for all multiplets we set
$\phi^{t/b}=0$. In \fig{fig:vlq_pt_mhz} we show (a)~the $p_T$ and
(b)~the $M_{\zH}$ distributions of the gluon initiated process in
comparison to the \sm{} expectation. In the $p_T$ spectra we see an
overall enhancement of the \sm{} distribution which is more distinct in
the boosted regime, i.e.\ $p_T \gtrsim 200$\,GeV. The threshold of the
top quark at $p_T\sim 150$\,GeV and the threshold of the \vlq{}s at
about $p_T\sim 590$\,GeV is visible for all representations and more
pronounced for $(T)$ and $(XT)$ than the other studied cases.

The invariant mass is more sensitive to threshold effects than the
transverse momentum, thus showing distinct peaks at $M_{\zH{}} \approx 2
m_t$ and at twice the \vlq{} masses. It shares the same features as the
$p_T$ distributions, i.e.\ the \vlq{} contributions lead to an overall
enhancement in comparison to the \sm{}, which is more distinct in the
boosted regime. Since the \vlq{} masses are not degenerate for the
scenario of the $(TBY)$ triplet studied here, we observe in total three
thresholds, one for the top quark, one for the $B$ quark at about
$900$\,GeV, and one for the $T$ quark at about $1.2$\,TeV.

%- }}}
%- {{{ section{Dimension-6 operators}

\subsection{Higher-dimensional operators}
\label{sec:dim6}

Finally, we describe the implementation of higher-dimensional operators,
which is independent of a concrete model implementation. In the \sm{},
our setup allows to set bounds on the implemented higher-dimensional
operators. On the other hand, our implementation can also be used in the
\mssm{} or the \thdm{} for the production of a \cp{}-even Higgs boson
and a $Z$ boson, where additionally one \dimfive{} operator coupling two
gluons to a pseudoscalar can be taken into account.  In our choice of
  the \dimsix{} operators, we adopt the so-called gauge basis of
  \citeres{Contino:2013kra,Dawson:1998py} with their corresponding
          {\code FeynRules} implementation~\cite{Alloul:2013naa}, but
          restrict ourselves to the operators which involve
          third-generation quarks.  We allow these operators to be added
          to any implemented model, i.e.\ the \sm{}, the \thdm{} and the
          \mssm{}, but note that we only implemented operators relevant
          for the production of the light \sm{}-like \cp{}-even Higgs
          boson $\ggzhsmall{}$.  The effective Lagrangian reads
\begin{align}
{\cal L} = {\cal L}_{\rm\sm/\thdm/\mssm} + \sum_{i=1}^6{\cal L}_i\,,
\end{align}
where the sum is over the following higher-dimensional operators:
\begin{equation}
  \begin{split}
{\cal L}_1 &= i\frac{c_{HQ}}{v^2}\left[\bar{Q}_L\gamma^\mu Q_L\right]\left[\Phi^\dagger\overleftrightarrow{D}_\mu\Phi\right]\,,\\
{\cal L}_2 &= i\frac{c_{Ht}}{v^2}\left[\bar{t}_R\gamma^\mu t_R\right]\left[\Phi^\dagger \overleftrightarrow{D}_\mu\Phi\right]\,,\\
{\cal L}_3 &= i\frac{4c^\prime_{HQ}}{v^2}\left[\bar{Q}_L\gamma^\mu T_{2k} Q_L\right]\left[\Phi^\dagger T_{2k}\overleftrightarrow{D}_\mu\Phi\right]\,,\\
{\cal L}_4 &= \frac{4g_sc_{tG}}{M_W^2}Y^\text{\sm}_t \Phi^\dagger
\bar{Q}_L\gamma^{\mu\nu}T_a t_R G_{\mu\nu}^a+\text{h.c.}\,,\\
{\cal L}_5 &= -\frac{c_t}{v^2}Y_t\Phi^\dagger\Phi\Phi^\dagger \bar{Q}_Lt_R+\text{h.c.}\,,\\
{\cal L}_6 &= \frac{g_s^2 c_{A}}{32 \pi^2 v}G^{a,\mu\nu}\tilde{G}^a_{\mu\nu}A\,.
    \label{eq:dim6}
  \end{split}
\end{equation}
Herein, we define the dual gluon field-strength tensor
$\tilde{G}^a_{\mu\nu} =
\frac{1}{2}\epsilon_{\mu\nu\alpha\beta}G^{a,\alpha\beta}$
($\epsilon_{0123}=+ 1$), $\gamma^{\mu\nu} =
\frac{i}{4}[\gamma^\mu,\gamma^\nu]$, and $T_{2k}=\sigma_k/2$, with the
Pauli matrices $\sigma_k$. The Hermitian derivative operator
$\overleftrightarrow{D}_\mu$ is given by
$\Phi^\dagger\overleftrightarrow{D}_\mu\Phi=\Phi^\dagger D_\mu
\Phi-(D_\mu \Phi^\dagger)\Phi$.

\eqn{eq:dim6} is set up as an extension of the \sm\ Lagrangian, which
means that, in the \sm\ mode of \vhnnlo{}, $\Phi$ is the \sm\ Higgs
doublet. However, \vhnnlo{} also allows to use \eqn{eq:dim6} for
\cp-even Higgs production in \bsm\ models, in which case \textit{the
  same} Feynman rules for \eqn{eq:dim6} as those derived in the \sm\ are
assumed, with the Higgs particle being interpreted as the one specified
in \blockentry{VHATNNLO}{2}. Note in particular that $Y^\text{\sm}_t$ in
\eqn{eq:dim6} is \textit{always} replaced by $\sqrt{2}m_t/v$, while
$Y_t$ is set to the Yukawa coupling of the model specified in
\blockentry{VHATNNLO}{2}{}.  All other couplings of the Higgs are taken
into account according to the model under consideration.

\begin{figure}
\begin{center}
\begin{tabular}{ccc}
      \includegraphics[height=.14\textheight]{feynmandiagrams/ggZH_Dim1.pdf} &
      \includegraphics[height=.14\textheight]{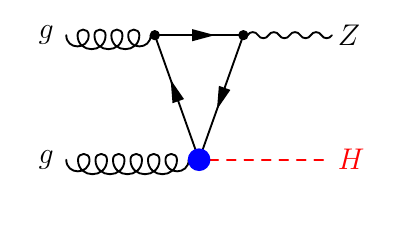} &
      \includegraphics[height=.14\textheight]{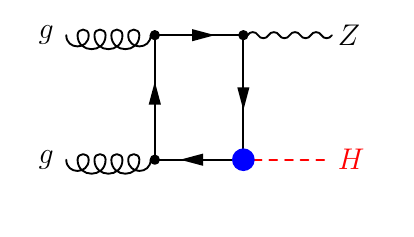} \\[-0.5cm]
      (a) & (b) & (c)\\
      \includegraphics[height=.14\textheight]{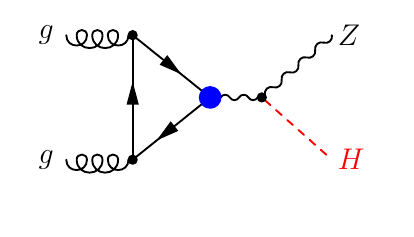} &
      \includegraphics[height=.14\textheight]{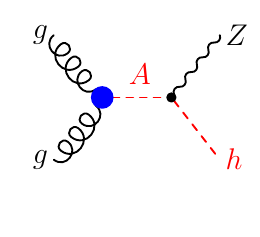}\\[-0.5cm]
      (d) & (e) 
\end{tabular}
    \parbox{\textwidth}{
      \caption[]{\label{fig:dim6feynman}\sloppy Sample Feynman diagrams
        for contributions to $\ggzh{}$ through \dimsix{} operators.  The
        Feynman diagrams (a), (c) and (d) can alternatively have
        modified $gq\bar{q}$ (through ${\cal L}_4$) or $Zq\bar{q}$
        couplings (through ${\cal L}_{1,2,3}$).  (e) shows the
        contribution of the \dimfive{} operator ${\cal L}_6$ for
        $\ggzhsmall{}$ in a \thdm{}/the \mssm{}.}}
\end{center}
\end{figure}

Except for ${\cal L}_4$, which differs by the sign, our notation follows
\citere{Alloul:2013naa}.  \fig{fig:dim6feynman} depicts selected Feynman
diagrams arising from the inclusion of higher dimensional operators. The
operator ${\cal L}_6$ is a \dimfive{} operator which can have a
significant impact in extended Higgs sectors and is therefore added to
the Lagrangian.  In the \thdm{} or the \mssm{}, it corresponds to an
effective coupling among two gluons and the pseudoscalar~$A$ at \lo,
relevant for the contribution $gg\to A\to \zh{}$.  This involves the
$g_Z^{Ah}$ coupling, see \eqn{eq:gahz}, as well as finite width effects
for the pseudoscalar~$A$, see the discussion in \sct{sec:2hdm}.

The coefficients of the operators in ${\cal L}_i$ are specified through
entries $i$ in {\tt Block DIM} in the \vhnnlo{} input file. We list them
in \tab{tab:dim6}.  We note that by default all coefficients are
understood to be constants.  However, by specifying an input scale
through \blockentry{DIM}{11}{}, the running of $c_{tG}$ to the
renormalization scale following \citere{Zhang:2014rja} is included.  The
strong coupling constant~$g_s=\sqrt{4\pi\alpha_s}$ is evaluated at the
renormalization scale~$\muR$. In contrast to other input blocks which
parametrize new physics, \vhnnlo\ inserts the default values given in
\tab{tab:dim6}; i.e.\ any unspecified coefficient of \eqn{eq:dim6} is
set to zero.  The range of values for the coefficients is unrestricted
in principle, but it is clear that very large values are potentially
non-perturbative, and thus the results of \vhnnlo\ will be unreliable.

\begin{table}
\begin{center}
\begin{tabular}{|>{\tt}c>{\tt}ccl|}
\hline
\multicolumn{4}{|c|}{\tt Block DIM}\\
\hline
\textrm{entry} & \textrm{default} & \textrm{range} & meaning\\
\hline
 1 & $0.$ & see text & $c_{HQ}$\\
 2 & $0.$ & see text & $c_{Ht}$\\
 3 & $0.$ & see text & $c^\prime_{HQ}$ \\
 4 & $0.$ & see text & $c_{tG}$ \\
 5 & $0.$ & see text & $c_t$ \\
 6 & $0.$ & see text & $c_{A}$\\
 10 & $0$ & \texttt{\{0,1\}} & add terms of order $\mathcal{O}(c_i^2 +
 c_ic_j)$: \{no, yes\}\\
 11 & -1. & \texttt{-1.,>0.} & scale of $c_{tG}$ [GeV] (no running:
 \texttt{-1.})\\
\hline
\end{tabular}
\caption{\label{tab:dim6} Coefficients of \dimsix{} operators are specified in {\tt Block DIM}.}
\end{center}
\end{table}

The effect of the \dimsix{} operators is two-fold in general: on the one
hand, they lead to additional vertices which do not occur in the model
under consideration; on the other hand, they may modify the coupling
constants of the existing vertices. \vhnnlo{} takes both of these
effects into account.  Some of these operators have been already
implemented in \citere{Bylund:2016phk} with a different
normalization. To compare our results to Tab.\ 7 of
\citere{Bylund:2016phk} one has to transform the effective couplings as
follows:
\begin{equation}
\begin{aligned}
c_{HQ} & \to c_{\phi Q}^{(1)}\frac{2m_t^2}{\Lambda^2}\,,\qquad c_{Ht}  \to c_{\phi t}\frac{2m_t^2}{\Lambda^2}\,,\\
c^\prime_{HQ} & \to c_{\phi Q}^{(3)}\frac{2m_t^2}{\Lambda^2}\,,\qquad c_{tG}  \to c_{tG}\frac{M_W^2}{2\Lambda^2}\,,
\end{aligned}
\label{eq:eff_coupl_transform}
\end{equation}
where $\Lambda$ is the energy scale of new physics. On the right-hand
side of \eqn{eq:eff_coupl_transform}, the couplings are normalized and
named as in \citere{Bylund:2016phk}. For $\Lambda = 1$\,TeV, $m_t =
173.3$\,GeV, $M_W = 79.824$\,GeV and $c_{\phi Q}^{(1)} = c_{\phi t} =
c_{\phi Q}^{(3)} = c_{tG} = 1$ we obtain
\begin{equation}
c_{HQ} = c_{Ht} = c^\prime_{HQ} = 0.0600658\quad\mathrm{and}\quad c_{tG} = 0.00318594\,.
\label{eq:transformed_eff_couplings}
\end{equation}

The total cross section can now be written as
\begin{equation}
\sigma = \sigma_{\text{\sm{}}} + \sigma^{(1)} + \sigma^{(2)}\,,
\end{equation}
where $\sigma_{\text{\sm{}}}$ is the total cross section of the Standard
Model, $\sigma^{(1)}$ contains the interference of the \sm{} amplitudes
which involves one modified vertex, and $\sigma^{(2)}$ contains the
interference of amplitudes of different operators, where at most one
modified vertex is inserted, or the square of an amplitude with one
modified vertex. Again, the amplitudes have been calculated with the
help of {\tt FormCalc}.

\begin{figure}[htb!]
  \begin{center}
    \begin{tabular}{cc}
      \includegraphics[width=0.49\textwidth]{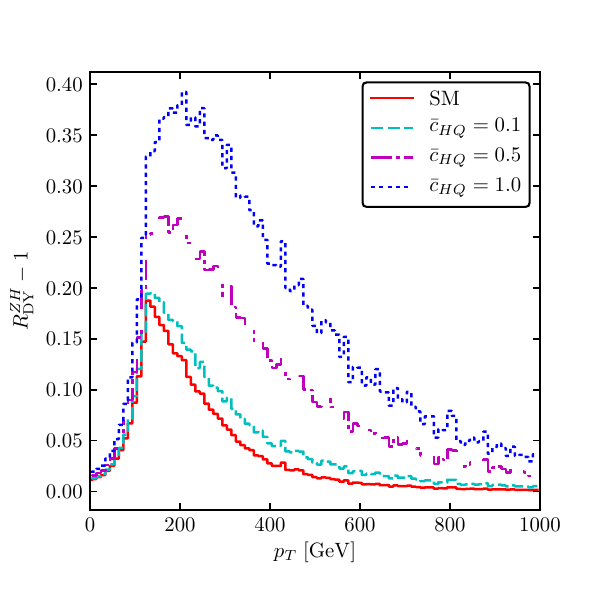} &
      \includegraphics[width=0.49\textwidth]{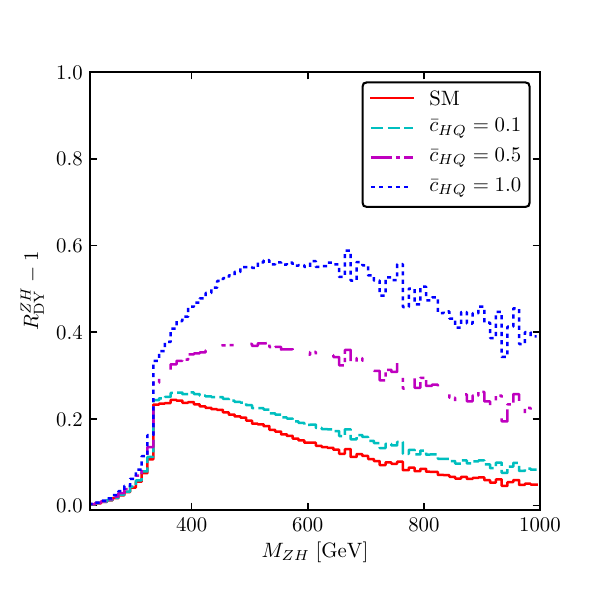}
      \\[-0.5cm]
      (a) & (b) \\[-0.1cm]
      \includegraphics[width=0.49\textwidth]{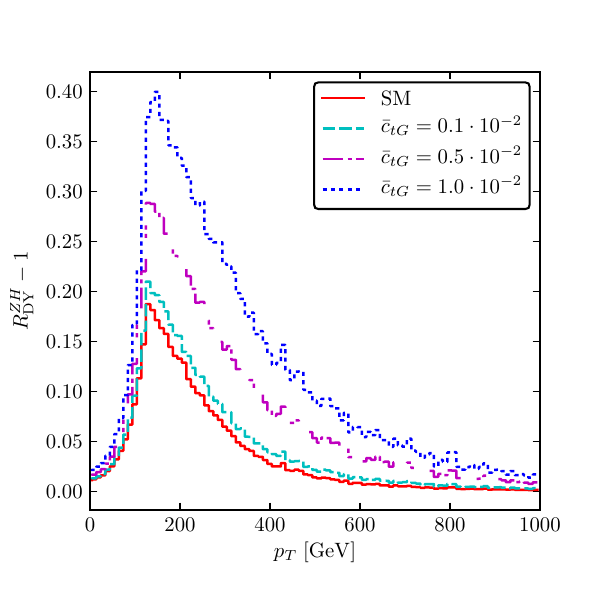} &
      \includegraphics[width=0.49\textwidth]{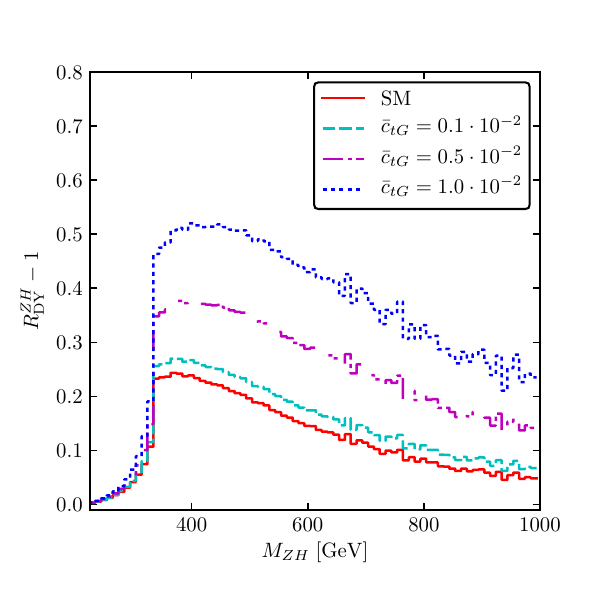}
      \\[-0.5cm] (c) & (d)
      %% \\[-0.1cm]
      %% \includegraphics[width=0.49\textwidth]{figures/ct_dy_ratio.pdf} &
      %% \includegraphics[width=0.49\textwidth]{figures/ct_dy_mhv_ratio.pdf}
      %% \\[-0.5cm] (e) & (f) \\[-0.1cm]
    \end{tabular}
    \parbox{\textwidth}{
      \caption[]{\label{fig:dim6-dist}\sloppy Example of the impact of
        different \dimsix{} operators on kinematic distributions while
        only including the interference term of the new operators and
        the \sm{}.}}
  \end{center}
\end{figure}

In \tab{tab:eff_xs_comparison} we show the results for the total cross
section of gluon initiated $\zh$ production at $\sqrt{s} = 8$ and
$13$\,TeV by taking into account $\mathcal{L}_{1,2,4,5}$, with the
couplings of \eqn{eq:transformed_eff_couplings}, and $c_t = 0.1$.  The
effect of $\mathcal{L}_3$ is not shown since it leads to the same
results as $\mathcal{L}_2$. We only enable one \dimsix{} operator at a
time. Furthermore, we fix the scales to $\muR=\muF = 125$\,GeV, use the
\lo{} \abbrev{MSTW}2008~\cite{Martin:2009iq} parton distribution
functions, and set $m_b=0$.  We observe full agreement with
\citere{Bylund:2016phk} within the numerical accuracy. As is well-known,
the largest effects of such operators typically occur at high transverse
momenta or high invariant masses, where the validity of the effective
field theory description becomes questionable. In the case of $\zh$
production, however, the operators under consideration only affect the
$\ggzh$ component of the cross section, which is known to affect mostly
the kinematic region at and slightly above the top quark threshold. In
order to investigate the impact of the \dimsix\ operators in more
detail, we employ the ratio of the full cross section to its \dy-like
component
$R_\text{\dy}^{\zH}\equiv\sigma^{\zH}/\sigma^{\zH}_\text{\dy}$, from
which one obtains
\begin{equation}
  \begin{split}
\sigma_{\ggzh}/\sigma_\text{\dy}^{\zH}=R_\text{\dy}^{\zH}-1\,.
    \label{eq:}
  \end{split}
\end{equation}
This quantity was shown in \citere{Harlander:2018yns} to be particularly
suited for a data-driven extraction of the gluon-initiated component
$\sigma_{\ggzh}$ from experiment.  Using the example of $\mathcal{L}_1$
and $\mathcal{L}_4$, \fig{fig:dim6-dist} shows that large effects of the
\dimsix\ operators can already be observed in the few-hundred-GeV region
of both $p_T$ and $M_{\zH}$, i.e., well in the validity range of the
effective theory. This observation adds to the virtues of the $\zH$
process in the search of New Physics.\footnote{We thank an anonymous
  referee of \abbrev{JHEP} for suggesting to include this study in the
  paper.}

\begin{table}
\begin{center}
\begin{tabular}{|>{$}c<{$}|>{$}c<{$}|>{$}l<{$}|>{$}r<{$}>{$}r<{$}>{$}r<{$}>{$}r<{$}|}
\hline
\sqrt{s} / \text{TeV} & \sigma_{\text{\sm{}}} / \text{fb} &
&\mathcal{L}_1 & \mathcal{L}_2 & \mathcal{L}_4 & \mathcal{L}_5 \\ \hline
\multirow{4}{*}{8} & \multirow{4}{*}{29.19}& \sigma^{(1)} / \text{fb}& 1.71 & -1.72 & 10.37 & -4.31\\
&&\sigma^{(2)}  / \text{fb}& 0.0471 & 0.0473 & 1.604 & 0.292\\
&&\sigma^{(1)}/\sigma_{\text{\sm{}}} & 0.059 & -0.059 & 0.355 & -0.147\\
&&\sigma^{(2)}/\sigma^{(1)} & 0.028 & -0.028 & 0.155 & -0.068 \\ \hline
\multirow{4}{*}{13} & \multirow{4}{*}{94.12}& \sigma^{(1)} / \text{fb}& 5.87 & -5.92 & 34.47 & -14.98\\
&&\sigma^{(2)}  / \text{fb}& 0.176 & 0.185 & 6.18 & 1.17\\
&&\sigma^{(1)}/\sigma_{\text{\sm{}}} & 0.062 & -0.063 & 0.366 & -0.159\\
&&\sigma^{(2)}/\sigma^{(1)} & 0.029 & -0.031 & 0.179 & -0.078 \\ \hline
\end{tabular}
\caption{\label{tab:eff_xs_comparison} Cross sections for $\ggzh{}$ production at the LHC at $\sqrt{s} = 8$\,TeV
and $\sqrt{s}=13$\,TeV for the \sm{} and the \dimsix{} operators in analogy to \citere{Bylund:2016phk}.}
\end{center}
\end{table}

%- {{{ fig:dim6-dist:

%- }}}

\clearpage

%- }}}

%- }}}
%- {{{ section{Conclusions}

\section{Conclusions}
\label{sec:conclusions}

We presented version {\tt 2} of the code \vhnnlo{}, which allows to
study various new-physics aspects in \higgsstrahlung{}. In detail we
described the general structure of the code and its control through an
input file containing {\tt SLHA}-inspired blocks.  We implemented the
Higgs sectors of a \thdm{}, which---in type~II---also allows for the
calculation of \mssm{} cross sections for \higgsstrahlung{}. For this
purpose, we added the relevant squark amplitudes to the gluon-initiated
contribution of \higgsstrahlung{}. In addition, the bottom-quark
initiated contribution is given at leading-order in perturbation theory
and resonances of (pseudo)scalars are included in both gluon- and
bottom-quark initiated contributions. We demonstrated their relevance
for light Higgs production in the \mssm{}.  The shape of their
interferences with the non-resonant Feynman diagrams can be studied, but
a thorough analysis is left for future work.  \cp{} mixing among the
three \thdm{} scalars can be taken into account.  Vector-like quarks can
be studied in the \sm{}, which---depending on the representation and the
mixing angles---can have a significant effect on the inclusive cross
section and the kinematical distributions. \vhnnlo{} can also provide
the transverse momentum distribution of the final state Higgs boson, and
invariant mass distributions of the two-particle final state. For the
\dy{}-like component, this requires a link to {\tt MCFM}. We showed
threshold effects of vector-like quarks and squarks in both
distributions.  Lastly, beyond such concrete model implementations,
\vhnnlo{} now includes higher-dimensional operators. We compared our
implementation to the literature and found agreement. It turns out that
their effect is particularly pronounced directly above the top-quark
threshold, which is well below the cut-off which is typically assumed
for the underlying effective field theory.  Apart from the already
mentioned internal link to {\tt MCFM}, the new version of \vhnnlo{} can
also be linked to \fh{} and \thdmc{} for the calculation of Higgs masses
and mixings. Other dependencies on external codes are described
throughout the manuscript.

\section*{Acknowledgments}
We thank Oliver Brein and Tom Zirke for their work on earlier versions
of the code. Thanks to Henning Bahl, Tim Stefaniak, and Jonas Wittbrodt
for comments and for pointing out a crucial typo in the paper and the
code. We are indebted to Eleni Vryonidou and Cen Zhang for their support
in tracing a bug in our implementation of higher-dimensional operators,
and helpful clarifications concerning \citere{Bylund:2016phk}. S.L. is
supported by the DFG project ``Precision Calculations in the Higgs
Sector - Paving the Way to the New Physics Landscape'' (ID: MU
3138/1-1). R.V.H. and J.K. are supported by BMBF under contract
05H15PACC1.

\appendix
\section{Installation and execution of \vhnnlo{}}
\label{sec:installation}

\vhnnlo{} can be downloaded from \citere{vhnnlowebpage}.  After
unpacking it, the user has to run {\tt ./configure}, which determines
local Fortran and C++ compilers and library dependencies.  Thereafter,
the user should open the {\tt Makefile} of \vhnnlo{} and adjust the
paths to the various needed external codes, see
Appendix~\ref{sec:lhapdfcubalt}.  Then {\tt make} will result in an
executable, which is placed in the newly generated {\tt /bin}
folder. This executable is run by typing\\ {\tt ./x.vhnnlo input.in
  output.out},\\ where {\tt input.in} has to be an input file, which
contains the various blocks described throughout the manuscript. Example
input files for the various new-physics models and for distributions can
be found in the folder {\tt /example}.  If the name of the output file
{\tt output.out} is not specified, it is named {\tt out.vh}.  In case
\vhnnlo{} is recompiled we recommend to type {\tt make recompile} to
make sure that a few dependencies are newly set up. If this does not
result in the desired behavior, a previous {\tt make (some)clean} can
be helpful.  If \vhnnlo{} should be linked to \thdmc{} or \fh{}, the
{\tt Makefile} has to be processed with an additional {\tt predef=2HDMC}
or {\tt FH}.  If \vhnnlo{} was compiled beforehand, it has to be
accompanied by {\tt recompile}, i.e.\ {\tt make recompile predef=2HDMC}
or {\tt FH}.  Note that adding {\tt GGZH=NO} allows to
compile without the gluon-induced component. In that case, also
the link to {\tt LoopTools} is not required.

\section{Links to external codes}
\label{sec:externalcodes}

First, \vhnnlo{} has to be linked to {\tt LHAPDF} for the usage of
modern \pdf{} sets, the {\tt CUBA} package for numerical integration,
{\tt LoopTools} for the calculation of loop integrals and potentially
{\tt MCFM} for the calculation of kinematic distributions.  As we
described in \sct{sec:2hdm} and \sct{sec:mssm} \vhnnlo{} can be linked
to external codes for the calculation of Higgs boson masses and decay
widths. Those are \thdmc{} for the \thdm{} and \fh{} for the \mssm{}.
We subsequently describe the various
needed and potential links to external codes.

\subsection{{\tt LHAPDF}, {\tt CUBA} and {\tt LoopTools}}
\label{sec:lhapdfcubalt}

\vhnnlo{} needs to be linked to {\tt LHAPDF}~\cite{Buckley:2014ana},
which allows to make use of up-to-date \pdf\ sets. We recommend to use
{\tt LHAPDF} version {\tt 6.2} or higher.  Also older versions $\geq$
{\tt 6} work, but might produce an unpleasant flow of {\tt LHAPDF}
screen messages.  After downloading~\cite{lhapdfwebpage} and installing
{\tt LHAPDF} into a local directory, the user has to specify the paths
to the {\tt LHAPDF} installation folder in the \vhnnlo{} Makefile
variables {\tt LHAPDFDIR}, {\tt LHAPDFINCDIR} and {\tt LHAPDFLIBDIR}, where the latter
two folders have to include relevant header files and contain the {\tt LHAPDF} library, respectively.
If {\tt LHAPDF} is installed in a global directory like {\tt /usr/share}, etc., the
specification of the paths might be obsolete. Note that the \pdf\ sets,
which were specified in the input file of \vhnnlo{}, need to be
installed, i.e.\ downloaded from the {\tt LHAPDF}
webpage~\cite{lhapdfwebpage}.  This can for example be done through the
{\tt LHAPDF} script {\tt lhapdf install}.

\begin{table}[htb]
\begin{center}
\begin{tabular}{|>{\tt}c>{\tt}cl|}
  \hline
  \multicolumn{3}{|c|}{\tt Block VEGAS}\\
  \hline
  \textrm{entry} & \textrm{default} & meaning\\
  \hline
        10 &       10000   & \dy\ starting points\\
        11 &        5000   & \dy\ increase points\\
        12 &       20000   & \dy\ minimal points\\
        13 &     1000000   & \dy\ maximal points\\
        20 &       10000   & $gg$ starting points\\
        21 &           0   & $gg$ increase points\\
        22 &       10000   & $gg$ minimal points\\
        23 &       70000   & $gg$ maximal points\\
        30 &       10000   & $b\bar b$ starting points\\
        31 &        5000   & $b\bar b$ increase points\\
        32 &      200000   & $b\bar b$ minimal points\\
        33 &      500000   & $b\bar b$ maximal points\\
  \hline
\end{tabular}
\end{center}
\vspace{-0.6cm}
\caption[]{Integration parameters.}
\label{tab:vegas}
\end{table}

For numerical integration \vhnnlo{} makes use of the {\tt CUBA}
package~\cite{Hahn:2004fe}\footnote{\texttt{CUBA} is available from
  \citere{cubawebpage}.} which needs to be installed separately. The
paths to the {\tt CUBA} installation directory is then specified in the
variable {\tt CUBADIR} in the \texttt{Makefile} of \vhnnlo{}.  For the
numerical integration, \vhnnlo\ uses \texttt{CUBA}'s implementation of
{\tt VEGAS} algorithm. The central integration parameters can be
controlled through {\tt Block VEGAS} in the input file of \vhnnlo, see
\tab{tab:vegas}. These are the number of evaluations the numerical
integration is starting with, the increase in the number of evaluations,
and the minimal and maximal number of evaluations. They are specified in
\blockentry{VEGAS}{10+10$\cdot i$}{}, \blockentry{VEGAS}{11+10$\cdot
  i$}{}, \blockentry{VEGAS}{12+10$\cdot i$}{} and
\blockentry{VEGAS}{13+10$\cdot i$}{}, respectively, where
$i\in\{0,1,2\}$ labels the \dy{}-like component, the gluon- and the
bottom-quark initiated contribution, respectively.

Finally, for the evaluation of loop integrals, \vhnnlo{} requires a link
to {\tt LoopTools}~\cite{vanOldenborgh:1989wn,Hahn:1998yk}, which can be
obtained from \citere{looptoolswebpage}. We recommend to use {\tt
  LoopTools} version {\tt 2.13} or higher.\footnote{If older versions of
  {\tt LoopTools} are used, the flag {\tt -ff2c} has to be added to {\tt
    FFLAGS} in the \texttt{Makefile} of \vhnnlo{} before compiling.}
The link to the {\tt LoopTools} installation directory and the library
is specified in the variables {\tt LTDIR} and {\tt LTLIBDIR} of the
\vhnnlo{} Makefile, respectively. If \vhnnlo{} is compiled with {\tt GGHZ=NO},
i.e. the gluon-induced component is not included, the link to {\tt LoopTools}
is not needed.

\subsection{{\tt MCFM}}
\label{sec:mcfm}

{\tt MCFM} can be downloaded from its webpage \cite{mcfmwebpage}.  It
has to be compiled and installed separately.  Since the implementation
of \vhnnlo{} is currently not compatible with {\tt OpenMP}, the {\tt
  MCFM} library should be built and linked without {\tt OpenMP}, if a
link to \vhnnlo{} should be established.  \vhnnlo{} can be linked to
{\tt MCFM} {\tt 8} by adjusting the path specified in the variable
{\tt MCFMDIR} and setting the corresponding flag {\tt MCFM=YES} in the Makefile
of \vhnnlo{}. Since {\tt MCFM} needs some additional files in the
working directory, one has to call {\tt make MCFM} to copy them into the
\vhnnlo{} directory. The interface to {\tt MCFM} does not rely on input
files, thus all parameters needed by {\tt MCFM} are set internally,
except the input given in the input block {\tt DISTRIB}. This includes
the start, end and bin width of the histogram, cuts on the invariant
mass of $M_{\vH{}}$ and the integration points for {\tt VEGAS}, see
\tab{tab:distrib}. All \sm{} parameters, the factorization and
renormalization scale and the parton distribution function are set to
the same values as given in the blocks {\tt SMPARAMS}, {\tt SCALES} and
{\tt PDFSPEC}, respectively. To produce results comparable to \vhnnlo{}
we disable the decays of $H$, $Z$, and $W$. This disables automatically
any jets so that all input regarding jets is filled with dummy
variables. The remaining input values for {\tt MCFM},
which are automatically set by \vhnnlo{}, are
summarized in \tab{tab:mcfm_default_input}.

\begin{table}[htb]
  \begin{center}
    \begin{tabular}{|c|l|l|}
      \hline
      parameter &  value & meaning \\
      \hline
      {\tt ewscheme} & 3 & EW scheme \\
      {\tt aemmz\_inp} & from \vhnnlo{} & $\alpha_\mathrm{EM}(M_Z)$ \\
      {\tt xw\_inp} & from \vhnnlo{} & $s_W^2$ \\
      {\tt nproc} & 101 ($\zH{}$), 96 ($W^-H$), 91 ($W^+H$) & Process number \\
      {\tt zerowidth} &  {\tt .true.} & Remove decay width\\
      {\tt removebr} & {\tt .true.} & Disable decay\\
      \hline
    \end{tabular}
  \end{center}
  \vspace{-0.6cm}
  \caption{Input parameters for {\tt MCFM} set in \vhnnlo{} automatically.}
  \label{tab:mcfm_default_input}
\end{table}

\subsection{\thdmc{}, \fh{} and {\tt SLHA}-style Higgs sector input}
\label{sec:thdmcfhfs}

After downloading the code \thdmc{} from the webpage~\cite{2hdmcwebpage}
and installing it, the path to the \thdmc{} installation needs to be
supplied through the variable {\tt 2HDMCPATH} in the \texttt{Makefile}
of \vhnnlo{}.  Compiling \vhnnlo\ with {\tt make predef=2HDMC}
establishes the link between \vhnnlo{} and \thdmc{}. If \vhnnlo{} was
already compiled before without the link, one should do {\tt make
  recompile predef=2HDMC}. Note that only recent versions of \thdmc{}
allow to include the $H_2$ basis.

\fh{} can be obtained from its webpage~\cite{fhwebpage}.  After
installation, \vhnnlo{} can be linked to \fh{} by specifying the full
path in {\tt FHPATH} in the \texttt{Makefile} of \vhnnlo{}.  In
addition, since different versions of \fh\ differ in their input and/or
output format, \vhnnlo\ requires the version number of \fh{} in the
variable {\tt FHVERSION}. \vhnnlo\ has been tested with \fh{} {\tt 2.9}
through {\tt 2.14}, which is the latest publicly available version.

Another option is to use \vhnnlo{} through an additional {\tt
  SLHA}-style spectrum file. Its name is provided in the \vhnnlo{} input
  file through \blockentry{SPECTRUMFILE}{1}{}.
  The additional {\tt SLHA}-style spectrum file needs to contain the following blocks:
\begin{itemize}
\item
  {\tt Block MASS}, for the on-shell masses of the gluino, $h$, $H$, $A$
boson, stop and sbottom quarks,
\item {\tt Block AU} and {\tt Block AD},
for the soft-breaking trilinear stop-Higgs and sbottom-Higgs coupling,
respectively,
\item
  {\tt Block STOPMIX} and {\tt Block SBOTMIX} to evaluate the stop and
  sbottom mixing angle,
\item {\tt Block ALPHA}, for the scalar
  mixing angle~$\alpha$,
\item
  {\tt Block HMIX} to get the
  values of the $\mu$ parameter and $\tan\beta$.
\end{itemize}
Note that these blocks are part of the standard output of typical
\susy\ spectrum generators such as
\texttt{SoftSusy}\,\cite{Allanach:2001kg},
\texttt{SPheno}\,\cite{Porod:2003um,Porod:2011nf}, or
\texttt{FlexibleSUSY}\,\cite{Athron:2014yba,Athron:2016fuq,Athron:2017fvs}.
Combining the latter with \texttt{Himalaya}\,\cite{Harlander:2017kuc}
allows to run \vhnnlo\ in the \mssm\ by taking into account three-loop
corrections to the light Higgs boson
mass\,\cite{Harlander:2008ju,Kant:2010tf}.
For now, \vhnnlo{} does not read in the Higgs boson decay width
from the spectrum file, since not all external codes provide such numbers.
Thus, the user needs to fill the entries \blockentry{MASS}{250/350/360}{}
in the \vhnnlo{} input file
with the Higgs boson decay widths of the light \cp{} even,
the heavy \cp{} even Higgs boson and the pseudoscalar, respectively.
We refer to the example \vhnnlo{} input file together with a
{\tt FlexibleSUSY} spectrum file in folder {/example}.

%- }}}

%- }}} body:
%- {{{ bibliography:

\def\app#1#2#3{{\it Act.~Phys.~Pol.~}\jref{\bf B #1}{#2}{#3}}
\def\apa#1#2#3{{\it Act.~Phys.~Austr.~}\jref{\bf#1}{#2}{#3}}
\def\annphys#1#2#3{{\it Ann.~Phys.~}\jref{\bf #1}{#2}{#3}}
\def\cmp#1#2#3{{\it Comm.~Math.~Phys.~}\jref{\bf #1}{#2}{#3}}
\def\cpc#1#2#3{{\it Comp.~Phys.~Commun.~}\jref{\bf #1}{#2}{#3}}
\def\epjc#1#2#3{{\it Eur.\ Phys.\ J.\ }\jref{\bf C #1}{#2}{#3}}
\def\fortp#1#2#3{{\it Fortschr.~Phys.~}\jref{\bf#1}{#2}{#3}}
\def\ijmpc#1#2#3{{\it Int.~J.~Mod.~Phys.~}\jref{\bf C #1}{#2}{#3}}
\def\ijmpa#1#2#3{{\it Int.~J.~Mod.~Phys.~}\jref{\bf A #1}{#2}{#3}}
\def\jcp#1#2#3{{\it J.~Comp.~Phys.~}\jref{\bf #1}{#2}{#3}}
\def\jetp#1#2#3{{\it JETP~Lett.~}\jref{\bf #1}{#2}{#3}}
\def\jphysg#1#2#3{{\small\it J.~Phys.~G~}\jref{\bf #1}{#2}{#3}}
\def\jhep#1#2#3{{\small\it JHEP~}\jref{\bf #1}{#2}{#3}}
\def\mpl#1#2#3{{\it Mod.~Phys.~Lett.~}\jref{\bf A #1}{#2}{#3}}
\def\nima#1#2#3{{\it Nucl.~Inst.~Meth.~}\jref{\bf A #1}{#2}{#3}}
\def\npb#1#2#3{{\it Nucl.~Phys.~}\jref{\bf B #1}{#2}{#3}}
\def\nca#1#2#3{{\it Nuovo~Cim.~}\jref{\bf #1A}{#2}{#3}}
\def\plb#1#2#3{{\it Phys.~Lett.~}\jref{\bf B #1}{#2}{#3}}
\def\prc#1#2#3{{\it Phys.~Reports }\jref{\bf #1}{#2}{#3}}
\def\prd#1#2#3{{\it Phys.~Rev.~}\jref{\bf D #1}{#2}{#3}}
\def\pR#1#2#3{{\it Phys.~Rev.~}\jref{\bf #1}{#2}{#3}}
\def\prl#1#2#3{{\it Phys.~Rev.~Lett.~}\jref{\bf #1}{#2}{#3}}
\def\pr#1#2#3{{\it Phys.~Reports }\jref{\bf #1}{#2}{#3}}
\def\ptp#1#2#3{{\it Prog.~Theor.~Phys.~}\jref{\bf #1}{#2}{#3}}
\def\ppnp#1#2#3{{\it Prog.~Part.~Nucl.~Phys.~}\jref{\bf #1}{#2}{#3}}
\def\rmp#1#2#3{{\it Rev.~Mod.~Phys.~}\jref{\bf #1}{#2}{#3}}
\def\sovnp#1#2#3{{\it Sov.~J.~Nucl.~Phys.~}\jref{\bf #1}{#2}{#3}}
\def\sovus#1#2#3{{\it Sov.~Phys.~Usp.~}\jref{\bf #1}{#2}{#3}}
\def\tmf#1#2#3{{\it Teor.~Mat.~Fiz.~}\jref{\bf #1}{#2}{#3}}
\def\tmp#1#2#3{{\it Theor.~Math.~Phys.~}\jref{\bf #1}{#2}{#3}}
\def\yadfiz#1#2#3{{\it Yad.~Fiz.~}\jref{\bf #1}{#2}{#3}}
\def\zpc#1#2#3{{\it Z.~Phys.~}\jref{\bf C #1}{#2}{#3}}
\def\ibid#1#2#3{{ibid.~}\jref{\bf #1}{#2}{#3}}
\def\otherjournal#1#2#3#4{{\it #1}\jref{\bf #2}{#3}{#4}}
\newcommand{\jref}[3]{{\bf #1}, #3 (#2)}
\newcommand{\hepph}[1]{{\tt [hep-ph/#1]}}
\newcommand{\mathph}[1]{{\tt [math-ph/#1]}}
\newcommand{\arxiv}[2]{{\tt arXiv:#1}}
\newcommand{\bibentry}[4]{#1, {\it #2}, #3\ifthenelse{\equal{#4}{}}{}{, }#4.}

%- }}}


\begin{thebibliography}{99}
%
% DO NOT EDIT THIS FILE!
%
% It is generated automatically.
%
% vhnnlov2_ref.tex -- generated by sortref-2.5  
% ((C) R. Harlander, http://www.robert-harlander.de/software/)
% on Mi 2. Mai 11:21:45 CEST 2018
%
% (deleting this header may lead to severe nose bleeding)
%

%1
\bibitem{vhnnlowebpage}    
  \url{https://web.physik.rwth-aachen.de/~harlander/software/vh@nnlo/}

%2
\bibitem{Aad:2012tfa}
  G.~Aad {\it et al.} [ATLAS Collaboration],
  Phys.\ Lett.\ B {\bf 716} (2012) 1
  [arXiv:1207.7214].
  %%CITATION = doi:10.1016/j.physletb.2012.08.020;%%

%3
\bibitem{Chatrchyan:2012xdj}
  S.~Chatrchyan {\it et al.} [CMS Collaboration],
  Phys.\ Lett.\ B {\bf 716} (2012) 30
  [arXiv:1207.7235].
  %%CITATION = doi:10.1016/j.physletb.2012.08.021;%%

%4
\bibitem{Aad:2015zhl}
  G.~Aad {\it et al.} [ATLAS and CMS Collaborations],
  Phys.\ Rev.\ Lett.\  {\bf 114} (2015) 191803
  [arXiv:1503.07589].
  %%CITATION = doi:10.1103/PhysRevLett.114.191803;%%

%5
\bibitem{Khachatryan:2016vau}
  G.~Aad {\it et al.} [ATLAS and CMS Collaborations],
  JHEP {\bf 1608} (2016) 045
  [arXiv:1606.02266].
  %%CITATION = doi:10.1007/JHEP08(2016)045;%%

%6
\bibitem{Aaboud:2017xsd}
  M.~Aaboud {\it et al.} [ATLAS Collaboration],
  JHEP {\bf 1712} (2017) 024
  [arXiv:1708.03299].
  %%CITATION = doi:10.1007/JHEP12(2017)024;%%

%7
\bibitem{Sirunyan:2017elk}
  A.~M.~Sirunyan {\it et al.} [CMS Collaboration],
  arXiv:1709.07497.
  %%CITATION = ARXIV:1709.07497;%%

%8
\bibitem{Brein:2012ne}
  O.~Brein, R.V.~Harlander, and T.J.E.~Zirke,
  Comput.\ Phys.\ Commun.\  {\bf 184} (2013) 998
  [arXiv:1210.5347].
  %%CITATION = doi:10.1016/j.cpc.2012.11.002;%%

%9
\bibitem{Han:1991ia}
  T.~Han and S.~Willenbrock,
  Phys.\ Lett.\ B {\bf 273} (1991) 167.
  %%CITATION = doi:10.1016/0370-2693(91)90572-8;%%

%10
\bibitem{Brein:2003wg}
  O.~Brein, A.~Djouadi, and R.V.~Harlander,
  Phys.\ Lett.\ B {\bf 579} (2004) 149
  [hep-ph/0307206].
  %%CITATION = doi:10.1016/j.physletb.2003.10.112;%%

%11
\bibitem{Ciccolini:2003jy}
  M.L.~Ciccolini, S.~Dittmaier, and M.~Kr{\"a}mer,
  Phys.\ Rev.\ D {\bf 68} (2003) 073003
  [hep-ph/0306234].
  %%CITATION = doi:10.1103/PhysRevD.68.073003;%%

%12
\bibitem{Hamberg:1990np}
  R.~Hamberg, W.L.~van Neerven, and T.~Matsuura,
  Nucl.\ Phys.\ B {\bf 359} (1991) 343
   Erratum: [Nucl.\ Phys.\ B {\bf 644} (2002) 403].
  %%CITATION = doi:10.1016/S0550-3213(02)00814-3, 10.1016/0550-3213(91)90064-5;%%

%13
\bibitem{Dawson:2012gs}
  S.~Dawson, T.~Han, W.K.~Lai, A.K.~Leibovich, and I.~Lewis,
  Phys.\ Rev.\ D {\bf 86} (2012) 074007
  [arXiv:1207.4207].
  %%CITATION = doi:10.1103/PhysRevD.86.074007;%%

%14
\bibitem{Brein:2011vx}
  O.~Brein, R.V.~Harlander, M.~Wiesemann, and T.~Zirke,
  Eur.\ Phys.\ J.\ C {\bf 72} (2012) 1868
  [arXiv:1111.0761].
  %%CITATION = doi:10.1140/epjc/s10052-012-1868-6;%%

%15
\bibitem{Hahn:2000kx}
  T.~Hahn,
  Comput.\ Phys.\ Commun.\  {\bf 140} (2001) 418
  [hep-ph/0012260].
  %%CITATION = doi:10.1016/S0010-4655(01)00290-9;%%

%16
\bibitem{Hahn:1998yk}
  T.~Hahn and M.~Perez-Victoria,
  Comput.\ Phys.\ Commun.\  {\bf 118} (1999) 153
  [hep-ph/9807565].
  %%CITATION = doi:10.1016/S0010-4655(98)00173-8;%%

%17
\bibitem{Kniehl:2011aa}
  B.A.~Kniehl and C.P.~Palisoc,
  Phys.\ Rev.\ D {\bf 85} (2012) 075027
  [arXiv:1112.1575].
  %%CITATION = doi:10.1103/PhysRevD.85.075027;%%

%18
\bibitem{Altenkamp:2012sx}
  L.~Altenkamp, S.~Dittmaier, R.V.~Harlander, H.~Rzehak, and T.J.E.~Zirke,
  JHEP {\bf 1302} (2013) 078
  [arXiv:1211.5015].
  %%CITATION = doi:10.1007/JHEP02(2013)078;%%

%19
\bibitem{Harlander:2014wda}
  R.V.~Harlander, A.~Kulesza, V.~Theeuwes, and T.~Zirke,
  JHEP {\bf 1411} (2014) 082
  [arXiv:1410.0217].
  %%CITATION = doi:10.1007/JHEP11(2014)082;%%

%20
\bibitem{Harlander:2013mla}
  R.V.~Harlander, S.~Liebler, and T.~Zirke,
  JHEP {\bf 1402} (2014) 023
  [arXiv:1307.8122].
  %%CITATION = doi:10.1007/JHEP02(2014)023;%%

%21
\bibitem{Gunion:1989we}
  J.F.~Gunion, H.E.~Haber, G.L.~Kane, and S.~Dawson,
  Front.\ Phys.\  {\bf 80} (2000) 1.
  %%CITATION = FRPHA,80,1;%%

%22
\bibitem{Akeroyd:1996he}
  A.G.~Akeroyd,
  Phys.\ Lett.\ B {\bf 377} (1996) 95
  [hep-ph/9603445].
  %%CITATION = doi:10.1016/0370-2693(96)00330-9;%%

%23
\bibitem{Akeroyd:1998ui}
  A.G.~Akeroyd,
  J.\ Phys.\ G {\bf 24} (1998) 1983
  [hep-ph/9803324].
  %%CITATION = doi:10.1088/0954-3899/24/11/001;%%

%24
\bibitem{Aoki:2009ha}
  M.~Aoki, S.~Kanemura, K.~Tsumura, and K.~Yagyu,
  Phys.\ Rev.\ D {\bf 80} (2009) 015017
  [arXiv:0902.4665].
  %%CITATION = doi:10.1103/PhysRevD.80.015017;%%

%25
\bibitem{Branco:2011iw}
  G.C.~Branco, P.M.~Ferreira, L.~Lavoura, M.N.~Rebelo, M.~Sher, and J.P.~Silva,
  Phys.\ Rept.\  {\bf 516} (2012) 1
  [arXiv:1106.0034].
  %%CITATION = doi:10.1016/j.physrep.2012.02.002;%%

%26
\bibitem{Craig:2012vn}
  N.~Craig, and S.~Thomas,
  JHEP {\bf 1211} (2012) 083
  [arXiv:1207.4835].
  %%CITATION = doi:10.1007/JHEP11(2012)083;%%

%27
\bibitem{Djouadi:2005gj}
  A.~Djouadi,
  Phys.\ Rept.\  {\bf 459} (2008) 1
  [hep-ph/0503173].
  %%CITATION = doi:10.1016/j.physrep.2007.10.005;%%

%28
\bibitem{Hespel:2015zea}
  B.~Hespel, F.~Maltoni, and E.~Vryonidou,
  JHEP {\bf 1506} (2015) 065
  [arXiv:1503.01656].
  %%CITATION = doi:10.1007/JHEP06(2015)065;%%

%29
\bibitem{Kao:1991xg}
  C.~Kao,
  Phys.\ Rev.\ D {\bf 46} (1992) 4907.
  %%CITATION = doi:10.1103/PhysRevD.46.4907;%%

%30
\bibitem{Yin:2002sq}
  J.~Yin, W.G.~Ma, R.Y.~Zhang, and H.S.~Hou,
  Phys.\ Rev.\ D {\bf 66} (2002) 095008.
  %%CITATION = doi:10.1103/PhysRevD.66.095008;%%

%31
\bibitem{Kao:2003jw}
  C.~Kao, G.~Lovelace, and L.H.~Orr,
  Phys.\ Lett.\ B {\bf 567} (2003) 259
  [hep-ph/0305028].
  %%CITATION = doi:10.1016/j.physletb.2003.06.042;%%

%32
\bibitem{Kao:2004vp}
  C.~Kao and S.~Sachithanandam,
  Phys.\ Lett.\ B {\bf 620} (2005) 80
  [hep-ph/0411331].
  %%CITATION = doi:10.1016/j.physletb.2005.06.016;%%

%33
\bibitem{Yang:2003kr}
  L.L.~Yang, C.S.~Li, J.J.~Liu, and L.G.~Jin,
  J.\ Phys.\ G {\bf 30} (2004) 1821
  [hep-ph/0312179].
  %%CITATION = doi:10.1088/0954-3899/30/12/005;%%

%34
\bibitem{Li:2005qna}
  Q.~Li, C.S.~Li, J.J.~Liu, L.G.~Jin, and C.-P.~Yuan,
  Phys.\ Rev.\ D {\bf 72} (2005) 034032
  [hep-ph/0501070].
  %%CITATION = doi:10.1103/PhysRevD.72.034032;%%

%35
\bibitem{Alwall:2014hca}
  J.~Alwall {\it et al.},
  JHEP {\bf 1407} (2014) 079
  [arXiv:1405.0301].
  %%CITATION = doi:10.1007/JHEP07(2014)079;%%

%36
\bibitem{Bylund:2016phk}
  O.~Bessidskaia Bylund, F.~Maltoni, I.~Tsinikos, E.~Vryonidou, and C.~Zhang,
  JHEP {\bf 1605} (2016) 052
  [arXiv:1601.08193].
  %%CITATION = doi:10.1007/JHEP05(2016)052;%%

%37
\bibitem{Aguilar-Saavedra:2013qpa}
  J.A.~Aguilar-Saavedra, R.~Benbrik, S.~Heinemeyer, and M.~P\'erez-Victoria,
  Phys.\ Rev.\ D {\bf 88} (2013) no.9,  094010
  [arXiv:1306.0572].
  %%CITATION = doi:10.1103/PhysRevD.88.094010;%%

%38
\bibitem{Ferrera:2011bk}
  G.~Ferrera, M.~Grazzini, and F.~Tramontano,
  Phys.\ Rev.\ Lett.\  {\bf 107} (2011) 152003
  [arXiv:1107.1164].
  %%CITATION = doi:10.1103/PhysRevLett.107.152003;%%

%39
\bibitem{Ferrera:2014lca}
  G.~Ferrera, M.~Grazzini, and F.~Tramontano,
  Phys.\ Lett.\ B {\bf 740} (2015) 51
  [arXiv:1407.4747].
  %%CITATION = doi:10.1016/j.physletb.2014.11.040;%%

%40
\bibitem{Campbell:2016jau}
  J.M.~Campbell, R.K.~Ellis, and C.~Williams,
  JHEP {\bf 1606} (2016) 179
  [arXiv:1601.00658].
  %%CITATION = doi:10.1007/JHEP06(2016)179;%%

%41
\bibitem{Denner:2011id}
  A.~Denner, S.~Dittmaier, S.~Kallweit, and A.~M\"uck,
  JHEP {\bf 1203} (2012) 075
  [arXiv:1112.5142].
  %%CITATION = doi:10.1007/JHEP03(2012)075;%%

%42
\bibitem{Butterworth:2008iy}
  J.~M.~Butterworth, A.~R.~Davison, M.~Rubin and G.~P.~Salam,
  Phys.\ Rev.\ Lett.\  {\bf 100} (2008) 242001
  [arXiv:0802.2470].
  %%CITATION = doi:10.1103/PhysRevLett.100.242001;%%

%43
\bibitem{Englert:2013vua}
  C.~Englert, M.~McCullough, and M.~Spannowsky,
  Phys.\ Rev.\ D {\bf 89} (2014) no.1,  013013
  [arXiv:1310.4828].
  %%CITATION = doi:10.1103/PhysRevD.89.013013;%%

%44
\bibitem{Campbell:2011bn}
  J.M.~Campbell, R.K.~Ellis, and C.~Williams,
  JHEP {\bf 1107} (2011) 018
  [arXiv:1105.0020].
  %%CITATION = doi:10.1007/JHEP07(2011)018;%%

%45
\bibitem{Campbell:1999ah}
  J.M.~Campbell and R.K.~Ellis,
  Phys.\ Rev.\ D {\bf 60} (1999) 113006
  [hep-ph/9905386].
  %%CITATION = doi:10.1103/PhysRevD.60.113006;%%

%46
\bibitem{Campbell:2015qma}
  J.M.~Campbell, R.K.~Ellis, and W.T.~Giele,
  Eur.\ Phys.\ J.\ C {\bf 75} (2015) no.6,  246
  [arXiv:1503.06182].
  %%CITATION = doi:10.1140/epjc/s10052-015-3461-2;%%

%47
\bibitem{Boughezal:2016wmq}
  R.~Boughezal, J.M.~Campbell, R.K.~Ellis, C.~Focke, W.~Giele, X.~Liu, F.~Petriello, and C.~Williams,
  Eur.\ Phys.\ J.\ C {\bf 77} (2017) no.1,  7
  [arXiv:1605.08011].
  %%CITATION = doi:10.1140/epjc/s10052-016-4558-y;%%

%48
\bibitem{Kniehl:1990iva}
  B.A.~Kniehl,
  Phys.\ Rev.\ D {\bf 42} (1990) 2253.
  %%CITATION = doi:10.1103/PhysRevD.42.2253;%%

%49
\bibitem{Dicus:1988yh}
  D.A.~Dicus and C.~Kao,
  Phys.\ Rev.\ D {\bf 38} (1988) 1008
   Erratum: [Phys.\ Rev.\ D {\bf 42} (1990) 2412].
  %%CITATION = doi:10.1103/PhysRevD.38.1008, 10.1103/PhysRevD.42.2412;%%

%50
\bibitem{Hasselhuhn:2016rqt}
  A.~Hasselhuhn, T.~Luthe and M.~Steinhauser,
  JHEP {\bf 1701} (2017) 073
  [arXiv:1611.05881].
  %%CITATION = doi:10.1007/JHEP01(2017)073;%%

%51
\bibitem{Skands:2003cj}
  P.Z.~Skands {\it et al.},
  JHEP {\bf 0407} (2004) 036
  [hep-ph/0311123].
  %%CITATION = doi:10.1088/1126-6708/2004/07/036;%%

%52
\bibitem{Allanach:2008qq}
  B.C.~Allanach {\it et al.},
  Comput.\ Phys.\ Commun.\  {\bf 180} (2009) 8
  [arXiv:0801.0045].
  %%CITATION = doi:10.1016/j.cpc.2008.08.004;%%

%53
\bibitem{Buckley:2014ana}
  A.~Buckley, J.~Ferrando, S.~Lloyd, K.~Nordstr\"om, B.~Page,
  M.~R\"ufenacht, M.~Sch\"onherr, and G.~Watt,
  Eur.\ Phys.\ J.\ C {\bf 75} (2015) 132
  [arXiv:1412.7420].
  %%CITATION = doi:10.1140/epjc/s10052-015-3318-8;%%

%54
\bibitem{Ferreira:2017bnx}
  P.~M.~Ferreira, S.~Liebler and J.~Wittbrodt,
  Phys.\ Rev.\ D {\bf 97} (2018) no.5,  055008
  [arXiv:1711.00024].
  %%CITATION = doi:10.1103/PhysRevD.97.055008;%%

%55
\bibitem{Harlander:2012pb}
  R.V.~Harlander, S.~Liebler, and H.~Mantler,
  Comput.\ Phys.\ Commun.\  {\bf 184} (2013) 1605
  [arXiv:1212.3249].
  %%CITATION = doi:10.1016/j.cpc.2013.02.006;%%

%56
\bibitem{Harlander:2016hcx}
  R.V.~Harlander, S.~Liebler, and H.~Mantler,
  Comput.\ Phys.\ Commun.\  {\bf 212} (2017) 239
  [arXiv:1605.03190].
  %%CITATION = doi:10.1016/j.cpc.2016.10.015;%%

%57
\bibitem{Jung:2015sna}
  S.~Jung, Y.W.~Yoon, and J.~Song,
  Phys.\ Rev.\ D {\bf 93} (2016) no.5,  055035
  [arXiv:1510.03450].
  %%CITATION = doi:10.1103/PhysRevD.93.055035;%%

%58
\bibitem{Greiner:2015ixr}
  N.~Greiner, S.~Liebler, and G.~Weiglein,
  Eur.\ Phys.\ J.\ C {\bf 76} (2016) no.3,  118
  [arXiv:1512.07232].
  %%CITATION = doi:10.1140/epjc/s10052-016-3965-4;%%

%59
\bibitem{Hespel:2014sla}
  B.~Hespel, D.~Lopez-Val, and E.~Vryonidou,
  JHEP {\bf 1409} (2014) 124
  [arXiv:1407.0281].
  %%CITATION = doi:10.1007/JHEP09(2014)124;%%

%60
\bibitem{Eriksson:2009ws}
  D.~Eriksson, J.~Rathsman, and O.~St{\aa}l,
  Comput.\ Phys.\ Commun.\  {\bf 181} (2010) 189
  [arXiv:0902.0851].
  %%CITATION = doi:10.1016/j.cpc.2009.09.011;%%

%61
\bibitem{Eriksson:2010zzb}
  D.~Eriksson, J.~Rathsman, and O.~St{\aa}l,
  Comput.\ Phys.\ Commun.\  {\bf 181} (2010) 833.
  %%CITATION = doi:10.1016/j.cpc.2009.12.016;%%

%62
\bibitem{Haber:2015pua}
  H.E.~Haber and O.~St{\aa}l,
  Eur.\ Phys.\ J.\ C {\bf 75} (2015) no.10,  491
   Erratum: [Eur.\ Phys.\ J.\ C {\bf 76} (2016) no.6,  312]
  [arXiv:1507.04281].
  %%CITATION = doi:10.1140/epjc/s10052-015-3697-x, 10.1140/epjc/s10052-016-4151-4;%%

%63
\bibitem{Brooijmans:2018xbu}
  G.~Brooijmans {\it et al.},
  arXiv:1803.10379.
  %%CITATION = ARXIV:1803.10379;%%

%64
\bibitem{Heinemeyer:1998yj}
  S.~Heinemeyer, W.~Hollik and G.~Weiglein,
  Comput.\ Phys.\ Commun.\  {\bf 124} (2000) 76
  [hep-ph/9812320].
  %%CITATION = doi:10.1016/S0010-4655(99)00364-1;%%

%65
\bibitem{Heinemeyer:1998np}
  S.~Heinemeyer, W.~Hollik and G.~Weiglein,
  Eur.\ Phys.\ J.\ C {\bf 9} (1999) 343
  [hep-ph/9812472].
  %%CITATION = doi:10.1007/s100529900006, 10.1007/s100520050537;%%

%66
\bibitem{Degrassi:2002fi}
  G.~Degrassi, S.~Heinemeyer, W.~Hollik, P.~Slavich and G.~Weiglein,
  Eur.\ Phys.\ J.\ C {\bf 28} (2003) 133
  [hep-ph/0212020].
  %%CITATION = doi:10.1140/epjc/s2003-01152-2;%%

%67
\bibitem{Frank:2006yh}
  M.~Frank, T.~Hahn, S.~Heinemeyer, W.~Hollik, H.~Rzehak and G.~Weiglein,
  JHEP {\bf 0702} (2007) 047
  [hep-ph/0611326].
  %%CITATION = doi:10.1088/1126-6708/2007/02/047;%%

%68
\bibitem{Hahn:2013ria}
  T.~Hahn, S.~Heinemeyer, W.~Hollik, H.~Rzehak and G.~Weiglein,
  Phys.\ Rev.\ Lett.\  {\bf 112} (2014) no.14,  141801
  [arXiv:1312.4937].
  %%CITATION = doi:10.1103/PhysRevLett.112.141801;%%

%69
\bibitem{Bahl:2016brp}
  H.~Bahl and W.~Hollik,
  Eur.\ Phys.\ J.\ C {\bf 76} (2016) no.9,  499
  [arXiv:1608.01880].
  %%CITATION = doi:10.1140/epjc/s10052-016-4354-8;%%

%70
\bibitem{Bahl:2017aev}
  H.~Bahl, S.~Heinemeyer, W.~Hollik and G.~Weiglein,
  Eur.\ Phys.\ J.\ C {\bf 78} (2018) no.1,  57
  [arXiv:1706.00346].
  %%CITATION = doi:10.1140/epjc/s10052-018-5544-3;%%

%71
\bibitem{Banks:1987iu}
  T.~Banks,
  Nucl.\ Phys.\ B {\bf 303} (1988) 172.
  %%CITATION = doi:10.1016/0550-3213(88)90222-2;%%

%72
\bibitem{Hall:1993gn}
  L.J.~Hall, R.~Rattazzi, and U.~Sarid,
  Phys.\ Rev.\ D {\bf 50} (1994) 7048
  [hep-ph/9306309].
  %%CITATION = doi:10.1103/PhysRevD.50.7048;%%

%73
\bibitem{Hempfling:1993kv}
  R.~Hempfling,
  Phys.\ Rev.\ D {\bf 49} (1994) 6168.
  %%CITATION = doi:10.1103/PhysRevD.49.6168;%%

%74
\bibitem{Carena:1994bv}
  M.~Carena, M.~Olechowski, S.~Pokorski, and C.E.M.~Wagner,
  Nucl.\ Phys.\ B {\bf 426} (1994) 269
  [hep-ph/9402253].
  %%CITATION = doi:10.1016/0550-3213(94)90313-1;%%

%75
\bibitem{Carena:1999py}
  M.~Carena, D.~Garcia, U.~Nierste, and C.E.M.~Wagner,
  Nucl.\ Phys.\ B {\bf 577} (2000) 88
  [hep-ph/9912516].
  %%CITATION = doi:10.1016/S0550-3213(00)00146-2;%%

%76
\bibitem{Carena:2000uj}
  M.~Carena, D.~Garcia, U.~Nierste, and C.E.M.~Wagner,
  Phys.\ Lett.\ B {\bf 499} (2001) 141
  [hep-ph/0010003].
  %%CITATION = doi:10.1016/S0370-2693(01)00009-0;%%

%77
\bibitem{fhwebpage}
  \url{http://feynhiggs.de}

%78
\bibitem{delAguila:2000aa}
  F.~del Aguila, M.~Perez-Victoria, and J.~Santiago,
  Phys.\ Lett.\ B {\bf 492} (2000) 98
  [hep-ph/0007160].
  %%CITATION = doi:10.1016/S0370-2693(00)01071-6;%%

%79
\bibitem{delAguila:2000rc}
  F.~del Aguila, M.~Perez-Victoria, and J.~Santiago,
  JHEP {\bf 0009} (2000) 011
  [hep-ph/0007316].
  %%CITATION = doi:10.1088/1126-6708/2000/09/011;%%

%80
\bibitem{Christensen:2008py}
  N.D.~Christensen and C.~Duhr,
  Comput.\ Phys.\ Commun.\  {\bf 180} (2009) 1614
  [arXiv:0806.4194].
  %%CITATION = doi:10.1016/j.cpc.2009.02.018;%%

%81
\bibitem{Alloul:2013bka}
  A.~Alloul, N.D.~Christensen, C.~Degrande, C.~Duhr, and B.~Fuks,
  Comput.\ Phys.\ Commun.\  {\bf 185} (2014) 2250
  [arXiv:1310.1921].
  %%CITATION = doi:10.1016/j.cpc.2014.04.012;%%

%82
\bibitem{Contino:2013kra}
  R.~Contino, M.~Ghezzi, C.~Grojean, M.~M{\"u}hlleitner, and M.~Spira,
  JHEP {\bf 1307} (2013) 035
  [arXiv:1303.3876].
  %%CITATION = doi:10.1007/JHEP07(2013)035;%%

%83
\bibitem{Dawson:1998py}
  S.~Dawson, S.~Dittmaier, and M.~Spira,
  Phys.\ Rev.\ D {\bf 58} (1998) 115012
  [hep-ph/9805244].
  %%CITATION = doi:10.1103/PhysRevD.58.115012;%%

%84
\bibitem{Alloul:2013naa}
  A.~Alloul, B.~Fuks, and V.~Sanz,
  JHEP {\bf 1404} (2014) 110
  [arXiv:1310.5150].
  %%CITATION = doi:10.1007/JHEP04(2014)110;%%

%85
\bibitem{Zhang:2014rja}
  C.~Zhang,
  Phys.\ Rev.\ D {\bf 90} (2014) no.1,  014008
  [arXiv:1404.1264].
  %%CITATION = doi:10.1103/PhysRevD.90.014008;%%

%86
\bibitem{Martin:2009iq}
  A.D.~Martin, W.J.~Stirling, R.S.~Thorne, and G.~Watt,
  Eur.\ Phys.\ J.\ C {\bf 63} (2009) 189
  [arXiv:0901.0002].
  %%CITATION = doi:10.1140/epjc/s10052-009-1072-5;%%

%87
\bibitem{Harlander:2018yns}
  R.V.~Harlander, J.~Klappert, C.~Pandini and A.~Papaefstathiou,
  arXiv:1804.02299 [hep-ph].
  %%CITATION = ARXIV:1804.02299;%%

%88
\bibitem{lhapdfwebpage}    
  \url{https://lhapdf.hepforge.org/}

%89
\bibitem{Hahn:2004fe}
  T.~Hahn,
  Comput.\ Phys.\ Commun.\  {\bf 168} (2005) 78
  [hep-ph/0404043].
  %%CITATION = doi:10.1016/j.cpc.2005.01.010;%%

%90
\bibitem{cubawebpage}  
  \url{http://www.feynarts.de/cuba/}

%91
\bibitem{vanOldenborgh:1989wn}
  G.J.~van Oldenborgh and J.A.M.~Vermaseren,
  Z.\ Phys.\ C {\bf 46} (1990) 425.
  %%CITATION = doi:10.1007/BF01621031;%%

%92
\bibitem{looptoolswebpage}    
  \url{http://www.feynarts.de/looptools/}

%93
\bibitem{mcfmwebpage}
  \url{https://mcfm.fnal.gov/}

%94
\bibitem{2hdmcwebpage}
  \url{http://2hdmc.hepforge.org}

%95
\bibitem{Allanach:2001kg}
  B.C.~Allanach,
  Comput.\ Phys.\ Commun.\  {\bf 143} (2002) 305
  [hep-ph/0104145].
  %%CITATION = doi:10.1016/S0010-4655(01)00460-X;%%

%96
\bibitem{Porod:2003um}
  W.~Porod,
  Comput.\ Phys.\ Commun.\  {\bf 153} (2003) 275
  [hep-ph/0301101].
  %%CITATION = doi:10.1016/S0010-4655(03)00222-4;%%

%97
\bibitem{Porod:2011nf}
  W.~Porod and F.~Staub,
  Comput.\ Phys.\ Commun.\  {\bf 183} (2012) 2458
  [arXiv:1104.1573].
  %%CITATION = doi:10.1016/j.cpc.2012.05.021;%%

%98
\bibitem{Athron:2014yba}
  P.~Athron, J.H.~Park, D.~St{\"o}ckinger, and A.~Voigt,
  Comput.\ Phys.\ Commun.\  {\bf 190} (2015) 139
  [arXiv:1406.2319].
  %%CITATION = doi:10.1016/j.cpc.2014.12.020;%%

%99
\bibitem{Athron:2016fuq}
  P.~Athron, J.H.~Park, T.~Steudtner, D.~St\"ockinger, and A.~Voigt,
  JHEP {\bf 1701} (2017) 079
  [arXiv:1609.00371].
  %%CITATION = doi:10.1007/JHEP01(2017)079;%%

%100
\bibitem{Athron:2017fvs}
  P.~Athron, M.~Bach, D.~Harries, T.~Kwasnitza, J.H.~Park, D.~St{\"o}ckinger, A.~Voigt, and J.~Ziebell,
  arXiv:1710.03760.
  %%CITATION = ARXIV:1710.03760;%%

%101
\bibitem{Harlander:2017kuc}
  R.V.~Harlander, J.~Klappert, and A.~Voigt,
  Eur.\ Phys.\ J.\ C {\bf 77} (2017) no.12,  814
  [arXiv:1708.05720].
  %%CITATION = doi:10.1140/epjc/s10052-017-5368-6;%%

%102
\bibitem{Harlander:2008ju}
  R.V.~Harlander, P.~Kant, L.~Mihaila, and M.~Steinhauser,
  Phys.\ Rev.\ Lett.\  {\bf 100} (2008) 191602
   [Phys.\ Rev.\ Lett.\  {\bf 101} (2008) 039901]
  [arXiv:0803.0672].
  %%CITATION = doi:10.1103/PhysRevLett.101.039901, 10.1103/PhysRevLett.100.191602;%%

%103
\bibitem{Kant:2010tf}
  P.~Kant, R.V.~Harlander, L.~Mihaila, and M.~Steinhauser,
  JHEP {\bf 1008} (2010) 104
  [arXiv:1005.5709].
  %%CITATION = doi:10.1007/JHEP08(2010)104;%%


\end{thebibliography}
\end{document}